\documentclass[aps,twocolumn,pra,superscriptaddress,articlepaper,nofootinbib,longbibliography]{revtex4-2}
\usepackage{physics}
\usepackage{mathtools}
\usepackage{amsmath, amsthm, amssymb}
\usepackage{upgreek}
\usepackage{booktabs}
\usepackage{url}
\usepackage{float}
\usepackage{enumitem}
\AtBeginDocument{
\heavyrulewidth=.08em
\lightrulewidth=.05em
\cmidrulewidth=.03em
\belowrulesep=.65ex
\belowbottomsep=0pt
\aboverulesep=.4ex
\abovetopsep=0pt
\cmidrulesep=\doublerulesep
\cmidrulekern=.5em
\defaultaddspace=.5em
}

\usepackage{txfonts}
\usepackage[pdftex]{graphicx}
\usepackage[usenames, dvipsnames, svgnames, table]{xcolor}
\usepackage[colorlinks=true, citecolor=Blue,linkcolor=BrickRed, urlcolor=magenta]{hyperref}
\usepackage{bm}
\usepackage{placeins} % To use \FloatBarrier to output all floats which are not yet typeset
\usepackage{setspace}
\usepackage[final]{changes}
\usepackage{multirow}
\usepackage{rotating}
\usepackage[normalem]{ulem}
\usepackage{soul}

\usepackage{cleveref}
\crefname{section}{Section}{Sections}
\crefname{appendix}{Appendix}{Appendices}
\crefname{figure}{Figure}{Figures}
\crefname{table}{Table}{Tables}
\crefname{equation}{Eq.}{Eqs.}
\crefname{assumption}{Assumption}{Assumptions}

\usepackage{siunitx}

\newcommand{\mrm}[1]{\mathrm{#1}}

\newcommand{\mcal}[1]{\mathcal{#1}}

\newcommand{\prtHz}{\per\sqrt{\mrm\Hz}}

\newcommand{\e}{\mathrm{e}} %roman exp
\begin{document}
% \title{Systematic search of laser modulation noise coupling in heterodyne interferometry and its application to time-delay interferometry}
% \title{Systematic search of laser and phase modulation noise coupling in heterodyne interferometry}
\title{Searching systematically for coupling of laser and phase-modulation noise in heterodyne interferometry}

\author{Kohei Yamamoto}
\email{y9m9k0h@gmail.com}
\affiliation{Center for Space Sciences and Technology, University of Maryland, Baltimore County, 1000 Hilltop Circle, Baltimore, MD 21250, USA}
\affiliation{NASA Goddard Space Flight Center, 8800 Greenbelt Road, Greenbelt, MD 20771, USA}
% \affiliation{Gravitational Astrophysics Lab, NASA/GSFC, 8800 Greenbelt Road, Greenbelt, MD 20771, USA}
\affiliation{Center for Research and Exploration in Space Science and Technology, NASA/GSFC, 8800 Greenbelt Road, Greenbelt, MD 20771, USA}

\author{Olaf Hartwig}
\affiliation{Max-Planck-Institut f\"ur Gravitationsphysik (Albert-Einstein-Institut), Callinstra\ss e 38, 30167 Hannover, Germany}
\affiliation{Leibniz Universität Hannover, Institut für Gravitationsphysik, Callinstra\ss e 38, 30167 Hannover, Germany}

\author{Lennart Wissel}
\affiliation{Max-Planck-Institut f\"ur Gravitationsphysik (Albert-Einstein-Institut), Callinstra\ss e 38, 30167 Hannover, Germany}
\affiliation{Leibniz Universität Hannover, Institut für Gravitationsphysik, Callinstra\ss e 38, 30167 Hannover, Germany}

\author{Holly Leopardi}
\affiliation{NASA Goddard Space Flight Center, 8800 Greenbelt Road, Greenbelt, MD 20771, USA}
% \affiliation{Gravitational Astrophysics Lab, NASA/GSFC, 8800 Greenbelt Road, Greenbelt, MD 20771, USA}

% \author{Christoph Bode}
% \affiliation{Max-Planck-Institut f\"ur Gravitationsphysik (Albert-Einstein-Institut), Callinstra\ss e 38, 30167 Hannover, Germany}
% \affiliation{Leibniz Universität Hannover, Institut für Gravitationsphysik, Callinstra\ss e 38, 30167 Hannover, Germany}

\author{Kenji Numata}
\affiliation{NASA Goddard Space Flight Center, 8800 Greenbelt Road, Greenbelt, MD 20771, USA}
% \affiliation{Lasers and Electro-optics Branch, NASA/GSFC, 8800 Greenbelt Road, Greenbelt, MD 20771, USA}

\author{Ryan Derosa}
\affiliation{NASA Goddard Space Flight Center, 8800 Greenbelt Road, Greenbelt, MD 20771, USA}
% \affiliation{Gravitational Astrophysics Lab, NASA/GSFC, 8800 Greenbelt Road, Greenbelt, MD 20771, USA}

\begin{abstract}
Heterodyne interferometry for precision science often comes with an optical phase modulation, for example, for intersatellite clock noise transfer for gravitational wave (GW) detectors in space, exemplified by the Laser Interferometer Space Antenna (LISA).
The phase modulation potentially causes various noise couplings to the final phase extraction of heterodyne beatnotes by a phasemeter.
In this paper, in the format of space-based GW detectors, we establish an analytical framework to systematically search for the coupling of various noises from the heterodyne and modulation frequency bands, which are relatively unexplored so far.
In addition to the noise caused by the phase modulation, the high-frequency laser phase noise is also discussed in the same framework.
The analytical result is also compared with a numerical experiment to confirm that our framework successfully captures the major noise couplings.
We also demonstrate a use case of this study by taking the LISA-like parameters as an example, which enables us to derive requirements on the level of the laser and phase modulation noises in the high frequency regimes.
\end{abstract}

\maketitle

%%%%%%%%%%%%%%%%%%%%%%%%%%%%%%%%%%%%%%%%%%%%%%%%%%%%%%%%%
\section{Introduction}\label{sec:intro}
Heterodyne interferometry is an optical technique widely used in precision science~\cite{LPF2016,GRACEFO2019,Shaddock2007,Numata2008,DongNguyen2020}.
Essential ingredients are two electromagnetic fields with a frequency offset, the so-called heterodyne frequency, a beam combiner to generate their optical beatnote at the heterodyne frequency, a photoreceiver (PR) to convert the optical power to an electronic voltage signal, and a phasemeter to measure the phase of the heterodyne beatnote.
The interferometer used as a transducer reports the observable of interest in an experimental setup, often spatial displacement, via the phase signal with very high precision and/or accuracy.

This technique is often combined with an optical modulation to transfer or acquire additional information, for example, intersatellite clock synchronization or absolute ranging for spaceborne gravitational-wave detectors like the Laser Interferometer Space Antenna (LISA)~\cite{LisaRed}, TianQin~\cite{TianQin}, or Taiji~\cite{Taiji}.
LISA aims to observe gravitational waves (GWs) in the millihertz regime by configuring interspacecraft heterodyne interferometers with three spacecraft separated by around 2.5\,million kilometer.
For the purpose of differential clock noise reduction, LISA phase-modulates an outgoing beam with an onboard system clock around \SI{2.4}{\giga\Hz} and extracts the phase of the heterodyne beatnotes not only between beam carriers that have the information of GWs but also between the clock sidebands in which the onboard clock difference is encoded~\cite{Hartwig2021,Yamamoto2022,Yang2023}.
All heterodyne frequencies in the constellation are adjusted within \SIrange{5}{30}{\mega\Hz} via laser offset lock, so-called frequency planning~\cite{Heinzel2024,Zhang2022,Li2026}.

Since we modulate the phase of the primary laser carrier, this could naturally cause additional noise couplings in the metrology chain.
From the perspective of the clock noise transfer, the coupling of the noise in the observation band, namely \SI{0.1}{\milli\Hz} to \SI{1}{\Hz}, has been studied in great depth~\cite{Barke2010,Numata2023,Zeng2023,Xu2024}.
However, the working principle of the phasemeter is known to also cause the unavoidable coupling of noises at the frequencies related to heterodyne beatnotes, e.g., the integer multiple of the heterodyne frequency~\cite{Wissel2022,Wissel2023,BodePhD}.
The combination of the multiple heterodyne beatnotes and the phasemeter architecture in LISA could generate numerous noise couplings from the heterodyne band.

In this paper, we systematically search for the coupling of modulation noises above the GW observation band to the final phase extraction of individual heterodyne beatnotes by the phasemeter.
In addition to the noise caused by phase modulation, the coupling of the laser phase noise is also discussed in the same analytical framework.
The noise frequency band of interest is not only the heterodyne band but also the modulation band (the MHz and GHz regime for LISA, respectively).
The noise couplings found analytically are compared with those found by numerical simulation to confirm that our analytical framework successfully captures major noise contributions.
Finally, we demonstrate a use case of this study by applying it to the LISA-like set of requirements, which enables us to derive new requirements on the high-frequency noise sources in the system.

%%%%%%%%%%%%%%%%%%%%%%%%%%%%%%%%%%%%%%%%%%%%%%%%%%%%%%%%%
\section{Framework}\label{sec:framework}
This section sets up a basic framework, which will then be used for systematic search for noise couplings in the following sections.
In \cref{sub:framework_modulation}, we first formulate a modulation signal with different noises (relative amplitude noise $a_i$, phase noise $\theta_i$, and additive noise $v_i$).
In \cref{sub:framework_beam}, we then model the phase-modulated electromagnetic field representing a laser beam to the first-order modulation sidebands: an upper sideband (usb) and a lower sideband (lsb).
In \cref{sub:framework_beatnote}, a PR signal with three heterodyne beatnotes (carrier-carrier, usb-usb, and lsb-lsb beatnotes) is described.
A general expression of readout noise at beatnote phase extraction by a phasemeter in \cref{sub:framework_readout} will be applied to the PR signal throughout the paper.
In \cref{sub:framework_strategy}, we wrap up this section by clarifying our strategy of the search for the noise coupling based on the framework.

\subsection{Modulation signal}\label{sub:framework_modulation}
We consider the case of modulating a beam phase with a sinusoidal signal.
Let us formulate the modulation signal $V_i$ in units of voltage that drives an electro-optic modulator (EOM) as
\begin{align}
    V_i(\tau) &= a_0(1+a_i(\tau))\cos(\omega_{\mrm{m},i}\tau + \phi^\mrm{sb}_{\mrm{tar},i}(\tau) + \theta_i(\tau)) + v_i(\tau),
    \label{eq:Vmi}
\end{align}
where $a_0$ is a nominal modulation amplitude;
$\omega_{\mrm{m},i}$ is an angular frequency of the sinusoidal modulation signal, related to the frequency $f_{\mrm{m},i}$ by $\omega_{\mrm{m},i}=2\pi f_{\mrm{m},i}$;
$\phi^\mrm{sb}_{\mrm{tar},i}$ is the target phase to encode onto modulation sidebands;
% $\phi_{m,i}$ is a nominal modulation phase, which is, for example, a clock noise for LISA;
$a_i$ is a modulation relative amplitude noise;
$\theta_i$ is a modulation additive phase noise;
$v_i$ is a modulation additive voltage noise.
$i$ is an index to label a modulation signal and an associated laser beam ($i\in \{i,j\}$).

%%%%%%%%%%%%%%%%%%%%%%%%%%%%%%%%%%%%%%%%%%%%%%%%%%%%%%%%%
\subsection{Laser beam}\label{sub:framework_beam}
A non-modulated electromagnetic field representing a laser beam can be formulated as
\begin{align}
    E_i = A_i\cdot \exp(j\omega_i\tau + j\phi^\mrm{car}_{\mrm{tar},i}(\tau) + jp_i(\tau)),
    \label{eq:Ei}
\end{align}
where $A_i$ is a nominal amplitude;
$\omega_i$ is a laser angular frequency, related to the frequency $f_i$ by $\omega_i=2\pi f_i$;
$\phi^\mrm{car}_{\mrm{tar},i}$ is the target phase that we aim to measure with a phasemeter at the end;
$p_i$ is a laser phase noise in radian.
In this paper, we assume that $A_i$ is constant.
The noise coupling of the beam amplitude, caused by the relative intensity noise (RIN), can be found in~\cite{Wissel2022,Wissel2023}, but also, we discuss the relation of $a_i$ to laser RIN in \cref{app:rin}.

Next, we phase-modulate the beam in \cref{eq:Ei} using the modulation signal in \cref{eq:Vmi}.
We consider the EOM as a pure linear converter of an input voltage to a phase with a conversion factor of $\pi/V_\mrm{\pi}$
% \begin{align}
%     C_\mrm{eom} &= \pi/V_\mrm{\pi},
%     \label{eq:Ceom}
% \end{align}
where $V_\mrm{\pi}$ is determined by the EOM specification.
Therefore, the actual modulation phase becomes
\begin{align}
    \phi_i(\tau) &= \frac{\pi}{V_\mrm{\pi}}V_i(\tau).
    \label{eq:phi_i}
\end{align}
As a result, the phase-modulated laser beam reads as follows:
\begin{align}
    E_i &= A_i\cdot \exp(j\omega_i\tau + j\phi^\mrm{car}_{\mrm{tar},i} + jp_i + j\phi_i)
    \nonumber\\
    &= A_i\cdot \exp\left(j\omega_i\tau + j\phi^\mrm{car}_{\mrm{tar},i} + j(p_i+ n_i) \right.
    \nonumber\\
    &\hspace{15mm}\left. + jm_i\cos(\omega_{\mrm{m},i}\tau + \phi^\mrm{sb}_{\mrm{tar},i} + \theta_i)\right),
    \label{eq:Ei_phim}
\end{align}
since we define
\begin{align}
    m_i &= m_0(1+a_i),
    \label{eq:m_i}\\
    m_0 &= \frac{\pi}{V_\mrm{\pi}}a_0,
    \label{eq:m_0}\\
    n_i &= \frac{\pi}{V_\mrm{\pi}}v_i.
    \label{eq:n_i}
\end{align}
where $m_0$ is a modulation depth in radian.
Here and throughout the rest of the paper, we omit the time arguments for simplicity.

We apply the Jacobi-Anger expansion to \cref{eq:Ei_phim}, to first-order modulation sidebands,
\begin{align}
    E_i &= A_i \cdot \exp\left(j\omega_i\tau + j\phi^\mrm{car}_{\mrm{tar},i} + j(p_i  + n_i)\right)
    \nonumber\\
    &\hspace{0mm}\cdot\left(J_0(m_i) + jJ_1(m_i)\e^{j(\omega_{\mrm{m},i}\tau + \phi^\mrm{sb}_{\mrm{tar},i} + \theta_i)} + jJ_1(m_i)\e^{-j(\omega_{\mrm{m},i}\tau + \phi^\mrm{sb}_{\mrm{tar},i} + \theta_i)}\right),
    \label{eq:Ei_phim_sb}
\end{align}
where $J_n(x)$ is the Bessel function of the first kind.
% For now, we stop the expansion to the first-order modulation sidebands, which result in optical beatnotes whose phase to extract as is described below.
Noise couplings due to the higher-order modulation sidebands would not be dominant; however, we particularly discuss the noise coupling caused by the second-order modulation sidebands in \cref{sec:mod_band} assuming that the nominal modulation depth $m_0$ is not so small.

It is obvious that the laser phase noise $p_i$ and the modulation additive voltage noise $n_i$ are indistinguishable.
Hence, for the rest of this paper, we only consider the laser phase noise $p_i$.
However, it is important to note that any result of $p_i$ also applies to $n_i$.

%%%%%%%%%%%%%%%%%%%%%%%%%%%%%%%%%%%%%%%%%%%%%%%%%%%%%%%%%
\subsection{Beatnote signal}\label{sub:framework_beatnote}
We finally formulate the interference between two laser beams $E_i$ and $E_j$.
The PR output signal $s_\mrm{pr}$ is expressed below in units of W with some constant factors omitted, such as a responsivity (A/W) and a transimpedance-amplifier gain (V/A):
\begin{widetext}
    \begin{align}
        s_\mrm{pr} &= |E_i + E_j|^2/2
        \nonumber\\
        &= (|E_i|^2 + |E_j|^2)/2 + \Re[E_iE^*_j]
        \nonumber\\
        &= (P_i + P_j)/2 + A_iA_j
        \nonumber\\
        &\hspace{5mm} \cdot \Re\left[\exp(j\omega_\mrm{het}\tau + \phi^\mrm{car}_\mrm{tar} + jp_\Delta)\cdot\left(J_0(m_i)J_0(m_j) + J_1(m_i)J_1(m_j)\e^{j(\Delta\omega_\mrm{m}\tau + \phi^\mrm{sb}_\mrm{tar} + \theta_\Delta)} + J_1(m_i)J_1(m_j)\e^{-j(\Delta\omega_\mrm{m}\tau + \phi^\mrm{sb}_\mrm{tar} + \theta_\Delta)}\right)\right]
        \nonumber\\
        &= (P_i + P_j)/2
        \nonumber\\
        &\hspace{5mm} + A_iA_jJ_0(m_i)J_0(m_j)\cos\left(\omega_\mrm{het}\tau + \phi^\mrm{car}_\mrm{tar} + p_\Delta\right)
        \nonumber\\
        &\hspace{5mm} + A_iA_jJ_1(m_i)J_1(m_j)\cos\left((\omega_\mrm{het}+\Delta\omega_\mrm{m})\tau + \phi^\mrm{usb}_\mrm{tar} + p_\Delta + \theta_\Delta\right)
        \nonumber\\
        &\hspace{5mm} + A_iA_jJ_1(m_i)J_1(m_j)\cos\left((\omega_\mrm{het}-\Delta\omega_\mrm{m})\tau - \phi^\mrm{lsb}_\mrm{tar} + p_\Delta - \theta_\Delta\right),
        \label{eq:spr}
    \end{align}
\end{widetext}
where the target phase differences to measure are defined as follows:
\begin{align}
    \phi^x_\mrm{tar} &= \phi^x_{\mrm{tar},i}-\phi^x_{\mrm{tar},j}\hspace{5mm}x\in\{\mrm{car},\mrm{sb}\},
    \label{eq:phixtar}\\
    \phi^\mrm{usb}_\mrm{tar} &= \phi^\mrm{sb}_\mrm{tar} + \phi^\mrm{car}_\mrm{tar},
    \label{eq:phiusbtar}\\
    \phi^\mrm{lsb}_\mrm{tar} &= \phi^\mrm{sb}_\mrm{tar} - \phi^\mrm{car}_\mrm{tar},
    \label{eq:philsbtar}
\end{align}
\cref{eq:spr} plays a pivotal role throughout this paper.
For simplicity, the same physical quantities from individual beams are bundled as its difference as follows:
$P_i = A^2_i$: single beam power;
$\omega_\mrm{het} = \omega_i - \omega_j$: heterodyne frequency;
$\Delta\omega_\mrm{m} = \omega_{\mrm{m},i} - \omega_{\mrm{m},j}$: modulation frequency offset (e.g., nominally around \SI{1}{\mega\Hz} in LISA);
$p_\Delta = p_i - p_j$: laser phase noise difference;
% $n_\Delta = n_i - n_j$: modulation additive noise difference;
$\theta_\Delta = \theta_i - \theta_j$: modulation phase noise difference.
Note that throughout the paper, we assume $\Delta\omega_\mrm{m}>0$: $\omega_{\mrm{m},i}>\omega_{\mrm{m},j}$.

The PR signal in \cref{eq:spr} has four constituents: non-interfering components, a carrier-carrier beatnote, a usb-usb beatnote, and a lsb-lsb beatnote.
As we model an EOM as a pure phase modulator, neglecting residual amplitude modulation, any modulation noise never changes the laser total power, and correspondingly also not the power of any non-interfering components as well.
Therefore, we discuss only the interfering components for the rest of the paper.

\subsection{Strategy}\label{sub:framework_strategy}
Let us wrap up this section by providing a general description of our strategy to analyze the noise coupling.

First, following~\cite{Wissel2022}, we model a given noise $x_i$ as a monotonic tone at the angular frequency of $\omega_\mrm{n}$ as
\begin{align}
    x_i(\tau;\omega_\mrm{n}) &= m_x \cos(\omega_\mrm{n} \tau + \rho),
    \label{eq:x_i}
\end{align}
where $m_x$ is the noise (long-term average) amplitude and $\rho$ is a given random phase.
We will only consider the first-order coupling of $m_x$.
As discussed later, some noises are expected to have coherence; therefore, it is important to base different noises on the consistent form.
In contrast to the time-domain analysis, the noises have to be evaluated in amplitude spectral density (ASD) at the end to project it onto a setup sensitivity, which is usually expressed in the frequency domain.
As presented in \cref{app:conversion}, we can easily translate the time-domain coupling factor to that in ASD by multiplying by a factor of $\sqrt{2}$; therefore, we will stick to the time-domain analysis.

% Second, following \cref{sub:framework_readout} that describes the phasemeter readout noise, we commonly tackle individual noise couplings aiming to end up with the form in \cref{eq:s_add}, which we recall here:
Second, \cref{sub:framework_readout} derives the phasemeter readout noise for the input signal in \cref{eq:s_add}, which we cite here:
\begin{align}
    s = k\cos(\omega \tau + \phi_k) + l\cos\left((\omega + \epsilon)\tau + \phi_l\right),
    \label{eq:s_add_cite}
\end{align}
where $\epsilon\ll\omega$ and $l \ll k$.
According to \cref{eq:IQdemod,eq:IQerr}, when the phasemeter extracts the target phase $\phi_k$, the actual phasemeter readout $\phi_\mathrm{read}$ is affected by the second monotonic noise term as:
\begin{align}
    \phi_\mathrm{read} &\approx \phi_k + l/k\cdot\sin(\epsilon\tau + \phi_l-\phi_k).
    \label{eq:IQdemod_cite}
\end{align}
The noise appears as the tone at $\epsilon$ in the observation band.
Hence, to analyze individual noise sources and their coupling to the phasemeter readout, we first transform equations to the same algebraic structure as \cref{eq:s_add_cite}, second derive the condition on the noise frequency $\omega_\mrm{n}$ that gives the noise term the frequency of $\omega+\epsilon\approx\omega$ with $\epsilon$ neglected for simplicity, and third apply \cref{eq:IQdemod_cite} to determine the coupled noise.

\begin{figure}
    \centering
    \includegraphics[width=8.6cm]{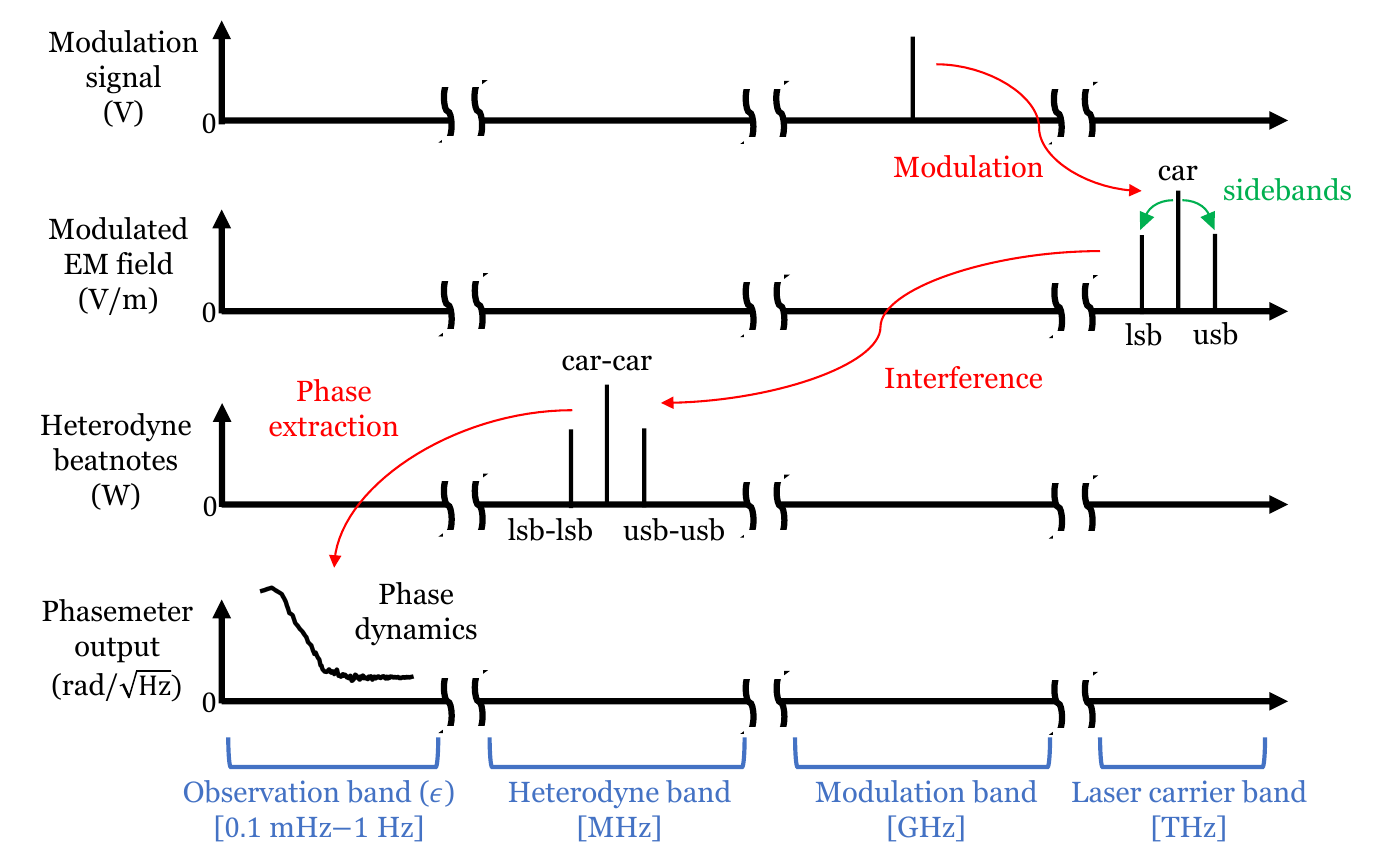}
    \caption{Illustration of the frequency bands discussed in this paper.
    The respective frequency regimes in the LISA case are written in brackets.
    Throughout the paper, $\epsilon$ represents a (angular) frequency in the observation band, phase in which is to be extracted by a phasemeter.
    Note that noises are not shown here.
    }
    \label{fig:fbands}
\end{figure}

Third, the whole noise analysis will be categorized into two parts in terms of frequency regime: heterodyne-band noise in \cref{sec:het_band} and modulation-band noise in \cref{sec:mod_band}.
The former is a noise around the heterodyne frequency $\omega_\mrm{het}$ (namely, the MHz regime for LISA), while the latter is a noise around the modulation frequency $\omega_{\mrm{m},i}$ (namely, the GHz regime for LISA).
We will show, in \cref{sec:mod_band}, that the modulation-band noise can be virtually down-converted to the heterodyne frequency band, with which the result of the heterodyne-band noise in \cref{sec:het_band} can be combined.
In \cref{fig:fbands}, we illustrate the relation between the frequency bands.

Finally, a certain type of noise ($p_i$, $a_i$, $\theta_i$, or $n_i$) has incoherent contributions from individual laser beams in an indistinguishable way, as shown in \cref{eq:spr}. Hence, we can consider only one of the laser contributions without the loss of generality. To estimate the total coupling factor, we can simply multiply the resulting noise coupling factor by $\sqrt{2}$ representing the incoherent sum between the two beams.
Note that actual space GW detectors frequency-lock one laser to another~\cite{Heinzel2024,Zhang2022,Li2026}; however, the locking bandwidth is typically dozens kilohertz, while most of the noise we discuss here is in the megahertz or gigahertz regime. Hence, we expect that the incoherent-noise assumption between lasers would still be valid even with laser locking for the coupling-factor estimation.

%%%%%%%%%%%%%%%%%%%%%%%%%%%%%%%%%%%%%%%%%%%%%%%%%%%%%%%%%
\section{Heterodyne-band noises}\label{sec:het_band}
\begin{figure}
    \centering
    \includegraphics[width=8.6cm]{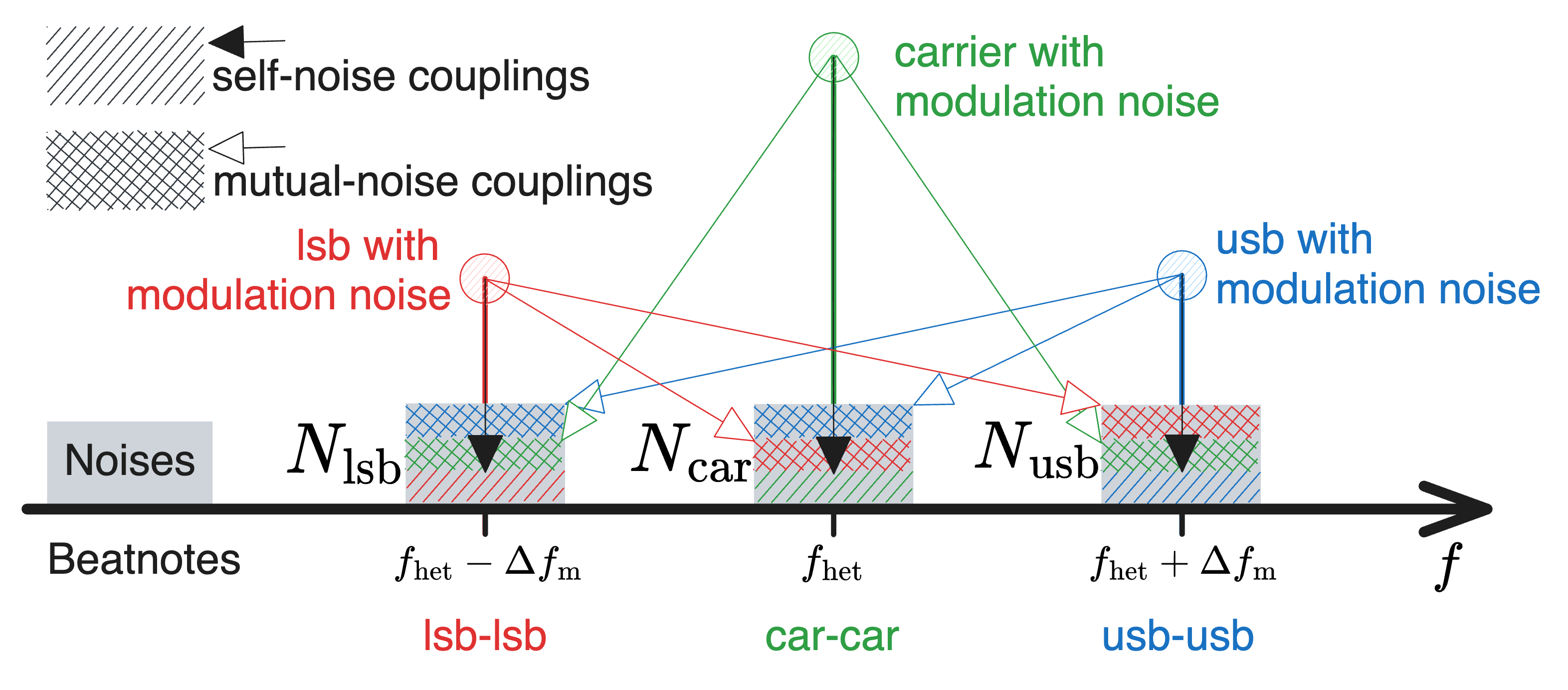}
    \caption{Illustration of the heterodyne-band noises from \cref{sec:het_band}. It shows the different frequency bands and noise couplings into the various beatnotes. Note that the noises, labeled $N_\text{lsb/car/usb}$ (with their width defined by the phasemeter bandwith) and representing $a_i, p_i, \theta_i\, \&\, n_i$, are schematically shown as an additive noise floor, whilst those (except for $a_i$, as visible in \cref{eq:spr}) are actually direct phase noises that can be expanded in frequency by the Jacobi-Anger expressions as explained in the corresponding  \cref{sub:het_band_self,sub:het_band_mutual}.}
    \label{fig:het-band-noises-illustration}
\end{figure}

As shown in \cref{eq:spr}, we have the three heterodyne beatnotes: the carrier-carrier beatnote at $\omega_\mrm{het}$ in the middle, and the usb-usb and lsb-lsb beatnotes symmetrically located around the carrier-carrier beatnote with the frequency offset of $\Delta \omega_\mrm{m}$.
In this section, we systematically derive noise couplings from the heterodyne band based on interactions among those beatnotes, and we categorize the coupling mechanisms into two for this purpose: \emph{self noise coupling} and \emph{mutual noise coupling}.
In the self noise coupling in \cref{sub:het_band_self}, we discuss the coupling of a noise in a certain beatnote to the phase extraction of itself.
On the other hand, the mutual noise coupling in \cref{sub:het_band_mutual} represents the coupling of a noise in a certain beatnote to the phase extraction of another beatnote.
In addition to the heterodyne-band noise, noise in the observation band will also be incorporated in the same framework. \Cref{fig:het-band-noises-illustration} shows an illustration for these two types of noise couplings.
All results are summarized in \cref{tab:noise_couplings} at the end.

\subsection{Self noise coupling}\label{sub:het_band_self}
We first provide a general description and then apply it to the individual beatnotes.
\\\\
\emph{General description: }
The self noise coupling considers only a single beatnote with amplitude and phase noise.
Hence, the signal form in the scope is
\begin{align}
    s &= A_\mrm{n}\cos(\omega_\mrm{A}\tau+\phi_\mrm{tar}+\phi_\mrm{n}),
    \label{eq:self_general}
\end{align}
where $A_\mrm{n} = A_0(1+a_\mrm{n})$ contains a nominal amplitude $A_0$ and a relative amplitude noise $a_\mrm{n}$, while $\phi_\mrm{n}$ is a phase noise added to a target phase $\phi_\mrm{tar}$.
$\omega_\mrm{A}$ is the beatnote frequency.
By definition, $\phi_\mrm{tar}$ is assumed to have a non-zero power only in the final observation band.

First, we derive a general coupling factor of $a_\mrm{n}$ by applying the monotonic noise representation in \cref{eq:x_i} as
\begin{align}
    A_\mrm{n} &= A_0(1+a_\mrm{n})
    \nonumber\\
    &= A_0(1+m_\mrm{n}\cos(\omega_\mrm{n}\tau+\rho)).
    \label{eq:A_n_self_general}
\end{align}
% The relative amplitude noise $a_\mrm{n}$ is replaced with a given noise amplitude $m_\mrm{n}$, a given angular frequency $\omega_\mrm{n}(>0)$, and a given random phase $\rho$.
If we neglect $\phi_\mrm{n}$ in \cref{eq:self_general} to focus on $A_\mrm{n}$ and substitute \cref{eq:A_n_self_general}, we get
\begin{widetext}
\begin{align}
    % s &= A_0(1+m_\mrm{n}\cos(\omega_\mrm{n}\tau+\rho))\cos(\omega_\mrm{A}\tau+\phi_\mrm{tar})
    % \nonumber\\
    s
    &= A_0\cos(\omega_\mrm{A}\tau+\phi_\mrm{tar}) + \frac{A_0m_\mrm{n}}{2}\left(\cos\left((\omega_\mrm{n}-\omega_\mrm{A})\tau-\phi_\mrm{tar}+\rho\right) + \cos\left((\omega_\mrm{n}+\omega_\mrm{A})\tau+\phi_\mrm{tar}+\rho\right)\right).
    \nonumber\\
    &\xrightarrow[\text{BPF at }\omega_A]{\omega_n= 2\omega_A} A_0\cos(\omega_\mrm{A}\tau+\phi_\mrm{tar}) + \frac{A_0m_\mrm{n}}{2}\cos\left(\omega_\mrm{A}\tau-\phi_\mrm{tar}+\rho\right).
    \label{eq:ampnoise_self_general}
\end{align}
\end{widetext}
Our interest is the noise term around the beat frequency $\omega_\mrm{A}$ as discussed in \cref{sub:framework_readout}, and only $\omega_\mrm{n}=2\omega_\mrm{A}$ results in an additional oscillating term at $\omega_\mrm{A}$.
Therefore, we substitute $\omega_n=2\omega_A$ and band-pass the signal around $\omega_A$ in the last line.
% As discussed in \cref{sub:framework_readout}, our interest is the noise term around $\omega_\mrm{A}$.
% Therefore, only $\omega_\mrm{n}=2\omega_\mrm{A}$, which results in an additional oscillating term at the beat frequency $\omega_\mrm{A}$, is important, since we define all (angular) frequencies by a positive value.
% Hence, the total signal bandpassed around $\omega_\mrm{A}$ reads
% \begin{align}
%     s_{\omega_\mrm{A}} &= A_0\cos(\omega_\mrm{A}\tau+\phi_\mrm{tar}) + \frac{A_0m_\mrm{n}}{2}\cos\left(\omega_\mrm{A}\tau-\phi_\mrm{tar}+\rho\right).
%     \label{eq:ampnoise_self_general_2wa}
% \end{align}
Because \cref{eq:ampnoise_self_general,eq:s_add_cite} have the same algebraic structure, the measured phase error can be derived by substituting $k=A_0$, $\phi_k=\phi_\mrm{tar}$, $l=\frac{A_0m_\mrm{n}}{2}$, and $\phi_l=\rho-\phi_\mrm{tar}$ into \cref{eq:IQdemod_cite},
\begin{align}
    \phi_\mrm{read} = \phi_\mrm{tar} + \frac{1}{2}m_\mrm{n}\sin(\rho-2\phi_\mrm{tar}).
    \label{eq:phi_read_self_ampnoise}
\end{align}
This is the same coupling mechanism as so-called ``2f-RIN" in the context of the laser RIN~\cite{Wissel2022}; see \cref{app:rin} for detailed discussions on the relation to laser RIN.

Second, we focus on the phase noise $\phi_\mrm{n}$ in \cref{eq:self_general} under $A_\mrm{n}=A_0$.
It is obvious from the simple addition of $\phi_\mrm{tar}+\phi_\mrm{n}$ in \cref{eq:self_general} that the noise power of $\phi_\mrm{n}$ in the observation band is indistinguishable from $\phi_\mrm{tar}$; hence, it appears the phase readout as is: $\cos\rho$ following the noise form in \cref{eq:x_i}.
To find other noise frequencies that have a coupling, let us transform \cref{eq:self_general} via Jacobi-Anger expansion as:
% \begin{widetext}
\begin{align}
    s &= A_0\cos(\omega_\mrm{A}\tau+\phi_\mrm{tar}+m_\mrm{n}\cos(\omega_\mrm{n}\tau+\rho))
    \nonumber\\
    % &= A_0\cos(\omega_\mrm{A}\tau+\phi_\mrm{tar})\cos(m_\mrm{n}\cos(\omega_\mrm{n}\tau+\rho))
    % \nonumber\\
    &\hspace{5mm}- A_0\sin(\omega_\mrm{A}\tau+\phi_\mrm{tar})\sin(m_\mrm{n}\cos(\omega_\mrm{n}\tau+\rho))
    \nonumber\\
    &\approx A_0J_0(m_\mrm{n})\cos(\omega_\mrm{A}\tau+\phi_\mrm{tar})
    \nonumber\\
    &\hspace{5mm}- 2A_0J_1(m_\mrm{n})\cos(\omega_\mrm{n}\tau+\rho)\sin(\omega_\mrm{A}\tau+\phi_\mrm{tar})
    \nonumber\\
    &\xrightarrow[\text{BPF at }\omega_A]{\omega_n= 2\omega_A} A_0J_0(m_\mrm{n})\cos(\omega_\mrm{A}\tau+\phi_\mrm{tar})
    \nonumber\\
    &\hspace{10mm}+ A_0J_1(m_\mrm{n})\cos(\omega_\mrm{A}\tau+\rho-\phi_\mrm{tar}-\frac{\pi}{2}).
    \label{eq:phasenoise_self_general0}
\end{align}
In the second line, we consider only $J_0(m_\mrm{n})$ and $J_1(m_\mrm{n})$, neglecting high-order Bessel functions because $J_k(m_\mrm{n})\sim O(m_\mrm{n}^k)$ under $m_\mrm{n}\ll1$.
The last line is the bandpassed signal around $\omega_\mrm{A}$ under $\omega_\mrm{n}=2\omega_\mrm{A}$.
% where in the last line, we consider only $J_0(m_\mrm{n})$ and $J_1(m_\mrm{n})$, neglecting high-order Bessel functions because $J_k(m_\mrm{n})\sim O(m_\mrm{n}^k)$ under $m_\mrm{n}\ll1$.
% Therefore, the bandpassed signal around $\omega_\mrm{A}$ has a noise term when $\omega_\mrm{n}=2\omega_\mrm{A}$ as:
% \begin{align}
%     s_{\omega_\mrm{A}} &\approx A_0J_0(m_\mrm{n})\cos(\omega_\mrm{A}\tau+\phi_\mrm{tar})
%     \nonumber\\
%     &\hspace{5mm}+ A_0J_1(m_\mrm{n})\cos(\omega_\mrm{A}\tau+\rho-\phi_\mrm{tar}-\frac{\pi}{2}).
%     \label{eq:phasenoise_self_general}
% \end{align}
% \end{widetext}
% In \cref{eq:phasenoise_self_general0}, the sinusoidal noise term is converted to the cosine with the phase shift by $\pi/2$ just to directly relate this equation to \cref{eq:s_add_cite}.
The sinusoidal noise term is converted to the cosine with the phase shift by $\pi/2$ just to directly relate this equation to \cref{eq:s_add_cite}.
Because \cref{eq:phasenoise_self_general0,eq:s_add_cite} has the same algebraic structure, the resulting phase error can be derived by substituting $k=A_0J_0(m_\mrm{n})\approx A_0$, $\phi_k=\phi_\mrm{tar}$, $l=A_0J_1(m_\mrm{n})\approx A_0m_\mrm{n}/2$, and $\phi_l=\rho-\phi_\mrm{tar}-\pi/2$ into \cref{eq:IQdemod_cite} under $m_\mrm{n}\ll1$ as:
\begin{align}
    \phi_\mrm{read} &= \phi_\mrm{tar} + \frac{1}{2}m_\mrm{n}\sin(\rho-2\phi_\mrm{tar}-\frac{\pi}{2})
    \nonumber\\
    &= \phi_\mrm{tar} - \frac{1}{2}m_\mrm{n}\cos(\rho-2\phi_\mrm{tar}).
    % &= \phi_\mrm{tar} - \frac{1}{2}m_\mrm{n}\cos(\rho-2\phi_\mrm{tar})
    \label{eq:phi_read_self_phasenoise}
\end{align}
This can be considered as a general derivation of what is called ``2f noise down-conversion" in \cite{BodePhD}.

\begin{table}[H]
    \centering
    \caption{\label{tab:self}
    General expression of the self noise coupling based on the signal form of $A_\mrm{n}\cos(\omega_\mrm{A}\tau+\phi_\mrm{tar}+\phi_\mrm{n})$ where $A_\mrm{n}=A_0(1+a_\mrm{n})$, as shown in \cref{eq:self_general}.
    $a_\mrm{n}$: relative amplitude noise; $\phi_\mrm{n}$: phase noise.
    $m_\mrm{n}$ is the noise amplitude.
    % The coupled noise is normalized by the original noise amplitude.
    \vspace{1.5mm}
    }
    \begin{tabular}{ccc}
    \toprule
    Noise type &  Frequency & Coupled noise
    \\
    \hline\\
    $a_\mrm{n}$ &  $2\omega_\mrm{A}$  & $\frac{m_\mrm{n}}{2}\sin(\rho-2\phi_\mrm{tar})$ 
    \vspace{2.0mm}\\
    $\phi_\mrm{n}$ &  $\epsilon$  & $\cos\rho$
    \vspace{2.0mm}\\
    &  $2\omega_\mrm{A}$  & $- \frac{m_\mrm{n}}{2}\cos(\rho-2\phi_\mrm{tar})$
    \\
    \bottomrule
    \end{tabular}
\end{table}
The general results are summarized in \cref{tab:self}.
The physical interpretation of the coupling paths is as follows: the phase noise at $\epsilon$ directly appears in the in-band phase measured by a phasemeter. In the meantime, the noise around $2\omega_A$ is wrapped around zero frequency to the beatnote frequency $\omega_A$, and disturb phase extraction.
Throughout the rest of this section, we apply these results to each beatnote in \cref{eq:spr} for the noise contribution of a single beam: beam $i$.
\\\\
\noindent\emph{Carrier-carrier beatnote: }
Let us extract the carrier-carrier beatnote from \cref{eq:spr}:
\begin{align}
    s^\mrm{car}_\mrm{pr} &= A_iA_jJ_0(m_i)J_0(m_0)\cos\left(\omega_\mrm{het}\tau + \phi^\mrm{car}_\mrm{tar} + p_i\right)
    \nonumber\\
    &\approx A_iA_jJ^2_0(m_0)\left(1 - \frac{m^2_0a_i}{2J_0(m_0)}\right)\cos\left(\omega_\mrm{het}\tau + \phi^\mrm{car}_\mrm{tar} + p_i\right).
    \label{eq:scar_prifo}
\end{align}
To consider the noise of one of the two beams, the modulation amplitude noise $a_i$ and the laser phase noise $p_i$ remain, while $a_j=p_j=0$.
The modulation phase noise $\theta_i$ is obviously not relevant, as the carrier-carrier beatnote does not include $\theta_i$.
The second line uses the following general relation:
\begin{align}
    J_0(x(1+\delta)) &\approx J_0(x) - \delta x J_1(x)
    \nonumber\\
    &\approx J_0(x)\left(1-\frac{x^2\delta}{2J_0(x)}\right),
    \label{eq:J0_delta}
\end{align}
where the first line results from $\delta\ll 1$, and the second line neglects $O(\delta x^3)$.
According to the summary in \cref{tab:self}, we need to consider only the noise at $2\omega_\mrm{het}$.

First, regarding the modulation amplitude noise $a_i$ under $a_i=m_\mrm{a}\cos(\omega_\mrm{n}\tau+\rho)$ following \cref{eq:x_i}, if \cref{eq:scar_prifo} is compared with \cref{eq:self_general}, one finds that $m_\mrm{n}$ in this case is $- \frac{m^2_0m_\mrm{a}}{2J_0(m_0)}$.
Therefore, we can derive the phasemeter output from its general form in \cref{eq:phi_read_self_ampnoise} as:
\begin{align}
    \phi^\mrm{car}_\mrm{read} &= \phi_\mrm{tar} - \frac{m^2_0}{4J_0(m_0)}\cdot m_\mrm{a}\sin(\rho-2\phi^\mrm{car}_\mrm{tar}).
    \label{eq:phi_read_2f_car}
\end{align}

Second, regarding the laser phase noise $p_i$ under $p_i=m_\mrm{p}\cos(\omega_\mrm{n}\tau+\rho)$ following \cref{eq:x_i}, the noise in the observation band appears in the phase readout as is according to \cref{tab:self}.
Concerning the noise at $2\omega_\mrm{het}$, if we apply \cref{eq:phi_read_self_phasenoise}, we get
\begin{align}
    \phi^\mrm{car}_\mrm{read} &= \phi_\mrm{tar} - \frac{1}{2}\cdot m_\mrm{p}\cos(\rho-2\phi^\mrm{car}_\mrm{tar}).
    \label{eq:phi_read_2f_car}
\end{align}

\noindent\emph{Sideband-sideband beatnotes: }
We now investigate the self noise coupling of $a_i$, $p_i$, and $\theta_i$ to the measured phase of the two sideband-sideband beatnotes in \cref{eq:spr} individually.
Similarly to \cref{eq:scar_prifo}, we only consider the noises of beam $i$, neglecting those of beam $j$.
Let us extract the sideband-sideband beatnotes in \cref{eq:spr} separately:
\begin{subequations}\label{eq:ssb_prifo}
    \begin{align}
        s^\mrm{usb}_\mrm{pr} &= A_iA_jJ_1(m_i)J_1(m_0)\cos\left((\omega_\mrm{het}+\Delta\omega_\mrm{m})\tau + \phi^\mrm{usb}_\mrm{tar} + p_i + \theta_i\right)
        \nonumber\\
        &\approx A_iA_jJ^2_1(m_0)(1+a_i)\cos\left((\omega_\mrm{het}+\Delta\omega_\mrm{m})\tau + \phi^\mrm{usb}_\mrm{tar} + p_i + \theta_i\right),
        \label{eq:susb_prifo}\\
        s^\mrm{lsb}_\mrm{pr} &= A_iA_jJ_1(m_i)J_1(m_0)\cos\left((\omega_\mrm{het}-\Delta\omega_\mrm{m})\tau - \phi^\mrm{lsb}_\mrm{tar} + p_i - \theta_i\right)
        \nonumber\\
        &\approx A_iA_jJ^2_1(m_0)(1+a_i)\cos\left((\omega_\mrm{het}-\Delta\omega_\mrm{m})\tau - \phi^\mrm{lsb}_\mrm{tar} + p_i - \theta_i\right),
        \label{eq:slsb_prifo}
    \end{align}
\end{subequations}
since in general
\begin{align}
    J_1(x(1+\delta)) &\approx J_1(x)(1+\delta)-\delta xJ_2(x)
    \nonumber\\
    &\approx J_1(x)(1+\delta),
    \label{eq:J1_delta}
\end{align}
where the first line results from $\delta\ll 1$, and the second line neglect $O(\delta x^3)$.
% Note that $\phi_\mrm{tar}$ in $s^\mrm{lsb}_\mrm{pr}$ has the negative sign because the target phase of the sideband phase extraction is generally the encoded clock noise, which appears with the opposite signs between the usb-usb and lsb-lsb sidebands by the same mechanism as $\theta_i$~\cite{Hartwig2021}.
According to the general expression summarized in \cref{tab:self}, the relevant noise frequency in the heterodyne band is the double of the target heterodyne frequency: $2(\omega_\mrm{het}+\Delta\omega_\mrm{m})$ for the usb-usb beatnote and $2(\omega_\mrm{het}-\Delta\omega_\mrm{m})$ for the lsb-lsb beatnote.

First, focusing on the modulation amplitude noise $a_i$ under $p_i=\theta_i=0$, the comparison of \cref{eq:ssb_prifo} with \cref{eq:self_general,eq:A_n_self_general} gives the phasemeter output via \cref{eq:phi_read_self_ampnoise} by
\begin{subequations}\label{eq:ampnoise_self_sb}
    \begin{align}
        \phi^\mrm{usb}_\mrm{read} &= \phi^\mrm{usb}_\mrm{tar} + \frac{1}{2}\cdot m_\mrm{a}\sin(\rho-2\phi^\mrm{usb}_\mrm{tar})\big|_{2(\omega_\mrm{het}+\Delta\omega_\mrm{m})},
        \label{eq:ampnoise_self_usb}\\
        \phi^\mrm{lsb}_\mrm{read} &= -\phi^\mrm{lsb}_\mrm{tar} + \frac{1}{2}\cdot m_\mrm{a}\sin(\rho+2\phi^\mrm{lsb}_\mrm{tar})\big|_{2(\omega_\mrm{het}-\Delta\omega_\mrm{m})}.
        \label{eq:ampnoise_self_lsb}
    \end{align}
\end{subequations}
``$\big|_\omega$" represents the original frequency of the noise source resulting in the coupled noise term.
In general, it is possible to linearly combine $\phi^\mrm{usb}_\mrm{read}$ and $\phi^\mrm{lsb}_\mrm{read}$ as:
\begin{align}
    \frac{\phi^\mrm{usb}_\mrm{read}-\phi^\mrm{lsb}_\mrm{read}}{2} &= \phi^\mrm{sb}_\mrm{tar} + \text{(noises)},
    \label{eq:sbcombi}
\end{align}
which may enhance a signal-to-noise ratio (SNR) depending on noise coherence. 
In the particular case of \cref{eq:ampnoise_self_sb}, the combination does not cancel the noise terms because they are incoherent due to the different original noise frequencies.

Second, when it comes to the laser phase noise $p_i$, the noise in the observation band directly couples to the phase readout as is: $\cos\rho$ following \cref{eq:x_i}.
However, it can be removed by the combination in \cref{eq:sbcombi}.
Concerning the heterodyne band, the noise components at $2(\omega_\mrm{het}+\Delta\omega_\mrm{m})$ for the usb-usb beatnote and $2(\omega_\mrm{het}-\Delta\omega_\mrm{m})$ for the lsb-lsb beatnote also couple separately, and, according to \cref{eq:phi_read_self_phasenoise}, the measured phases read
\begin{subequations}\label{eq:pnoise_self_sb}
    \begin{align}
        \phi^\mrm{usb}_\mrm{read} &= \phi^\mrm{usb}_\mrm{tar} - \frac{1}{2}\cdot m_\mrm{p}\cos(\rho-2\phi^\mrm{usb}_\mrm{tar})\big|_{2(\omega_\mrm{het}+\Delta\omega_\mrm{m})},
        \label{eq:pnoise_self_usb}\\
        \phi^\mrm{lsb}_\mrm{read} &= -\phi^\mrm{lsb}_\mrm{tar} - \frac{1}{2}\cdot m_\mrm{p}\cos(\rho+2\phi^\mrm{lsb}_\mrm{tar})\big|_{2(\omega_\mrm{het}-\Delta\omega_\mrm{m})}.
        \label{eq:pnoise_self_lsb}
    \end{align}
\end{subequations}
Similarly to \cref{eq:ampnoise_self_sb}, these cannot be canceled by the combination in \cref{eq:sbcombi}, as the original laser phase noises are in the different frequency regimes, therefore, incoherent.

Finally, concerning the modulation phase noise $\theta_i$ under $\theta_i=m_\mrm{\theta}\cos(\omega_\mrm{n}\tau+\rho)$ following \cref{eq:x_i}, the noise component in the observation band directly couples to the phase readout with a coupling factor of $\pm\cos\rho$.
Contrary to the laser phase noise $p_i$ discussed above, $\theta_i$ has the opposite signs between the usb-usb beatnote and the lsb-lsb beatnote in \cref{eq:ssb_prifo}, and cannot be suppressed by the combination in \cref{eq:sbcombi}.
Regarding the heterodyne band, the same logic as the laser phase noise above applies but with the sign flip for the lsb-lsb beatnote.
The phasemeter measurements with the noise in the relevant heterodyne-band frequencies are written by
\begin{subequations}\label{eq:thetanoise_self_sb}
    \begin{align}
        \phi^\mrm{usb}_\mrm{read} &= \phi^\mrm{usb}_\mrm{tar} - \frac{1}{2}\cdot m_\mrm{\theta}\cos(\rho-2\phi^\mrm{usb}_\mrm{tar})\big|_{2(\omega_\mrm{het}+\Delta\omega_\mrm{m})},
        \label{eq:thetanoise_self_usb}\\
        \phi^\mrm{lsb}_\mrm{read} &= -\phi^\mrm{lsb}_\mrm{tar} + \frac{1}{2}\cdot m_\mrm{\theta}\cos(\rho+2\phi^\mrm{lsb}_\mrm{tar})\big|_{2(\omega_\mrm{het}-\Delta\omega_\mrm{m})}.
        \label{eq:thetanoise_self_lsb}
    \end{align}
\end{subequations}
Since they are originally at the different frequencies, the noises will be incoherently added, if we combine $\phi^\mrm{usb}_\mrm{read}$ and $\phi^\mrm{lsb}_\mrm{read}$ as shown in \cref{eq:sbcombi}.

The results of the self noise coupling in this section are summarized in \cref{tab:noise_couplings} as its diagonal elements.

\subsection{Mutual noise coupling}\label{sub:het_band_mutual}
Similarly to \cref{sub:het_band_self}, we first provide a general description and then apply it to the individual beatnotes.
\\\\
\emph{General description: }
If a signal of interest contains two sinusoidal tones at different frequencies, we can generally formulate the signal as
\begin{align}
    s &= s_\mrm{A} + s_\mrm{B}
    \nonumber\\
    &= A_0\cos(\omega_\mrm{A}\tau+\phi_\mrm{tar}) + B_n\cos(\omega_\mrm{B}\tau+\phi_\mrm{B}+\phi_\mrm{n}),
    \label{eq:mutual_general}
\end{align}
where $\omega_\mrm{A}\neq\omega_\mrm{B}$, and a given phase $\phi_\mrm{B}$ of the noise sinusoid is considered.
The first term $s_\mrm{A}$ is the primary, whose phase $\phi_\mrm{tar}$ has to be measured, and the second term $s_\mrm{B}$ is the secondary considered as a noise source.
$b_n$ in $B_n(=B_0(1+b_n))$ and $\phi_\mrm{n}$ are the relative amplitude noise and the phase noise, respectively.
In the mutual noise coupling, the question to answer is how the noises in $s_\mrm{B}$ couples to the measurement of $\phi_\mrm{tar}$ in $s_\mrm{A}$.

First, regarding the relative amplitude noise $b_n$ ($=m_\mrm{n}\cos(\omega_\mrm{n}\tau+\rho)$) in $B_n$, the secondary noise tone can be written as
\begin{align}
    % s_\mrm{B} &= B_0(1+m_\mrm{n}\cos(\omega_\mrm{n}\tau+\rho))\cos(\omega_\mrm{B}\tau + \phi_\mrm{B})
    % \nonumber\\
    s_\mrm{B} &= B_0\cos(\omega_\mrm{B}\tau + \phi_\mrm{B}) + \frac{B_0m_\mrm{n}}{2}\cos\left((\omega_\mrm{n}-\omega_\mrm{B})\tau+\rho-\phi_\mrm{B}\right)
    \nonumber\\
    &\hspace{20mm}+ \frac{B_0m_\mrm{n}}{2}\cos\left((\omega_\mrm{n}+\omega_\mrm{B})\tau+\rho+\phi_\mrm{B}\right).
    \label{eq:ampnoise_mutual_general}
\end{align}
Our interest is the noise term at $\omega_\mrm{A}$; therefore, there are multiple solutions to $\omega_\mrm{n}$,
\begin{align}
    \omega_\mrm{n} - \omega_\mrm{B} &= \pm\omega_\mrm{A} \Rightarrow \omega_\mrm{n} = \pm\omega_\mrm{A} + \omega_\mrm{B},
    \label{eq:mutual_omegan1}\\
    \omega_\mrm{n} + \omega_\mrm{B} &= \pm\omega_\mrm{A} \Rightarrow \omega_\mrm{n} = \omega_\mrm{A} - \omega_\mrm{B},
    \label{eq:mutual_omegan2}
\end{align}
where \cref{eq:mutual_omegan2} only considers the positive sign of $\omega_\mrm{A}$ as we define all (angular) frequencies with $\omega>\SI{0}{Hz}$.
Therefore, for a certain pair of $s_\mrm{A}$ and $s_\mrm{B}$, there are two possible noise frequencies: $\omega_\mrm{A}+\omega_\mrm{B}$ and the positive one out of $-\omega_\mrm{A}+\omega_\mrm{B}$ or $\omega_\mrm{A}-\omega_\mrm{B}$.
% In either case, it has the same magnitude of the coupling factor as they share the amplitude of $B_0m_\mrm{n}/2$ according to \cref{eq:ampnoise_mutual_general}.
% However, depending on the relative sign between the given noise phase $\rho$ and the resulting monotonic phase ramp $\omega_\mrm{A}\tau$, the sign of the coupling factor becomes different.
The bandpassed signal around $\omega_\mrm{A}$ can be written separately for each case as
\begin{widetext}
\begin{align}
    s_{\omega_\mrm{A}} &=
    \begin{cases}
        A_0\cos(\omega_\mrm{A}\tau+\phi_\mrm{tar}) + \frac{B_0m_\mrm{n}}{2}\cos\left(\omega_\mrm{A}\tau+\rho-\phi_\mrm{B}\right) &\text{ if $\omega_\mrm{n}=\omega_\mrm{A}+\omega_\mrm{B}$,}
        \\
        A_0\cos(\omega_\mrm{A}\tau+\phi_\mrm{tar}) + \frac{B_0m_\mrm{n}}{2}\cos\left(\omega_\mrm{A}\tau+\rho+\phi_\mrm{B}\right) &\text{ if $\omega_\mrm{n}=\omega_\mrm{A}-\omega_\mrm{B}$,}
        \\
        A_0\cos(\omega_\mrm{A}\tau+\phi_\mrm{tar}) + \frac{B_0m_\mrm{n}}{2}\cos\left(\omega_\mrm{A}\tau-\rho+\phi_\mrm{B}\right) &\text{ if $\omega_\mrm{n}=-\omega_\mrm{A}+\omega_\mrm{B}$.}
    \end{cases}
    \label{eq:ampnoise_mutual_general}
\end{align}
Because \cref{eq:ampnoise_mutual_general,eq:s_add_cite} have the same algebraic form, the resulting phase errors can be derived by substituting $k=A_0$, $\phi_k=\phi_\mrm{tar}$, $l=B_0m_\mrm{n}/2$, and $\phi_l=\pm\rho\pm\phi_\mrm{B}$ into \cref{eq:IQdemod_cite},
\begin{align}
    \phi_\mrm{read} &=
    \begin{cases}
        \phi_\mrm{tar} + \frac{B_0}{2A_0}m_\mrm{n}\sin(\rho-\phi_\mrm{B}-\phi_\mrm{tar}) &\text{ if $\omega_\mrm{n}=\omega_\mrm{A}+\omega_\mrm{B}$,}
        \\
        \phi_\mrm{tar} + \frac{B_0}{2A_0}m_\mrm{n}\sin(\rho+\phi_\mrm{B}-\phi_\mrm{tar}) &\text{ if $\omega_\mrm{n}=\omega_\mrm{A}-\omega_\mrm{B}$,}
        \\
        \phi_\mrm{tar} - \frac{B_0}{2A_0}m_\mrm{n}\sin(\rho-\phi_\mrm{B}+\phi_\mrm{tar}) &\text{ if $\omega_\mrm{n}=-\omega_\mrm{A}+\omega_\mrm{B}$.}
    \end{cases}
    \label{eq:phi_read_mutual_ampnoise}
\end{align}

Second, regarding the phase noise $\phi_\mrm{n}$, we can follow the similar derivation in \cref{eq:phasenoise_self_general0} with the Jacobi-Anger expansion,
\begin{align}
    s_\mrm{B} &= B_0\cos(\omega_\mrm{B}\tau+\phi_\mrm{B}+m_\mrm{n}\cos(\omega_\mrm{n}\tau+\rho))
    \nonumber\\
    % &= B_0\cos(\omega_\mrm{B}\tau+\phi_\mrm{B})\cos(m_\mrm{n}\cos(\omega_\mrm{n}\tau+\rho)) - B_0\sin(\omega_\mrm{B}\tau+\phi_\mrm{B})\sin(m_\mrm{n}\cos(\omega_\mrm{n}\tau+\rho))
    % \nonumber\\
    &\approx B_0J_0(m_\mrm{n})\cos(\omega_\mrm{B}\tau+\phi_\mrm{B}) - 2B_0J_1(m_\mrm{n})\cos(\omega_\mrm{n}\tau+\rho)\sin(\omega_\mrm{B}\tau+\phi_\mrm{B})
    \nonumber\\
    &= B_0J_0(m_\mrm{n})\cos(\omega_\mrm{B}\tau+\phi_\mrm{B}) - B_0J_1(m_\mrm{n})\left(\cos((\omega_\mrm{B}+\omega_\mrm{n})\tau+\phi_\mrm{B}+\rho-\frac{\pi}{2}) + \cos((\omega_\mrm{B}-\omega_\mrm{n})\tau+\phi_\mrm{B}-\rho-\frac{\pi}{2})\right),
    % \nonumber\\
    % &\rightarrow B_0J_0(m_\mrm{n})\cos(\omega_\mrm{B}\tau) - B_0J_1(m_\mrm{n})\left(\cos((\omega_\mrm{B}+\omega_\mrm{n})\tau+\rho) - \cos((\omega_\mrm{B}-\omega_\mrm{n})\tau-\rho)\right)
    \label{eq:phasenoise_mutual_sB}
\end{align}
where in the second line, we neglect second- and high-order Bessel functions because $m_\mrm{n}\ll1$.
In the last line, similarly to \cref{eq:phasenoise_self_general0}, the noise terms are converted to the cosine with the phase shift by $\pi/2$ to directly relate this equation to \cref{eq:s_add_cite}.
This suggests that we need to consider the same noise frequency bands as the modulation amplitude noise above.
Therefore, the total signal around $\omega_\mrm{A}$ becomes
\begin{align}
    s_{\omega_\mrm{A}} &=
    \begin{cases}
        A_0\cos(\omega_\mrm{A}\tau+\phi_\mrm{tar}) - B_0J_1(m_\mrm{n})\cos(\omega_\mrm{A}\tau+\rho-\phi_\mrm{B}+\frac{\pi}{2}) &\text{ if $\omega_\mrm{n}=\omega_\mrm{A}+\omega_\mrm{B}$,}
        \\
        A_0\cos(\omega_\mrm{A}\tau+\phi_\mrm{tar}) - B_0J_1(m_\mrm{n})\cos(\omega_\mrm{A}\tau+\rho+\phi_\mrm{B}-\frac{\pi}{2}) &\text{ if $\omega_\mrm{n}=\omega_\mrm{A}-\omega_\mrm{B}$,}
        \\
        A_0\cos(\omega_\mrm{A}\tau+\phi_\mrm{tar}) - B_0J_1(m_\mrm{n})\cos(\omega_\mrm{A}\tau-\rho+\phi_\mrm{B}-\frac{\pi}{2}) &\text{ if $\omega_\mrm{n}=-\omega_\mrm{A}+\omega_\mrm{B}$.}
    \end{cases}
    \label{eq:phasenoise_mutual_general}
\end{align}
\end{widetext}
From \cref{eq:s_add_cite,eq:IQdemod_cite}, the resulting phase error can be derived by substituting $k=A_0$, $\phi_k=\phi_\mrm{tar}$, $l=- B_0J_1(m_\mrm{n})\approx - B_0m_\mrm{n}/2$, and $\phi_l=\pm\rho\pm\phi_\mrm{B}\pm\pi/2$ into \cref{eq:IQdemod_cite},
\begin{align}
    \phi_\mrm{read} &=
    \begin{cases}
        \phi_\mrm{tar} - \frac{B_0}{2A_0}m_\mrm{n}\cos(\rho-\phi_\mrm{B}-\phi_\mrm{tar}) &\text{ if $\omega_\mrm{n}=\omega_\mrm{A}+\omega_\mrm{B}$,}
        \\
        \phi_\mrm{tar} + \frac{B_0}{2A_0}m_\mrm{n}\cos(\rho+\phi_\mrm{B}-\phi_\mrm{tar}) &\text{ if $\omega_\mrm{n}=\omega_\mrm{A}-\omega_\mrm{B}$,}
        \\
        \phi_\mrm{tar} + \frac{B_0}{2A_0}m_\mrm{n}\cos(\rho-\phi_\mrm{B}+\phi_\mrm{tar}) &\text{ if $\omega_\mrm{n}=-\omega_\mrm{A}+\omega_\mrm{B}$.}
    \end{cases}
    \label{eq:phi_read_mutual_phasenoise}
\end{align}

\begin{table}
    \centering
    \caption{\label{tab:mutual}
     General expression of the mutual noise coupling based on the signal form of $A_0\cos(\omega_\mrm{A}\tau+\phi_\mrm{tar}) + B_n\cos(\omega_\mrm{B}\tau+\phi_\mrm{B}+\phi_\mrm{n})$ where $B_n=B_0(1+ b_n)$, as shown in \cref{eq:mutual_general}.
    $b_n$: relative amplitude noise; $\phi_\mrm{n}$: phase noise.
    $m_\mrm{n}$ is the noise amplitude.
    Only the positive solution of the noise frequency bands is valid out of $\omega_\mrm{A}-\omega_\mrm{B}$ and $-\omega_\mrm{A}+\omega_\mrm{B}$.
    \vspace{1.5mm}
    }
    \begin{tabular}{ccc}
    \toprule
    Noise type & Frequency & Coupled noise
    \\
    \hline\\
    $b_n$ &  $\omega_\mrm{A}+\omega_\mrm{B}$  & $ \frac{B_0m_\mrm{n}}{2A_0}\sin\left(\rho-\phi_\mrm{B}-\phi_\mrm{tar}\right)$ 
    \vspace{1.5mm}\\
    &  $\omega_\mrm{A}-\omega_\mrm{B}$  & $\frac{B_0m_\mrm{n}}{2A_0}\sin(\rho+\phi_\mrm{B}-\phi_\mrm{tar})$
    \vspace{1.5mm}\\
    &  $-\omega_\mrm{A}+\omega_\mrm{B}$  & $-\frac{B_0m_\mrm{n}}{2A_0}\sin(\rho-\phi_\mrm{B}+\phi_\mrm{tar})$
    \vspace{1.5mm}\\
    $\phi_\mrm{n}$ &  $\omega_\mrm{A}+\omega_\mrm{B}$  & $- \frac{B_0m_\mrm{n}}{2A_0}\cos(\rho-\phi_\mrm{B}-\phi_\mrm{tar})$
    \vspace{1.5mm}\\
    &  $\omega_\mrm{A}-\omega_\mrm{B}$  & $\frac{B_0m_\mrm{n}}{2A_0}\cos(\rho+\phi_\mrm{B}-\phi_\mrm{tar})$
    \vspace{1.5mm}\\
    &  $-\omega_\mrm{A}+\omega_\mrm{B}$  & $\frac{B_0m_\mrm{n}}{2A_0}\cos(\rho-\phi_\mrm{B}+\phi_\mrm{tar})$
    \vspace{1.5mm}\\
    \bottomrule
    \end{tabular}
\end{table}
The general results of the mutual noise coupling above are summarized in \cref{tab:mutual}.
The physical interpretation of the coupling paths is as follows: noise of $s_B$ at $\omega_A+\omega_B$ is wrapped around zero frequency to the beatnote frequency $\omega_A$, and disturb phase extraction of $s_A$.
In the meantime, noise of $s_B$ at $\mp\omega_A\pm\omega_B$ (only the positive one is valid) sits at the beatnote frequency $\omega_A$ without the wrapping because $\mp\omega_A\pm\omega_B$ is the simple difference between the beatnote frequency of $s_A$ and $s_B$.
Note that the self noise coupling in \cref{tab:self} can be understood as the special case of the mutual noise coupling where $\omega_\mrm{B}=\omega_\mrm{A}$ and $B_0=A_0$, except for the treatment of the noise in the observation band.
% Note that only the positive solution of the noise frequency bands is valid out of $\omega_\mrm{A}-\omega_\mrm{B}$ and $-\omega_\mrm{A}+\omega_\mrm{B}$.
\\\\
\noindent\emph{Carrier-carrier beatnote to sideband-sideband beatnotes:}
We can focus on the following part of the PR signal in \cref{eq:spr},
\begin{align}
    s^\mrm{c2sb}_\mrm{pr} &= A_iA_jJ^2_0(m_0)\left(1 - \frac{m^2_0a_i}{2J_0(m_0)}\right)\cos\left(\omega_\mrm{het}\tau + \phi^\mrm{car}_\mrm{tar} + p_i\right)
    \nonumber\\
    &\hspace{5mm} + A_iA_jJ^2_1(m_0)\cos\left((\omega_\mrm{het}+\Delta\omega_\mrm{m})\tau + \phi^\mrm{usb}_\mrm{tar}\right)
    \nonumber\\
    &\hspace{5mm} + A_iA_jJ^2_1(m_0)\cos\left((\omega_\mrm{het}-\Delta\omega_\mrm{m})\tau - \phi^\mrm{lsb}_\mrm{tar}\right),
    \label{eq:sc2sb_prifo}
\end{align}
where only the laser phase noise $p_i$ and the modulation amplitude noise $a_i$ in the carrier-carrier beatnote are considered, as the modulation phase noise $\theta_i$ does not appear in the carrier.
% The $a_i$ part in the carrier-carrier beatnote arises from $J_0(m_i)J_0(m_0)$, as derived in \cref{eq:scar_prifo}.
In this case, the carrier is interpreted as the offending noise term for the sideband terms: $\omega_\mrm{A}=\omega_\mrm{het}\pm\Delta\omega_\mrm{m}$ and $\omega_\mrm{B}=\omega_\mrm{het}$ in \cref{eq:mutual_general}.
Therefore, the relevant noise frequencies are
\begin{enumerate}[label=(\alph*)]
    \item The carrier-carrier to the usb-usb: $\omega_\mrm{n}=\omega_\mrm{A}+\omega_\mrm{B}=2\omega_\mrm{het}+\Delta\omega_\mrm{m}$ and $\omega_\mrm{n}=\omega_\mrm{A}-\omega_\mrm{B}=\Delta\omega_\mrm{m}$,
    \item The carrier-carrier to the lsb-lsb: $\omega_\mrm{n}=\omega_\mrm{A}+\omega_\mrm{B}=2\omega_\mrm{het}-\Delta\omega_\mrm{m}$ and $\omega_\mrm{n}=-\omega_\mrm{A}+\omega_\mrm{B}=\Delta\omega_\mrm{m}$.
\end{enumerate}
Therefore, we have three frequencies to investigate.
Regarding the noise at $\Delta\omega_\mrm{m}$, this suggests that it couples to both sideband phases, which becomes important when we combine the phases as shown in \cref{eq:sbcombi}.

First, regarding the modulation amplitude noise $a_i$, comparing \cref{eq:sc2sb_prifo} under $p_i=0$ with \cref{eq:ampnoise_mutual_general}, we can simply substitute quantities as follows: $A_0 = A_iA_jJ^2_1(m_0)$, $B_0 = A_iA_jJ^2_0(m_0)$, $m_\mrm{n} = - \frac{m^2_0m_\mrm{a}}{2J_0(m_0)}$, $\phi_\mrm{tar}=\phi^\mrm{usb}_\mrm{tar}\text{ or }-\phi^\mrm{lsb}_\mrm{tar}$, and $\phi_\mrm{B}=\phi^\mrm{car}_\mrm{tar}$.
%, where $m_\mrm{a}$ is a noise amplitude in the monotonic noise representation.
Therefore, the general expression of the phase readout in \cref{eq:phi_read_mutual_ampnoise} becomes
\begin{subequations}\label{eq:ampnoise_mutual_sb}
    \begin{align}
        \phi^\mrm{usb}_\mrm{read} &= 
        \phi^\mrm{usb}_\mrm{tar}
        - \frac{J_0(m_0)m^2_0}{4J^2_1(m_0)}\cdot m_\mrm{a}\sin(\rho-\phi^\mrm{sb}_\mrm{tar})\big|_{\Delta\omega_\mrm{m}}
        \nonumber\\
        &\hspace{9mm} - \frac{J_0(m_0)m^2_0}{4J^2_1(m_0)}\cdot m_\mrm{a}\sin(\rho-\phi^\mrm{usb}_\mrm{tar}-\phi^\mrm{car}_\mrm{tar})\big|_{2\omega_\mrm{het}+\Delta\omega_\mrm{m}}
        \label{eq:ampnoise_mutual_usb}\\
        \phi^\mrm{lsb}_\mrm{read} &=
        -\phi^\mrm{lsb}_\mrm{tar}
        + \frac{J_0(m_0)m^2_0}{4J^2_1(m_0)}\cdot m_\mrm{a}\sin(\rho-\phi^\mrm{sb}_\mrm{tar})\big|_{\Delta\omega_\mrm{m}}
        \nonumber\\
        &\hspace{11mm}- \frac{J_0(m_0)m^2_0}{4J^2_1(m_0)}\cdot m_\mrm{a}\sin(\rho+\phi^\mrm{lsb}_\mrm{tar}-\phi^\mrm{car}_\mrm{tar})\big|_{2\omega_\mrm{het}-\Delta\omega_\mrm{m}}
        \label{eq:ampnoise_mutual_lsb}
    \end{align}
\end{subequations}
The equations suggest that the modulation amplitude noise at $\Delta\omega_\mrm{m}$ coherently sums up in the combination in \cref{eq:sbcombi}, therefore, cannot be suppressed.
Also, other noises are in the different frequency bands; hence, they will be incoherently summed in the combination.

Second, to focus on the laser phase noise $p_i$, we compare \cref{eq:sc2sb_prifo} under $a_i=0$ with \cref{eq:phasenoise_mutual_general}.
It suggests substituting quantities as follows: $A_0 = A_iA_jJ^2_1(m_0)$, $B_0 = A_iA_jJ^2_0(m_0)$, $m_\mrm{n} = m_\mrm{p}$, $\phi_\mrm{tar}=\phi^\mrm{usb}_\mrm{tar}\text{ or }-\phi^\mrm{lsb}_\mrm{tar}$, and $\phi_\mrm{B}=\phi^\mrm{car}_\mrm{tar}$.
The measured sideband phases result in
\begin{subequations}\label{eq:phasenoise_mutual_sb}
    \begin{align}
        \phi^\mrm{usb}_\mrm{read} &=
        \phi^\mrm{usb}_\mrm{tar}
        + \frac{J^2_0(m_0)}{2J^2_1(m_0)}\cdot m_\mrm{p}\cos(\rho-\phi^\mrm{sb}_\mrm{tar})\big|_{\Delta\omega_\mrm{m}}
        \nonumber\\
        &\hspace{9mm} - \frac{J^2_0(m_0)}{2J^2_1(m_0)}\cdot m_\mrm{p}\cos(\rho-\phi^\mrm{usb}_\mrm{tar}-\phi^\mrm{car}_\mrm{tar})\big|_{2\omega_\mrm{het}+\Delta\omega_\mrm{m}},
        \label{eq:ampnoise_mutual_usb}\\
        \phi^\mrm{lsb}_\mrm{read} &=
        -\phi^\mrm{lsb}_\mrm{tar}
        + \frac{J^2_0(m_0)}{2J^2_1(m_0)}\cdot m_\mrm{p}\cos(\rho-\phi^\mrm{sb}_\mrm{tar})\big|_{\Delta\omega_\mrm{m}}
        \nonumber\\
        &\hspace{11mm} - \frac{J^2_0(m_0)}{2J^2_1(m_0)}\cdot m_\mrm{p}\cos(\rho+\phi^\mrm{lsb}_\mrm{tar}-\phi^\mrm{car}_\mrm{tar})\big|_{2\omega_\mrm{het}-\Delta\omega_\mrm{m}}.
        \label{eq:phasenoise_mutual_lsb}
    \end{align}
\end{subequations}
% When we combine $\phi^\mrm{usb}_\mrm{read}$ and $\phi^\mrm{lsb}_\mrm{read}$, the noises behaves as follows:
% \begin{itemize}
%     \item The noise at $\Delta\omega_\mrm{m}$ are coherently subtracted; hence, it can be removed completely in theory,
%     \item Other noises are in the different frequency bands; hence, they will incoherently get mixed.
% \end{itemize}
% Concerning Point 1, however, it is uncertain how well the coherence between the noise split to the frequencies of the usb-usb and lsb-lsb beatnotes is maintained through the metrology chain; hence, it would be a valid assumption for the requirement description that we can suppress the noise by a factor of $\sqrt{2}$ with the two sideband phases combined.
The equations suggest that the laser phase noise at $\Delta\omega_\mrm{m}$ could be subtracted completely in the sideband combination in \cref{eq:sbcombi}.
% , while other noises are originally in the different frequency bands, hence, they will be incoherent.
\\\\
\noindent\emph{Sideband-sideband beatnotes to carrier-carrier beatnote: }
If we consider the sideband-sideband beatnotes as noise sources and the carrier-carrier as the target beatnote, the PR signal in \cref{eq:spr} becomes
% \begin{widetext}
\begin{align}
    s^\mrm{sb2c}_\mrm{pr} &= A_iA_jJ^2_0(m_0)\cos\left(\omega_\mrm{het}\tau + \phi^\mrm{car}_\mrm{tar}\right)
    \nonumber\\
    &\hspace{0mm} + A_iA_jJ^2_1(m_0)\left(1+a_i\right)\cos\left((\omega_\mrm{het}+\Delta\omega_\mrm{m})\tau + \phi^\mrm{usb}_\mrm{tar} + p_i + \theta_i\right)
    \nonumber\\
    &\hspace{0mm} + A_iA_jJ^2_1(m_0)\left(1+a_i\right)\cos\left((\omega_\mrm{het}-\Delta\omega_\mrm{m})\tau - \phi^\mrm{lsb}_\mrm{tar} + p_i - \theta_i\right).
    \label{eq:ssb2c_prifo}
\end{align}
% \end{widetext}
% where $\phi_\mrm{tar}$ is the target phase to extract from the carrier-carrier beatnote.
% Note that in general, any phase in the carrier-carrier beatnote also appears in the sideband-sideband beatnotes.
% Therefore, the target phase $\phi_\mrm{tar}$ should also be included in the phase of the sideband-sideband beatnotes, even though they are not the target beatnotes here.
In this coupling from the sidebands to the carrier, $\omega_\mrm{A}=\omega_\mrm{het}$ and $\omega_\mrm{B}=\omega_\mrm{het}\pm\Delta\omega_\mrm{m}$ in \cref{eq:mutual_general}.
Therefore, the relevant noise frequencies are
\begin{enumerate}[label=(\alph*)]
    \item The usb-usb to the carrier-carrier: $\omega_\mrm{n}=\omega_\mrm{A}+\omega_\mrm{B}=2\omega_\mrm{het}+\Delta\omega_\mrm{m}$ and $\omega_\mrm{n}=-\omega_\mrm{A}+\omega_\mrm{B}=\Delta\omega_\mrm{m}$,
    \item The lsb-lsb to the carrier-carrier: $\omega_\mrm{n}=\omega_\mrm{A}+\omega_\mrm{B}=2\omega_\mrm{het}-\Delta\omega_\mrm{m}$ and $\omega_\mrm{n}=\omega_\mrm{A}-\omega_\mrm{B}=\Delta\omega_\mrm{m}$.
\end{enumerate}
Therefore, we have three frequencies to investigate and, for $\Delta\omega_\mrm{m}$, we also need to take into account the coherence between the usb-usb and lsb-lsb beatnotes.

First, regarding the modulation amplitude noise $a_i$, the comparison of \cref{eq:ssb2c_prifo} under $p_i=\theta_i=0$ with \cref{eq:ampnoise_mutual_general} suggests that we can substitute quantities as follows: $A_0 = A_iA_jJ^2_0(m_0)$, $B_0 = A_iA_jJ^2_1(m_0)$, $m_\mrm{n} = m_\mrm{a}$, $\phi_\mrm{tar}=\phi^\mrm{car}_\mrm{tar}$, $\phi_\mrm{B}=\phi^\mrm{usb}_\mrm{tar}$ or $-\phi^\mrm{lsb}_\mrm{tar}$.
Therefore, the general expression of the phase readout in \cref{eq:phi_read_mutual_ampnoise} gives
\begin{widetext}
\begin{align}
    \phi^\mrm{car}_\mrm{read} &= \phi^\mrm{car}_\mrm{tar} - \frac{J^2_1(m_0)}{2J^2_0(m_0)}\cdot m_\mrm{a}\left(\sin(\rho-\phi^\mrm{sb}_\mrm{tar})\big|_{\Delta\omega_\mrm{m}} - \sin(\rho-\phi^\mrm{usb}_\mrm{tar}-\phi^\mrm{car}_\mrm{tar})\big|_{2\omega_\mrm{het}+\Delta\omega_\mrm{m}}\right)
    \nonumber\\
    &\hspace{9mm} + \frac{J^2_1(m_0)}{2J^2_0(m_0)}\cdot m_\mrm{a}\left(\sin(\rho-\phi^\mrm{sb}_\mrm{tar})\big|_{\Delta\omega_\mrm{m}} + \sin(\rho + \phi^\mrm{lsb}_\mrm{tar} - \phi^\mrm{car}_\mrm{tar})\big|_{2\omega_\mrm{het}-\Delta\omega_\mrm{m}}\right)
    \nonumber\\
    &= \phi^\mrm{car}_\mrm{tar} + \frac{J^2_1(m_0)}{2J^2_0(m_0)}\cdot m_\mrm{a}\sin(\rho-\phi^\mrm{usb}_\mrm{tar}-\phi^\mrm{car}_\mrm{tar})\big|_{2\omega_\mrm{het}+\Delta\omega_\mrm{m}} + \frac{J^2_1(m_0)}{2J^2_0(m_0)}\cdot m_\mrm{a}\sin(\rho + \phi^\mrm{lsb}_\mrm{tar} - \phi^\mrm{car}_\mrm{tar})\big|_{2\omega_\mrm{het}-\Delta\omega_\mrm{m}},
    \label{eq:ampnoise_mutual_sb2c}
\end{align}
In the first line, the first and second parentheses are the noise coupling from the usb-usb and lsb-lsb beatnotes, respectively.
%The original noise frequencies are explicitly written as ``$\big|_\omega$" to express the coherence between noises.
The modulation amplitude noise at $\Delta\omega_\mrm{m}$ completely cancel between the usb-usb and lsb-lsb beatnotes in the second line.

Second, regarding the laser phase noise $p_i$, the comparison of \cref{eq:ssb2c_prifo} under $a_i=\theta_i=0$ with \cref{eq:phasenoise_mutual_general} suggests substituting quantities as follows: $A_0 = A_iA_jJ^2_0(m_0)$, $B_0 = A_iA_jJ^2_1(m_0)$, $m_\mrm{n} = m_\mrm{p}$, $\phi_\mrm{tar}=\phi^\mrm{car}_\mrm{tar}$, and $\phi_\mrm{B}=\phi^\mrm{usb}_\mrm{tar}$ or $-\phi^\mrm{lsb}_\mrm{tar}$.
Applying \cref{eq:phi_read_mutual_phasenoise}, the measured carrier-carrier beatnote phase is given by
\begin{align}
    \phi^\mrm{car}_\mrm{read} &= \phi^\mrm{car}_\mrm{tar} + \frac{J^2_1(m_0)}{2J^2_0(m_0)}\cdot m_\mrm{p}\left(\cos(\rho-\phi^\mrm{sb}_\mrm{tar})\big|_{\Delta\omega_\mrm{m}} - \cos(\rho-\phi^\mrm{usb}_\mrm{tar}-\phi^\mrm{car}_\mrm{tar})\big|_{2\omega_\mrm{het}+\Delta\omega_\mrm{m}}\right)
    \nonumber\\
    &\hspace{9mm}+ \frac{J^2_1(m_0)}{2J^2_0(m_0)}\cdot m_\mrm{p}\left(\cos(\rho-\phi^\mrm{sb}_\mrm{tar})\big|_{\Delta\omega_\mrm{m}} - \cos(\rho+\phi^\mrm{lsb}_\mrm{tar}-\phi^\mrm{car}_\mrm{tar})\big|_{2\omega_\mrm{het}-\Delta\omega_\mrm{m}}\right)
    \nonumber\\
    &= \phi^\mrm{car}_\mrm{tar} + \frac{J^2_1(m_0)}{J^2_0(m_0)}\cdot m_\mrm{p}\cos(\rho-\phi^\mrm{sb}_\mrm{tar})\big|_{\Delta\omega_\mrm{m}}
    \nonumber\\
    &\hspace{9mm}- \frac{J^2_1(m_0)}{2J^2_0(m_0)}\cdot m_\mrm{p}\cos(\rho-\phi^\mrm{usb}_\mrm{tar}-\phi^\mrm{car}_\mrm{tar})\big|_{2\omega_\mrm{het}+\Delta\omega_\mrm{m}} - \frac{J^2_1(m_0)}{2J^2_0(m_0)}\cdot m_\mrm{p}\cos(\rho+\phi^\mrm{lsb}_\mrm{tar}-\phi^\mrm{car}_\mrm{tar})\big|_{2\omega_\mrm{het}-\Delta\omega_\mrm{m}},
    \label{eq:phasenoise_mutual_sb2c}
\end{align}
In the first line, the first and second parentheses are the noise couplings from the usb-usb and lsb-lsb beatnotes, respectively.
The noise from $\Delta\omega_\mrm{m}$ coherently sums up between the usb-usb and lsb-lsb beatnotes.

Last but not least, regarding the modulation phase noise $\theta_i$, we can apply the almost same discussion as the laser phase noise in \cref{eq:phasenoise_mutual_sb2c}, but with the opposite sign for the noise couplings from the lsb-lsb beatnote.
Therefore, we get the measured carrier-carrier beatnote phase as,
\begin{align}
    \phi^\mrm{car}_\mrm{read} &= \phi^\mrm{car}_\mrm{tar} + \frac{J^2_1(m_0)}{2J^2_0(m_0)}\cdot m_\mrm{\theta}\left(\cos(\rho-\phi^\mrm{sb}_\mrm{tar})\big|_{\Delta\omega_\mrm{m}} - \cos(\rho-\phi^\mrm{usb}_\mrm{tar}-\phi^\mrm{car}_\mrm{tar})\big|_{2\omega_\mrm{het}+\Delta\omega_\mrm{m}}\right)
    \nonumber\\
    &\hspace{9mm} - \frac{J^2_1(m_0)}{2J^2_0(m_0)}\cdot m_\mrm{\theta}\left(\cos(\rho-\phi^\mrm{sb}_\mrm{tar})\big|_{\Delta\omega_\mrm{m}} - \cos(\rho+\phi^\mrm{lsb}_\mrm{tar}-\phi^\mrm{car}_\mrm{tar})\big|_{2\omega_\mrm{het}-\Delta\omega_\mrm{m}}\right)
    \nonumber\\
    &= \phi^\mrm{car}_\mrm{tar} - \frac{J^2_1(m_0)}{2J^2_0(m_0)}\cdot m_\mrm{\theta}\cos(\rho-\phi^\mrm{usb}_\mrm{tar}-\phi^\mrm{car}_\mrm{tar})\big|_{2\omega_\mrm{het}+\Delta\omega_\mrm{m}} + \frac{J^2_1(m_0)}{2J^2_0(m_0)}\cdot m_\mrm{\theta}\cos(\rho+\phi^\mrm{lsb}_\mrm{tar}-\phi^\mrm{car}_\mrm{tar})\big|_{2\omega_\mrm{het}-\Delta\omega_\mrm{m}},
    \label{eq:thetanoise_mutual_sb2c}
\end{align}
\end{widetext}
where the noise at $\Delta\omega_\mrm{m}$ cancels between the usb-usb and lsb-lsb beatnotes.
\\\\
\noindent\emph{Sideband-sideband beatnote to sideband-sideband beatnote: }
Finally, discussed is the coupling of the noise in one sideband-sideband beatnote to phase extraction of the other sideband-sideband beatnote.
Therefore, we can focus on the following part of the PR signal in \cref{eq:spr},
\begin{align}
    s^\mrm{sb2sb}_\mrm{pr} &= A_iA_jJ^2_1(m_0)\left(1+a_i\right)\cos\left((\omega_\mrm{het}+\Delta\omega_\mrm{m})\tau + \phi^\mrm{usb}_\mrm{tar} + p_i + \theta_i\right)
    \nonumber\\
    &\hspace{1mm} + A_iA_jJ^2_1(m_0)\left(1+a_i\right)\cos\left((\omega_\mrm{het}-\Delta\omega_\mrm{m})\tau - \phi^\mrm{lsb}_\mrm{tar} + p_i - \theta_i\right).
    \label{eq:ssb2sb_prifo}
\end{align}
In this type of noise coupling, $\omega_\mrm{A}=\omega_\mrm{het}\pm\Delta\omega_\mrm{m}$ and $\omega_\mrm{B}=\omega_\mrm{het}\mp\Delta\omega_\mrm{m}$ in \cref{eq:mutual_general}.
The relevant noise frequencies are
\begin{enumerate}[label=(\alph*)]
    \item The usb-usb to the lsb-lsb: $\omega_\mrm{n}=-\omega_\mrm{A}+\omega_\mrm{B}=2\Delta\omega_\mrm{m}$ and $\omega_\mrm{n}=\omega_\mrm{A}+\omega_\mrm{B}=2\omega_\mrm{het}$,
    \item The lsb-lsb to the usb-usb: $\omega_\mrm{n}=\omega_\mrm{A}-\omega_\mrm{B}=2\Delta\omega_\mrm{m}$ and $\omega_\mrm{n}=\omega_\mrm{A}+\omega_\mrm{B}=2\omega_\mrm{het}$.
\end{enumerate}
Therefore, we have two frequencies to investigate, both of which need to be carefully analyzed in terms of the coherence between the usb-usb and lsb-lsb beatnotes.
In addition, note that the noise at $2\omega_\mrm{het}$ also exists in the self noise coupling of the carrier-carrier beatnote, see discussions around \cref{eq:phi_read_2f_car}, which becomes important if we combine the sideband phases with the carrier.
% This suggests that we also need to take care of how the noise behaves once we combine the sideband phases with the carrier.

First, with respect to the modulation amplitude noise $a_i$, we compare \cref{eq:ssb2sb_prifo} under $p_i=0$ and $\theta_i=0$ with \cref{eq:ampnoise_mutual_general}.
It suggests that we can substitute quantities as follows: $A_0 = A_iA_jJ^2_1(m_0)$, $B_0 = A_iA_jJ^2_1(m_0)$, $m_\mrm{n} = m_\mrm{a}$, $\phi_\mrm{tar}=\phi^\mrm{usb}_\mrm{tar}\text{ or }-\phi^\mrm{lsb}_\mrm{tar}$, and $\phi_\mrm{B}=-\phi^\mrm{lsb}_\mrm{tar}\text{ or }\phi^\mrm{usb}_\mrm{tar}$.
Therefore, the general expression of the phase readout in \cref{eq:phi_read_mutual_ampnoise} gives
\begin{subequations}\label{eq:ampnoise_mutual_sb2sb}
    \begin{align}
        \phi^\mrm{usb}_\mrm{read} &= \phi^\mrm{usb}_\mrm{tar} + \frac{1}{2}\cdot m_\mrm{a}\sin(\rho-2\phi^\mrm{sb}_\mrm{tar})\big|_{2\Delta\omega_\mrm{m}}
        \nonumber\\
        &\hspace{9.5mm} + \frac{1}{2}\cdot m_\mrm{a}\sin(\rho-2\phi^\mrm{car}_\mrm{tar})\big|_{2\omega_\mrm{het}},
        \label{eq:ampnoise_mutual_lsb2usb}\\
        \phi^\mrm{lsb}_\mrm{read} &= -\phi^\mrm{lsb}_\mrm{tar} - \frac{1}{2}\cdot m_\mrm{a}\sin(\rho-2\phi^\mrm{sb}_\mrm{tar})\big|_{2\Delta\omega_\mrm{m}}
        \nonumber\\
        &\hspace{11.5mm} + \frac{1}{2}\cdot m_\mrm{a}\sin(\rho-2\phi^\mrm{car}_\mrm{tar})\big|_{2\omega_\mrm{het}},
        \label{eq:ampnoise_mutual_usb2lsb}
    \end{align}
\end{subequations}
Once we combine the two phases as \cref{eq:sbcombi}, the noise at $2\omega_\mrm{het}$ cancels, while the noise at $2\Delta\omega_\mrm{m}$ coherently sums up.

Second, we analyze the laser phase noise $p_i$ by comparing \cref{eq:ssb2sb_prifo} under $a_i=0$ and $\theta_i=0$ with \cref{eq:phasenoise_mutual_general}.
We can substitute quantities as follows: $A_0 = A_iA_jJ^2_1(m_0)$, $B_0 = A_iA_jJ^2_1(m_0)$, $m_\mrm{n} = m_\mrm{p}$, $\phi_\mrm{tar}=\phi^\mrm{usb}_\mrm{tar}\text{ or }-\phi^\mrm{lsb}_\mrm{tar}$, and $\phi_\mrm{B}=-\phi^\mrm{lsb}_\mrm{tar}\text{ or }\phi^\mrm{usb}_\mrm{tar}$, which result in
%where $a_i=m_\mrm{a}\cos(\omega_\mrm{n}\tau+\rho)$ is the monotonic noise representation like \cref{eq:p_i_2omegahet}.
\begin{subequations}\label{eq:phasenoise_mutual_sb2sb}
    \begin{align}
        \phi^\mrm{usb}_\mrm{read} &= \phi_\mrm{tar} + \frac{1}{2}\cdot m_\mrm{p}\cos(\rho-2\phi^\mrm{sb}_\mrm{tar})\big|_{2\Delta\omega_\mrm{m}} 
        \nonumber\\
        &\hspace{9mm} - \frac{1}{2}\cdot m_\mrm{p}\cos(\rho-2\phi^\mrm{car}_\mrm{tar})\big|_{2\omega_\mrm{het}},
        \label{eq:phasenoise_mutual_lsb2usb}\\
        \phi^\mrm{lsb}_\mrm{read} &= -\phi_\mrm{tar} + \frac{1}{2}\cdot m_\mrm{p}\cos(\rho-2\phi^\mrm{sb}_\mrm{tar})\big|_{2\Delta\omega_\mrm{m}} 
        \nonumber\\
        &\hspace{11.5mm} - \frac{1}{2}\cdot m_\mrm{p}\cos(\rho-2\phi^\mrm{car}_\mrm{tar})\big|_{2\omega_\mrm{het}}.
        \label{eq:phasenoise_mutual_usb2lsb}
    \end{align}
\end{subequations}
This suggests that the combination in \cref{eq:sbcombi} is insensitive to laser phase noise at those frequencies.

Last but not least, as for the modulation phase noise $\theta_i$, we can apply the almost same discussion as the laser phase noise above, but with the sign flip for the coupling of the noises from the lsb-lsb beatnote to the usb-usb beatnote.
Therefore, we get the measured sideband-sideband beatnote phases as,
\begin{subequations}\label{eq:thetanoise_mutual_sb2sb}
    \begin{align}
        \phi^\mrm{usb}_\mrm{read} &= \phi_\mrm{tar} - \frac{1}{2}\cdot m_\mrm{\theta}\cos(\rho-2\phi^\mrm{sb}_\mrm{tar})\big|_{2\Delta\omega_\mrm{m}}
        \nonumber\\
        &\hspace{9mm} + \frac{1}{2}\cdot m_\mrm{\theta}\cos(\rho-2\phi^\mrm{car}_\mrm{tar})\big|_{2\omega_\mrm{het}},
        \label{eq:thetanoise_mutual_lsb2usb}\\
        \phi^\mrm{lsb}_\mrm{read} &= -\phi_\mrm{tar} + \frac{1}{2}\cdot m_\mrm{\theta}\cos(\rho-2\phi^\mrm{sb}_\mrm{tar})\big|_{2\Delta\omega_\mrm{m}}
        \nonumber\\
        &\hspace{11.5mm} - \frac{1}{2}\cdot m_\mrm{\theta}\cos(\rho-2\phi^\mrm{car}_\mrm{tar})\big|_{2\omega_\mrm{het}}.
        \label{eq:thetanoise_mutual_usb2lsb}
    \end{align}
\end{subequations}
Therefore, the noises at the same frequencies coherently sum up in the combination in \cref{eq:sbcombi}; hence, they cannot be suppressed.

The results of the mutual noise coupling in this section are summarized in \cref{tab:noise_couplings} as its non-diagonal elements.
\begin{table*}
    \centering
    \caption{\label{tab:noise_couplings}
    Summary table of the heterodyne-band noise couplings in the time domain.
    They couple from the secondary beatnote as a noise source (rows) to the measured phase of the primary beatnote (columns).
    The couplings are shown in the form of ``noise type(angular frequency)[coupling factor]" where the coupling factor is defined by (coupled noise term)/(original noise amplitude).
    Only one of the two interfering beams is considered as a noise source.
    The time-domain coupling factors shown here can be converted to those in ASD by an additional factor of $\sqrt{2}$, as described in \cref{app:conversion}.
    The laser phase noise $p_i$, the modulation amplitude noise $a_i$, and the modulation phase noise $\theta_i$ are considered, while the modulation additive voltage noise $n_i$ has identical effects to $p_i$ as discussed at the end of \cref{sub:framework_beam}.
    By definition, the diagonal elements are the self noise couplings in \cref{sub:het_band_self}, while the non-diagonal elements are the mutual noise couplings in \cref{sub:het_band_mutual}.
    Symbols before some noise couplings: $\vee$ = noises that coherently cancel with the same numbering; $\wedge$ = noises that coherently sum up with the same numbering.
    For simplicity, we define the followings: $\tilde{\phi}^\mrm{usb}_\mrm{tar} = \phi^\mrm{usb}_\mrm{tar} + \phi^\mrm{car}_\mrm{tar}$ and $\tilde{\phi}^\mrm{lsb}_\mrm{tar} = \phi^\mrm{lsb}_\mrm{tar} - \phi^\mrm{car}_\mrm{tar}$.
    We only provide mathematical expressions here, which can be used also for subsequent theoretical work like time-delay interferometry, while leaving the substitution of concrete values for \cref{sec:verification,sec:application}.
    \vspace{1.5mm}
    }
    \begin{tabular}{cccc}
    \toprule\vspace{1.5mm}
     & car-car phase: $\phi^\mrm{car}_\mrm{read}$  & usb-usb phase: $\phi^\mrm{usb}_\mrm{read}$ & lsb-lsb phase: $\phi^\mrm{lsb}_\mrm{read}$ \\
    \hline \\\vspace{1.5mm}
    %%% car-car %%%%%%%%%%%%
    \multirow{4}{*}{car-car}
    % \multirow{4}{*}{}
        & $p_i(\epsilon)$ $\left[\cos\rho\right]$
        & $^{\vee1} p_i(\Delta \omega_\mrm{m})$ $\left[\frac{J^2_0(m_0)}{2J^2_1(m_0)}\cos(\rho-\phi^\mrm{sb}_\mrm{tar})\right]$
        & $^{\vee1} p_i(\Delta \omega_\mrm{m})$ $\left[\frac{J^2_0(m_0)}{2J^2_1(m_0)}\cos(\rho-\phi^\mrm{sb}_\mrm{tar})\right]$
        \\\vspace{1.5mm}
        & $p_i(2\omega_\mrm{het})$ $\left[-\frac{1}{2}\cos(\rho-2\phi^\mrm{car}_\mrm{tar})\right]$
        & $p_i(2\omega_\mrm{het} + \Delta \omega_\mrm{m})$ $\left[-\frac{J^2_0(m_0)}{2J^2_1(m_0)}\cos(\rho-\tilde{\phi}^\mrm{usb}_\mrm{tar})\right]$
        & $p_i(2\omega_\mrm{het} - \Delta \omega_\mrm{m})$ $\left[-\frac{J^2_0(m_0)}{2J^2_1(m_0)}\cos(\rho+\tilde{\phi}^\mrm{lsb}_\mrm{tar})\right]$
        \\\vspace{1.5mm}
        & $a_i(2\omega_\mrm{het})$ $\left[-\frac{m^2_0}{4J_0(m_0)}\sin(\rho-2\phi^\mrm{car}_\mrm{tar})\right]$
        & $^{\wedge1} a_i(\Delta \omega_\mrm{m})$ $\left[- \frac{J_0(m_0)m^2_0}{4J^2_1(m_0)}\sin(\rho-\phi^\mrm{sb}_\mrm{tar})\right]$
        & $^{\wedge1} a_i(\Delta \omega_\mrm{m})$ $\left[\frac{J_0(m_0)m^2_0}{4J^2_1(m_0)}\sin(\rho-\phi^\mrm{sb}_\mrm{tar})\right]$
        \\\vspace{1.5mm}
        & 
        & $a_i(2\omega_\mrm{het} + \Delta \omega_\mrm{m})$ $\left[- \frac{J_0(m_0)m^2_0}{4J^2_1(m_0)}\sin(\rho-\tilde{\phi}^\mrm{usb}_\mrm{tar})\right]$
        & $a_i(2\omega_\mrm{het} - \Delta \omega_\mrm{m})$ $\left[- \frac{J_0(m_0)m^2_0}{4J^2_1(m_0)}\sin(\rho+\tilde{\phi}^\mrm{lsb}_\mrm{tar})\right]$
        \\
    \midrule \\\vspace{1.5mm}
    %%% usb-usb %%%%%%%%%%%%
    \multirow{6}{*}{usb-usb}
    % \multirow{6}{*}{}
        & $^{\wedge2} p_i(\Delta \omega_\mrm{m})$ $\left[\frac{J^2_1(m_0)}{2J^2_0(m_0)}\cos(\rho-\phi^\mrm{sb}_\mrm{tar})\right]$
        & $^{\vee2} p_i(\epsilon)$ $\left[\cos\rho\right]$
        & $^{\vee3} p_i(2\Delta \omega_\mrm{m})$ $\left[\frac{1}{2}\cos(\rho-2\phi^\mrm{sb}_\mrm{tar})\right]$
        \\\vspace{1.5mm}
        & $p_i(2\omega_\mrm{het} + \Delta \omega_\mrm{m})$ $\left[-\frac{J^2_1(m_0)}{2J^2_0(m_0)}\cos(\rho-\tilde{\phi}^\mrm{usb}_\mrm{tar})\right]$
        & $p_i(2(\omega_\mrm{het}+\Delta \omega_\mrm{m}))$ $\left[-\frac{1}{2}\cos(\rho-2\phi^\mrm{usb}_\mrm{tar})\right]$
        &  $^{\vee4} p_i(2 \omega_\mrm{het})$ $\left[-\frac{1}{2}\cos(\rho-2\phi^\mrm{car}_\mrm{tar})\right]$
        \\\vspace{1.5mm}
        & $^{\vee5} a_i(\Delta \omega_\mrm{m})$ $\left[-\frac{J_1^2(m_0)}{2J^2_0(m_0)}\sin(\rho-\phi^\mrm{sb}_\mrm{tar})\right]$
        & $a_i(2(\omega_\mrm{het}+\Delta \omega_\mrm{m}))$ $\left[\frac{1}{2}\sin(\rho-2\phi^\mrm{usb}_\mrm{tar})\right]$
        & $^{\wedge3} a_i(2\Delta \omega_\mrm{m})$ $\left[-\frac{1}{2}\sin(\rho-2\phi^\mrm{sb}_\mrm{tar})\right]$
        \\\vspace{1.5mm}
        & $a_i(2\omega_\mrm{het}+\Delta \omega_\mrm{m})$ $\left[\frac{J^2_1(m_0)}{2J^2_0(m_0)}\sin(\rho-\tilde{\phi}^\mrm{usb}_\mrm{tar})\right]$
        & $^{\wedge4} \theta_i(\epsilon)$ $[\cos\rho]$
        & $^{\vee6} a_i(2 \omega_\mrm{het})$ $\left[\frac{1}{2}\sin(\rho-2\phi^\mrm{car}_\mrm{tar})\right]$
        \\\vspace{1.5mm}
        & $^{\vee7} \theta_i(\Delta \omega_\mrm{m})$ $\left[\frac{J^2_1(m_0)}{2J^2_0(m_0)}\cos(\rho-\phi^\mrm{sb}_\mrm{tar})\right]$
        & $\theta_i(2(\omega_\mrm{het}+\Delta \omega_\mrm{m}))$ $\left[-\frac{1}{2}\cos(\rho-2\phi^\mrm{usb}_\mrm{tar})\right]$
        & $^{\wedge5} \theta_i(2\Delta \omega_\mrm{m})$ $\left[\frac{1}{2}\cos(\rho-2\phi^\mrm{sb}_\mrm{tar})\right]$
        \\\vspace{1.5mm}
        & $\theta_i(2\omega_\mrm{het}+\Delta \omega_\mrm{m})$ $\left[-\frac{J^2_1(m_0)}{2J^2_0(m_0)}\cos(\rho-\tilde{\phi}^\mrm{usb}_\mrm{tar})\right]$
        &
        & $^{\wedge6} \theta_i(2 \omega_\mrm{het})$ $\left[-\frac{1}{2}\cos(\rho-2\phi^\mrm{car}_\mrm{tar})\right]$
        \\
    \midrule \\\vspace{1.5mm}
    %%% lsb-lsb %%%%%%%%%%%%
    \multirow{6}{*}{lsb-lsb}
    % \multirow{6}{*}{}
        & $^{\wedge2} p_i(\Delta \omega_\mrm{m})$ $\left[\frac{J^2_1(m_0)}{2J^2_0(m_0)}\cos(\rho-\phi^\mrm{sb}_\mrm{tar})\right]$
        & $^{\vee3} p_i(2\Delta \omega_\mrm{m})$ $\left[\frac{1}{2}\cos(\rho-2\phi^\mrm{sb}_\mrm{tar})\right]$
        &  $^{\vee2} p_i(\epsilon)$ $\left[\cos\rho\right]$
        \\\vspace{1.5mm}
        & $p_i(2\omega_\mrm{het} - \Delta \omega_\mrm{m})$ $\left[-\frac{J^2_1(m_0)}{2J^2_0(m_0)}\cos(\rho+\tilde{\phi}^\mrm{lsb}_\mrm{tar})\right]$
        & $^{\vee4} p_i(2 \omega_\mrm{het})$ $\left[-\frac{1}{2}\cos(\rho-2\phi^\mrm{car}_\mrm{tar})\right]$
        &  $p_i(2(\omega_\mrm{het}-\Delta \omega_\mrm{m}))$ $\left[-\frac{1}{2}\cos(\rho+2\phi^\mrm{lsb}_\mrm{tar})\right]$
        \\\vspace{1.5mm}
        & $^{\vee5} a_i(\Delta \omega_\mrm{m})$ $\left[\frac{J_1^2(m_0)}{2J^2_0(m_0)}\sin(\rho-\phi^\mrm{sb}_\mrm{tar})\right]$
        & $^{\wedge3} a_i(2\Delta \omega_\mrm{m})$ $\left[\frac{1}{2}\sin(\rho-2\phi^\mrm{sb}_\mrm{tar})\right]$
        & $a_i(2(\omega_\mrm{het}-\Delta \omega_\mrm{m}))$ $\left[\frac{1}{2}\sin(\rho+2\phi^\mrm{lsb}_\mrm{tar})\right]$
        \\\vspace{1.5mm}
        & $a_i(2\omega_\mrm{het}-\Delta \omega_\mrm{m})$ $\left[\frac{J^2_1(m_0)}{2J^2_0(m_0)}\sin(\rho+\tilde{\phi}^\mrm{lsb}_\mrm{tar})\right]$
        & $^{\vee6} a_i(2 \omega_\mrm{het})$ $\left[\frac{1}{2}\sin(\rho-2\phi^\mrm{car}_\mrm{tar})\right]$
        & $^{\wedge4} \theta_i(\epsilon)$ $\left[-\cos\rho\right]$
        \\\vspace{1.5mm}
        & $^{\vee7} \theta_i(\Delta \omega_\mrm{m})$ $\left[-\frac{J^2_1(m_0)}{2J^2_0(m_0)}\cos(\rho-\phi^\mrm{sb}_\mrm{tar})\right]$
        & $^{\wedge5} \theta_i(2\Delta \omega_\mrm{m})$ $\left[-\frac{1}{2}\cos(\rho-2\phi^\mrm{sb}_\mrm{tar})\right]$
        & $\theta_i(2(\omega_\mrm{het}-\Delta \omega_\mrm{m}))$ $\left[\frac{1}{2}\cos(\rho+2\phi^\mrm{lsb}_\mrm{tar})\right]$
        \\\vspace{1.5mm}
        & $\theta_i(2\omega_\mrm{het}-\Delta \omega_\mrm{m})$ $\left[\frac{J^2_1(m_0)}{2J^2_0(m_0)}\cos(\rho+\tilde{\phi}^\mrm{lsb}_\mrm{tar})\right]$
        & $^{\wedge6} \theta_i(2 \omega_\mrm{het})$ $\left[\frac{1}{2}\cos(\rho-2\phi^\mrm{car}_\mrm{tar})\right]$
        & 
        \\
    \bottomrule
    \end{tabular}
\end{table*}

%%%%%%%%%%%%%%%%%%%%%%%%%%%%%%%%%%%%%%%%%%%%%%%%%%%%%%%%%
\section{Modulation-band noises}\label{sec:mod_band}
Here we analyze the noise coupling from frequencies around integer multiples of the modulation angular frequency $\omega_{\mrm{m},i}$ $(=2\pi f_{\mrm{m},i})$ to phase extraction by a phasemeter.
Two types of analysis will be performed: \emph{trigonometric analysis} and \emph{sideband analysis}.
The trigonometric analysis in \cref{sub:mod_band_trig} focuses on the phase part of a single laser beam, even neglecting the fact that it is the phase of an electromagnetic field.
This approach is straightforward and can find most of major noise couplings.
On the other hand, the sideband analysis in \cref{sub:mod_band_sideband} expands a single laser beam by noise sidebands to capture noise coupling mechanisms, which the trigonometric analysis cannot find, especially the mixing between the modulation sidebands and the noise sidebands. 

\subsection{Trigonometric analysis}\label{sub:mod_band_trig}
Let us extract only the laser phase noise $p_i$ and the phase modulation signal in \cref{eq:Ei_phim} as
\begin{align}
    \phi_{m,i} = p_i + m_i \cos(\omega_{\mrm{m},i}\tau + \phi^\mrm{sb}_{\mrm{tar},i} + \theta_i).
    \label{eq:phi_m}
\end{align}
As noted at the end of \cref{sub:framework_beam}, $p_i$ can implicitly include the modulation voltage noise $n_i$.
% Note that we also explicitly write the target phase of the modulation signal $\phi_{\mrm{clk},i}$.
We will show that noise in the modulation-frequency regime can be virtually downconverted with an amplitude scaling factor and reinterpreted as noise in the heterodyne-band regime, which then couples as illustrated in \cref{fig:het-band-noises-illustration}.
This enables us to derive the total coupling factor as a product of the amplitude scaling factor arising from the frequency downconversion and the coupling factor of the heterodyne-band noise listed in \cref{tab:noise_couplings}.
Following the monotonic noise representation in \cref{eq:x_i}, we consider a given noise $x_i$ at the angular frequency of $\omega_\mrm{n} = n\omega_{\mrm{m},i} + \Delta\omega_x$ ($n\in \mathbb{N}$) where $\Delta\omega_x$ is a given offset frequency from the integer multiple of the modulation frequency.
By individually substituting $p_i$, $a_i$, and $\theta_i$ into $x_i$ in the monotonic noise representation, we discuss the different noise sources below.
\\\\
\noindent\emph{Noise frequency conversion: }
First, let us focus on $p_i$ under $a_i=\theta_i=0$ in \cref{eq:phi_m}.
By the harmonic addition theorem, \cref{eq:phi_m} under this condition becomes
\begin{align}
    \phi_{m,i} &= m_\mrm{p}\cos((n\omega_{\mrm{m},i} + \Delta\omega_\mrm{p})\tau + \rho) + m_0 \cos(\omega_{\mrm{m},i}\tau + \phi^\mrm{sb}_{\mrm{tar},i})
    \nonumber\\
    &= m'_i \cos(\omega_{\mrm{m},i}\tau + \phi'_i),
    \label{eq:phi_m_mp}
\end{align}
where the new modulation index $m'_i$ and phase $\phi'_i$ are  given by
\begin{align}
    m'_i &\approx m_0\sqrt{1 + 2\frac{m_\mrm{p}}{m_0}\cos((n\omega_{\mrm{m},i}\Delta\omega_\mrm{p})\tau+\rho-\phi^\mrm{sb}_{\mrm{tar},i})}
    \nonumber\\
    &\approx m_0\left(1 + \frac{m_\mrm{p}}{m_0}\cos((n\omega_{\mrm{m},i}+\Delta\omega_\mrm{p})\tau+\rho-\phi^\mrm{sb}_{\mrm{tar},i})\right),
    \label{eq:mi_mp}\\
    \tan\phi'_i &= \frac{m_0\sin\phi^\mrm{sb}_{\mrm{tar},i} + m_\mrm{p}\sin((n\omega_{\mrm{m},i}+\Delta\omega_\mrm{p})\tau+\rho)}{m_0\cos\phi^\mrm{sb}_{\mrm{tar},i} + m_\mrm{p}\cos((n\omega_{\mrm{m},i}+\Delta\omega_\mrm{p})\tau+\rho)}
    \nonumber\\
    \Rightarrow \phi'_i &\approx \phi^\mrm{sb}_{\mrm{tar},i} + \frac{m_\mrm{p}}{m_0}\sin((n\omega_{\mrm{m},i}+\Delta\omega_\mrm{p})\tau+\rho-\phi^\mrm{sb}_{\mrm{tar},i}).
    \label{eq:phii_mp}
\end{align}
Only the linear term of $m_\mrm{p}$ is considered, such as $m_\mrm{p}\ll1$ and $m_\mrm{p}/m_0\ll1$.
\cref{eq:mi_mp,eq:phii_mp} suggest that the laser noise at $n\omega_{\mrm{m},i} + \Delta\omega_\mrm{p}$ is virtually downconverted by $\omega_{\mrm{m},i}$ to the modulation amplitude and phase noise at $(n-1)\omega_{\mrm{m},i} + \Delta\omega_\mrm{p}$ with an amplitude scaling factor of $1/m_0$.
The physical interpretation of this result is as follows: the sinusoidal tone at $n\omega_{\mrm{m},i}+\Delta\omega_\mrm{p}$ looks as a single sideband at $(n-1)\omega_{\mrm{m},i}+\Delta\omega_\mrm{p}$ from the perspective of another tone at $\omega_{\mrm{m},i}$ (= modulation signal in this case).
As discussed in \cref{app:ssb}, the single sideband is  understood as a combination of the coherent amplitude and phase modulation with the relative phase shift of $\pi/2$, which is consistent with \cref{eq:mi_mp,eq:phii_mp}.

Second, regarding the modulation amplitude noise $a_i$, \cref{eq:phi_m} under $p_i=\theta_i=0$ becomes
\begin{widetext}
\begin{align}
    \phi_{m,i} &= m_0(1+m_\mrm{a}\cos((n\omega_{\mrm{m},i} + \Delta\omega_\mrm{a})\tau + \rho)) \cos(\omega_{\mrm{m},i}\tau + \phi^\mrm{sb}_{\mrm{tar},i})
    \nonumber\\
    &= m_0\cos(\omega_{\mrm{m},i}\tau + \phi^\mrm{sb}_{\mrm{tar},i}) + \frac{m_0m_\mrm{a}}{2}\left(\cos(((n-1)\omega_{\mrm{m},i} + \Delta\omega_\mrm{a})\tau + \rho-\phi^\mrm{sb}_{\mrm{tar},i}) + \cos(((n+1)\omega_{\mrm{m},i} + \Delta\omega_\mrm{a})\tau + \rho+\phi^\mrm{sb}_{\mrm{tar},i})\right).
    \label{eq:phii_ma}
\end{align}
Hence, the modulation amplitude noise $a_i$ at $n\omega_{\mrm{m},i} + \Delta\omega_\mrm{a}$ turns into the laser phase noise at $(n-1)\omega_{\mrm{m},i} + \Delta\omega_\mrm{a}$ (down-conversion) and $(n+1)\omega_{\mrm{m},i}+\Delta\omega_\mrm{a}$ (up-conversion) with an amplitude scaling factor of $m_0/2$.

Third, regarding the modulation phase noise $\theta_i$, \cref{eq:phi_m} under $p_i=a_i=0$ becomes
\begin{align}
    \phi_{m,i} &= m_0\cos(\omega_{\mrm{m},i}\tau + \phi^\mrm{sb}_{\mrm{tar},i} + m_\mrm{\theta}\cos((n\omega_{\mrm{m},i} + \Delta\omega_\mrm{\theta})\tau + \rho))
    \nonumber\\
    &\approx m_0\cos(\omega_{\mrm{m},i}\tau + \phi^\mrm{sb}_{\mrm{tar},i}) + \frac{m_0m_\mrm{\theta}}{2}\left(\sin(((n-1)\omega_{\mrm{m},i}+\Delta\omega_\mrm{\theta}) \tau + \rho -\phi^\mrm{sb}_{\mrm{tar},i})-\sin(((n+1)\omega_{\mrm{m},i}+\Delta\omega_\mrm{\theta})\tau+\phi^\mrm{sb}_{\mrm{tar},i}+\rho)\right)
    \label{eq:phii_mtheta}
\end{align}
where we consider only the first-order Jacobi Anger and Taylor expansion in terms of $m_\mrm{\theta}$.
Hence, the modulation phase noise $\theta_i$ at $n\omega_{\mrm{m},i} + \Delta\omega_\mrm{\theta}$ turns into the laser phase noise at $(n-1)\omega_{\mrm{m},i} + \Delta\omega_\mrm{\theta}$ (down-conversion) and $(n+1)\omega_{\mrm{m},i} + \Delta\omega_\mrm{\theta}$ (up-conversion) with an amplitude scaling factor of $m_0/2$.
\\\\
\noindent\emph{Cascaded analysis: }
The analysis above suggests that the laser phase noise $p_i$ is virtually down-converted by $\omega_{\mrm{m},i}$ to the modulation amplitude and phase noises ($a_i$ and $\theta_i$), and vice versa.
This implies that the down-conversions occur in a cascaded manner.
To describe this functionally, let us write a given noise $x_i$ in \cref{eq:x_i} as a function of the angular frequency $\omega_\mrm{n}$, parameterized by the noise amplitude $m_x$ and the random phase $\rho$ as:
 \begin{align}
    x_i(\omega_\mrm{n}; m_x, \rho) &= m_x \cos(\omega_\mrm{n} \tau + \rho).
    \label{eq:x_i_2}
\end{align}
Based on this, the process of the frequency downconversion of the laser phase noise $p_i$ can be expressed as:
\begin{align}
    p_i(n\omega_{\mrm{m},i}+\Delta\omega_\mrm{p}; m_\mrm{p}, \rho) &\rightarrow
    \begin{cases}
        a_i\left((n-1)\omega_{\mrm{m},i}+\Delta\omega_\mrm{p};\frac{m_\mrm{p}}{m_0},\rho-\phi^\mrm{sb}_{\mrm{tar},i}\right) &
        \\
        \theta_i\left((n-1)\omega_{\mrm{m},i}+\Delta\omega_\mrm{p};\frac{m_\mrm{p}}{m_0},\rho-\phi^\mrm{sb}_{\mrm{tar},i}-\frac{\pi}{2}\right). &
    \end{cases}
    \label{eq:pi2ai_2thetai}
\end{align}
The right arrow represents the virtual frequency conversion process, and the terms listed on the right-hand side are the resulting noises.
On the other hand, the modulation amplitude and phase noises at $n\omega_{\mrm{m},i}+\Delta\omega_\mrm{a}$ and $n\omega_{\mrm{m},i}+\Delta\omega_\mrm{\theta}$ turn into the laser noise as,
\begin{align}
    a_i(n\omega_{\mrm{m},i}+\Delta\omega_\mrm{a}; m_\mrm{a}, \rho) &\rightarrow p_i\left((n-1)\omega_{\mrm{m},i}+\Delta\omega_\mrm{a};\frac{m_0m_\mrm{a}}{2},\rho-\phi^\mrm{sb}_{\mrm{tar},i}\right)
    \nonumber\\
    &\hspace{5mm} + p_i\left((n+1)\omega_{\mrm{m},i}+\Delta\omega_\mrm{a};\frac{m_0m_\mrm{a}}{2},\rho+\phi^\mrm{sb}_{\mrm{tar},i}\right),
    \label{eq:ai2pi}
    \\
    \theta_i(n\omega_{\mrm{m},i}+\Delta\omega_\mrm{\theta}; m_\mrm{\theta}, \rho) &\rightarrow p_i\left((n-1)\omega_{\mrm{m},i}+\Delta\omega_\mrm{\theta};\frac{m_0m_\mrm{\theta}}{2},\rho-\phi^\mrm{sb}_{\mrm{tar},i}-\frac{\pi}{2}\right)
    \nonumber\\
    &\hspace{5mm} + p_i\left((n+1)\omega_{\mrm{m},i}+\Delta\omega_\mrm{a};\frac{m_0m_\mrm{\theta}}{2},\rho+\phi^\mrm{sb}_{\mrm{tar},i}+\frac{\pi}{2}\right).
    \label{eq:thetai2pi}
\end{align}
If we combine \cref{eq:pi2ai_2thetai,eq:ai2pi} for the coupling path $p_i\rightarrow a_i\rightarrow p_i$ and \cref{eq:pi2ai_2thetai,eq:thetai2pi} for the coupling path $p_i\rightarrow \theta_i\rightarrow p_i$, we get
\begin{align}
    p_i(n\omega_{\mrm{m},i}+\Delta\omega_\mrm{p}; m_\mrm{p}, \rho) &\rightarrow
    \begin{cases}
        a_i\left((n-1)\omega_{\mrm{m},i}+\Delta\omega_\mrm{p};\frac{m_\mrm{p}}{m_0},\rho-\phi^\mrm{sb}_{\mrm{tar},i}\right) &
        \\
        \theta_i\left((n-1)\omega_{\mrm{m},i}+\Delta\omega_\mrm{p};\frac{m_\mrm{p}}{m_0},\rho-\phi^\mrm{sb}_{\mrm{tar},i}-\frac{\pi}{2}\right), &
    \end{cases}
    \nonumber\\
    &\rightarrow
    p_i\left((n-2)\omega_{\mrm{m},i}+\Delta\omega_\mrm{p};\frac{m_\mrm{p}}{2},\rho-2\phi^\mrm{sb}_{\mrm{tar},i}\right) + p_i\left(n\omega_{\mrm{m},i}+\Delta\omega_\mrm{p};\frac{m_\mrm{p}}{2},\rho\right)
    \nonumber\\
    &\hspace{5mm} + p_i\left((n-2)\omega_{\mrm{m},i}+\Delta\omega_\mrm{p};\frac{m_\mrm{p}}{2},\rho-2\phi^\mrm{sb}_{\mrm{tar},i}-\pi\right) + p_i\left(n\omega_{\mrm{m},i}+\Delta\omega_\mrm{p};\frac{m_\mrm{p}}{2},\rho\right)
    \nonumber\\
    &= 0 + p_i(n\omega_{\mrm{m},i}+\Delta\omega_\mrm{p}; m_\mrm{p}, \rho)
    \label{eq:pi2aithetai2pi}
\end{align}
\end{widetext}
% where the up-conversion parts in \cref{eq:ai2pi,eq:thetai2pi} are neglected.
The resulting $p_i$ at $(n-2)\omega_{\mrm{m},i}+\Delta\omega_\mrm{p}$ via $a_i$ and $\theta_i$ at $(n-1)\omega_{\mrm{m},i}+\Delta\omega_\mrm{p}$ have a relative $\pi$ phase shift; therefore, they cancel and result in 0 in the last line.
The up-converted parts resulting from $a_i$ and $\theta_i$ just recovers the original laser phase noise at $n\omega_{\mrm{m},i}+\Delta\omega_\mrm{p}$.
Hence, in this framework, the laser phase noise $p_i$ can be down-converted only once, and if we try to downconvert it twice, the original laser phase noise is reproduced.

Let us start the same cascaded analysis with the modulation amplitude and phase noises.
Contrary to the laser phase noise down-conversion in \cref{eq:pi2aithetai2pi}, $a_i$ and $\theta_i$ are different noise sources.
Therefore, they are not expected to cancel even if we investigate them at the same frequency.
However, due to the cancellation at the second down-conversion of the laser phase noise above, the modulation amplitude and phase noises also have the limited number of down-conversions in this framework: up to two down-conversions.
The double down-conversions can be formulated as
\begin{widetext}
\begin{align}
    a_i(n\omega_{\mrm{m},i}+\Delta\omega_\mrm{a}; m_\mrm{a}, \rho) &\rightarrow p_i\left((n-1)\omega_{\mrm{m},i}+\Delta\omega_\mrm{a};\frac{m_0m_\mrm{a}}{2},\rho-\phi^\mrm{sb}_{\mrm{tar},i}\right)
    % \nonumber\\
    \rightarrow
    \begin{cases}
        a_i\left((n-2)\omega_{\mrm{m},i}+\Delta\omega_\mrm{a};\frac{m_\mrm{a}}{2},\rho-2\phi^\mrm{sb}_{\mrm{tar},i}\right) &
        \\
        \theta_i\left((n-2)\omega_{\mrm{m},i}+\Delta\omega_\mrm{a};\frac{m_\mrm{a}}{2},\rho-2\phi^\mrm{sb}_{\mrm{tar},i}-\frac{\pi}{2}\right), &
    \end{cases}
    \label{eq:ai2pi2ai}
    \\
    \theta_i(n\omega_{\mrm{m},i}+\Delta\omega_\mrm{\theta}; m_\mrm{\theta}, \rho) &\rightarrow p_i\left((n-1)\omega_{\mrm{m},i}+\Delta\omega_\mrm{\theta};\frac{m_0m_\mrm{\theta}}{2},\rho-\phi^\mrm{sb}_{\mrm{tar},i}-\frac{\pi}{2}\right)
    % \nonumber\\
    \rightarrow
    \begin{cases}
        a_i\left((n-2)\omega_{\mrm{m},i}+\Delta\omega_\mrm{\theta};\frac{m_\mrm{\theta}}{2},\rho-2\phi^\mrm{sb}_{\mrm{tar},i}-\frac{\pi}{2}\right) &
        \\
        \theta_i\left((n-2)\omega_{\mrm{m},i}+\Delta\omega_\mrm{\theta};\frac{m_\mrm{\theta}}{2},\rho-2\phi^\mrm{sb}_{\mrm{tar},i}-\pi\right). &
    \end{cases}
    \label{eq:thetai2pi2thetai}
\end{align}
\end{widetext}
The first right arrows only consider the frequency downconversion, neglecting the upconverted part in \cref{eq:ai2pi,eq:thetai2pi}
If we down-convert the resulting noise around $(n-2)\omega_{\mrm{m},i}$ once more, the same cancelation as \cref{eq:pi2aithetai2pi} happens again.

\begin{figure*}
    \centering
\includegraphics[width=12.9cm]{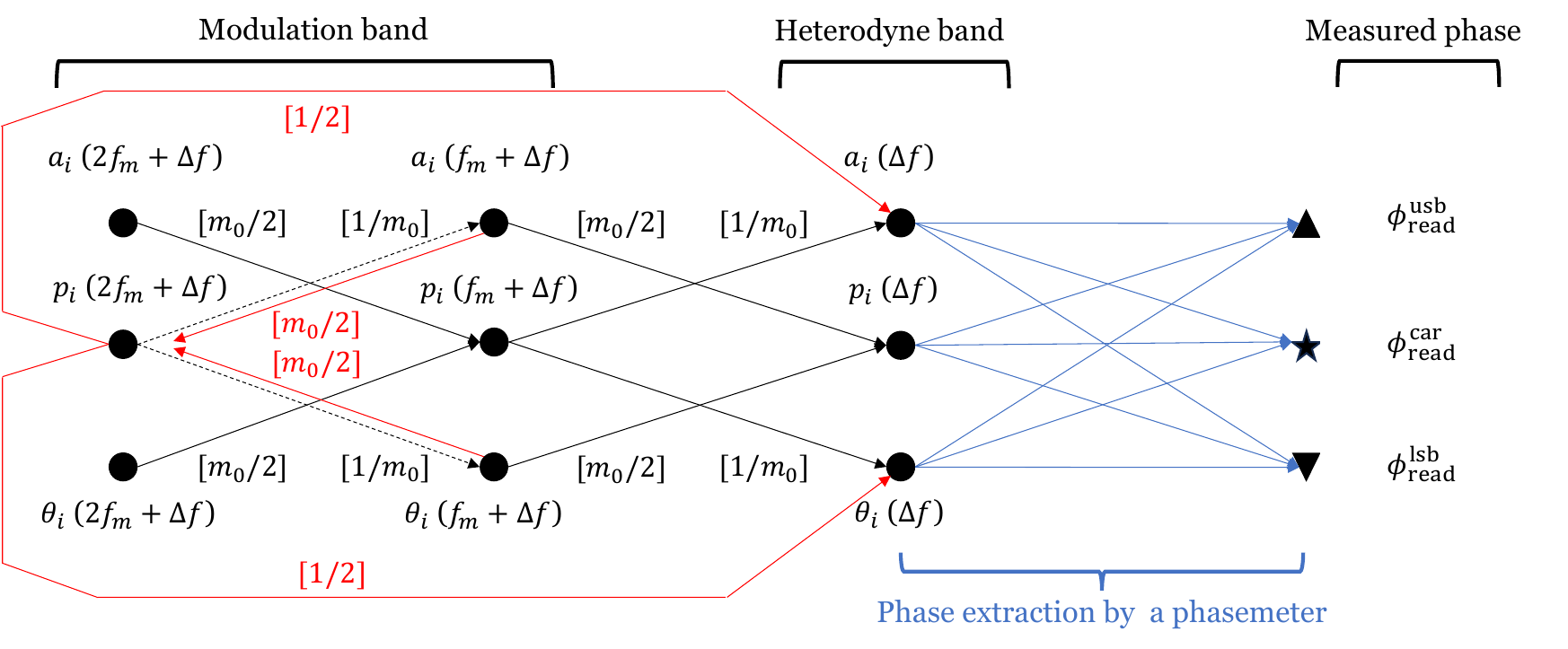}
    \caption{
    Diagram comprehensively visualizing coupling of the laser and modulation noise to phase extraction by a phasemeter.
    Blue arrows represent the coupling of noise at the heterodyne (or observation) band $\Delta f$, discussed in \cref{sec:het_band}.
    Black arrows represent the virtual frequency conversion of noises around the integer multiple of the modulation frequency $\omega_{\mrm{m},i}$ that are captured by the trigonometric analysis in \cref{sub:mod_band_trig}.
    Red arrows represent the modulation-band noise coupling, which can only be found in the sideband analysis in \cref{sub:mod_band_sideband}.
    The modulation-band noise coupling via the second-order modulation sidebands is not depicted.
    The noise scaling factors on individual paths are written in square brackets.
    The dashed arrows from $p_i(2f_m+\Delta f)$ to $a_i(f_m+\Delta f)$ and $\theta_i(f_m+\Delta f)$ represents the fact that the down-conversion exist but it does not couple to the heterodyne or observation band with further down conversions as derived in \cref{eq:pi2aithetai2pi}.
    }
    \label{fig:ghz_diagram}
\end{figure*}

The analysis of the virtual cascaded down-conversions is visualized in \cref{fig:ghz_diagram} in black.
The noise analysis in this section is naturally connected to that of the heterodyne-band noise via the downconversion.

\subsection{Sideband analysis}\label{sub:mod_band_sideband}
The trigonometric analysis in \cref{sub:mod_band_trig} captures many of the coupling paths from the modulation-band noise, however, not all of them.
There is a missing coupling mechanism via the mixing between the modulation sideband and phase noise sidebands that arises from the fact that \cref{eq:phi_m} is the phase of an electromagnetic field.
In this section, we search for the noise coupling via the interaction of the modulation and the laser phase noise by expanding the phase of the single electromagnetic field to sidebands.
The coupling of the modulation amplitude and phase noise is expected to be fully covered in \cref{sub:mod_band_trig}; hence, we focus on the laser phase noise in this section.

The electromagnetic field in \cref{eq:Ei_phim} with the laser phase noise $p_i$ under $p_i=m_\mrm{p}\cos((n\omega_{\mrm{m},i} + \Delta\omega_\mrm{p})\tau + \rho)$ ($n\in \mathbb{N}$) and $n_i=a_i=\theta_i=0$ reads
\begin{widetext}
    \begin{align}
        E_i &= A_i\cdot \exp\left(j\omega_i\tau+j\phi^\mrm{car}_{\mrm{tar},i}+jm_\mrm{p}\cos((n\omega_{\mrm{m},i} + \Delta\omega_\mrm{p})\tau + \rho)+jm_0\cos(\omega_{\mrm{m},i}\tau+\phi^\mrm{sb}_{\mrm{tar},i})\right)
        \nonumber\\
        &\approx A_i\cdot\e^{j\left(\omega_i\tau+\phi^\mrm{car}_{\mrm{tar},i}\right)}\cdot\left(J_0(m_0) + \sum^{\infty}_{k=1} j^k J_k(m_0) \left\{\e^{j k \left(\omega_{\mrm{m},i}\tau+\phi^\mrm{sb}_{\mrm{tar},i}\right)} +  \e^{-j k \left(\omega_{\mrm{m},i}\tau+\phi^\mrm{sb}_{\mrm{tar},i}\right)}\right\}\right)
        \nonumber\\
        &\hspace{23.5mm}\cdot\left(J_0(m_\mrm{p}) + jJ_1(m_\mrm{p})\e^{j((n\omega_{\mrm{m},i}+\Delta\omega_\mrm{p})\tau+\rho)} + jJ_1(m_\mrm{p})\e^{-j((n\omega_{\mrm{m},i}+\Delta\omega_\mrm{p})\tau+\rho)}\right)
        \nonumber\\
        &= A_i\cdot\e^{j\left(\omega_i\tau+\phi^\mrm{car}_{\mrm{tar},i}\right)}\cdot\left(\mcal{S}_0 + \sum^{\infty}_{k=1} \left(\mcal{S}_{+k}+\mcal{S}_{-k}\right)\right)\cdot\left(J_0(m_\mrm{p}) + \mcal{P}_{+n} + \mcal{P}_{-n}\right),
    \end{align}
\end{widetext}
where
\begin{subequations}\label{eq:Ssub}
    \begin{align}
        \mcal{S}_0 &\coloneqq J_0(m_0),
        \label{eq:S0}\\
        \mcal{S}_{+k} &\coloneqq j^k J_k(m_0) \e^{j k \left(\omega_{\mrm{m},i}\tau+\phi^\mrm{sb}_{\mrm{tar},i}\right)},
        \label{eq:Sk}\\
        \mcal{S}_{-k} &\coloneqq j^k J_k(m_0) \e^{-j k \left(\omega_{\mrm{m},i}\tau+\phi^\mrm{sb}_{\mrm{tar},i}\right)},
        \label{eq:S-k}\\
        \mcal{P}_{+n} &\coloneqq jJ_1(m_\mrm{p})\e^{j((n\omega_{\mrm{m},i}+\Delta\omega_\mrm{p})\tau+\rho)},
        \label{eq:Pn}\\
        \mcal{P}_{-n} &\coloneqq jJ_1(m_\mrm{p})\e^{-j((n\omega_{\mrm{m},i}+\Delta\omega_\mrm{p})\tau+\rho)}.
        \label{eq:P-n}
    \end{align}
\end{subequations}
$\mcal{S}$ represents the carrier or the modulation sidebands, and $\mcal{P}$ represents the noise sidebands.
Note that we for now consider the modulation sidebands higher than the first-order, while the noise sidebands sticks to its first-order terms.

The noise couplings missed by the trigonometric analysis in \cref{sub:mod_band_trig} is the interaction between $\mcal{S}_{\pm k}$ and $\mcal{P}$.
Of the possible combinations, those between $\mcal{S}_+$ and $\mcal{P}_+$ or $\mcal{S}_-$ and $\mcal{P}_-$ push the noise sideband far away from the carrier or the modulation sidebands, which is expected not to result in significant noise coupling at the end within the PR signal in \cref{eq:spr}.
Hence, let us focus on the following two terms,
\begin{subequations}\label{eq:Tsub}
    \begin{align}
        \mcal{T}^{-n}_k &= \mcal{P}_{-n}\mcal{S}_{+k}
        \nonumber\\
        &= j^{k+1} J_k(m_0)J_1(m_\mrm{p}) \e^{-j\left(((n-k)\omega_{\mrm{m},i}+\Delta\omega_\mrm{p})\tau-k\phi^\mrm{sb}_{\mrm{tar},i}+\rho\right)},
        \label{eq:Tk-n}\\
        \mcal{T}^{n}_{-k} &= \mcal{P}_{+n}\mcal{S}_{-k}
        \nonumber\\
        &= j^{k+1} J_k(m_0)J_1(m_\mrm{p}) \e^{j\left(((n-k)\omega_{\mrm{m},i}+\Delta\omega_\mrm{p})\tau-k\phi^\mrm{sb}_{\mrm{tar},i}+\rho\right)}.
        \label{eq:Tn-k}
    \end{align}
\end{subequations}
The superscript and subscript of $\mcal{T}$ represent the integer associated with the noise sideband $\mcal{P}$ and the modulation sideband $\mcal{S}$, respectively.

The division of $\mcal{T}$ by $\mcal{S}$ reveals how the noise sidebands get associated with individual beatnotes:
\begin{widetext}
\begin{subequations}\label{eq:Tnk_norm}
    \begin{align}
        \mcal{T}^{-n}_k/\mcal{S}_0 &= j^{k+1} \frac{J_k(m_0)J_1(m_\mrm{p})}{J_0(m_0)} \exp\left[-j(((n-k)\omega_{\mrm{m},i}+\Delta\omega_\mrm{p})\tau-k\phi^\mrm{sb}_{\mrm{tar},i}+\rho)\right],
        \label{eq:Tk-n_S0}\\
        \mcal{T}^{-n}_k/\mcal{S}_{+1} &= j^{k} \frac{J_k(m_0)J_1(m_\mrm{p})}{J_1(m_0)} \exp\left[-j(((n-k+1)\omega_{\mrm{m},i}+\Delta\omega_\mrm{p})\tau-(k-1)\phi^\mrm{sb}_{\mrm{tar},i}+\rho)\right],
        \label{eq:Tk-n_S+1}\\
        \mcal{T}^{-n}_k/\mcal{S}_{-1} &= j^{k} \frac{J_k(m_0)J_1(m_\mrm{p})}{J_1(m_0)} \exp\left[-j(((n-k-1)\omega_{\mrm{m},i}+\Delta\omega_\mrm{p})\tau-(k+1)\phi^\mrm{sb}_{\mrm{tar},i}+\rho)\right],
        \label{eq:Tk-n_S-1}\\
        \mcal{T}^n_{-k}/\mcal{S}_0 &= j^{k+1} \frac{J_k(m_0)J_1(m_\mrm{p})}{J_0(m_0)} \exp\left[j(((n-k)\omega_{\mrm{m},i}+\Delta\omega_\mrm{p})\tau-k\phi^\mrm{sb}_{\mrm{tar},i}+\rho)\right],
        \label{eq:Tn-k_S0}\\
        \mcal{T}^n_{-k}/\mcal{S}_{+1} &= j^{k} \frac{J_k(m_0)J_1(m_\mrm{p})}{J_1(m_0)} \exp\left[j(((n-k-1)\omega_{\mrm{m},i}+\Delta\omega_\mrm{p})\tau-(k+1)\phi^\mrm{sb}_{\mrm{tar},i}+\rho)\right],
        \label{eq:Tn-k_S+1}\\
        \mcal{T}^n_{-k}/\mcal{S}_{-1} &= j^{k} \frac{J_k(m_0)J_1(m_\mrm{p})}{J_1(m_0)} \exp\left[j(((n-k+1)\omega_{\mrm{m},i}+\Delta\omega_\mrm{p})\tau-(k-1)\phi^\mrm{sb}_{\mrm{tar},i}+\rho)\right].
        \label{eq:Tn-k_S-1}
    \end{align}
\end{subequations}
\end{widetext}
Each represents a noise single sideband around an individual tone, which is interpreted as the combination of coherent amplitude and phase noises as discussed in \cref{app:ssb}.
Hence, if the noise single sideband associated with $\mcal{S}_{k'} (k'\in\{0,+1,-1\})$ has the frequency of $\Delta\omega_\mrm{p}$ in the heterodyne band, it could finally couple to the phase measurement of heterodyne beatnotes through the amplitude and phase noises at $\Delta\omega_\mrm{p}$ of the beatnote composed of $\mcal{S}_{k'}$.
This suggests that our interest is the combination of $n$ and $k$ that ends up with the frequency of $\Delta\omega_\mrm{p}$ by removing $\omega_{\mrm{m},i}$.

Let us first stick to the first-order modulation sideband ($k=1$), and also consider only $n\in\{1,2\}$.
Regarding the coupling via the carrier, \cref{eq:Tk-n_S0,eq:Tn-k_S0} suggest that $n=k$ is required; therefore, the only possible pair is $n=k=1$.
Focusing on those terms, we get
\begin{align}
    \mcal{T}^{-1}_1/\mcal{S}_0 &+ \mcal{T}^1_{-1}/\mcal{S}_0
    \nonumber\\
    &= -\frac{J_1(m_0)}{J_0(m_0)}m_\mrm{p}\cos(\Delta\omega_\mrm{p}\tau-\phi^\mrm{sb}_{\mrm{tar},i}+\rho)
    \nonumber\\
    &\approx \frac{1}{m_0}\cdot\left(-\frac{m^2_0}{2J_0(m_0)}m_\mrm{p}\cos(\Delta\omega_\mrm{p}\tau-\phi^\mrm{sb}_{\mrm{tar},i}+\rho)\right).
    \label{eq:T1-1_sum}
\end{align}
This just reproduces the coupling path captured in \cref{sub:mod_band_trig}: $p_i(\omega_{\mrm{m},i} + \Delta\omega_\mrm{p})$ virtually turns to $a_i(\Delta\omega_\mrm{p})$ and $\theta_i(\Delta\omega_\mrm{p})$ (see \cref{eq:pi2ai_2thetai}), out of which only $a_i(\Delta\omega_\mrm{p})$ remains here as $\theta_i(\Delta\omega_\mrm{p})$ has no coupling path through the carrier.
The second line in \cref{eq:T1-1_sum} explicitly shows this by decomposing the result into the coupling factor $1/m_0$ arising from the frequency downconversion in \cref{eq:pi2ai_2thetai} and the form of the heterodyne-band modulation amplitude noise in the carrier-carrier beatnote in \cref{eq:scar_prifo,eq:sc2sb_prifo}.

On the other hand, regarding the noise coupling via the modulation sidebands, \cref{eq:Tk-n_S-1,eq:Tn-k_S+1} suggest that $k=1$ and $n=2$ are required, which give
\begin{subequations}\label{eq:T21_S1}
    \begin{align}
        \mcal{T}^2_{-1}/\mcal{S}_{+1} &= \frac{m_\mrm{p}}{2} \exp\left[j(\Delta\omega_\mrm{p}\tau-2\phi^\mrm{sb}_{\mrm{tar},i}+\rho+\pi/2)\right],
        \label{eq:T2-1_S+1}\\
        \mcal{T}^{-2}_1/\mcal{S}_{-1} &= \frac{m_\mrm{p}}{2} \exp\left[-j(\Delta\omega_\mrm{p}\tau-2\phi^\mrm{sb}_{\mrm{tar},i}+\rho-\pi/2)\right].
        \label{eq:T-2+1_S-1}
    \end{align}
\end{subequations}
According to \cref{app:ssb}, the upper single sideband is equivalent to the coherent amplitude and phase modulations with the latter delayed by $\pi/2$ against the former.
Therefore, the coupling paths can be written as
\begin{subequations}\label{eq:pi_2fm_viasb}
    \begin{align}
        &p_i\left(2\omega_{\mrm{m},i}+\Delta \omega_\mrm{p}; m_\mrm{p}, \rho\right)
        \nonumber\\
        &\xrightarrow[\text{via usb}]{\text{}} 
        \begin{cases}
            a_i\left(\Delta \omega_\mrm{p};\frac{m_\mrm{p}}{2},\rho-2\phi^\mrm{sb}_{\mrm{tar},i}+\pi/2\right)
            \\
            \theta_i\left(\Delta \omega_\mrm{p};\frac{m_\mrm{p}}{2},\rho-2\phi^\mrm{sb}_{\mrm{tar},i}\right)
        \end{cases}
        % \nonumber\\
        % &\hspace{10mm} = p_i\left(\omega_{\mrm{m},i}+\Delta\omega_\mrm{p}; \frac{m_0 m_\mrm{p}}{2}, \rho-\phi^\mrm{sb}_{\mrm{tar},i}+\frac{\pi}{2}\right),
        \label{eq:pi_2fm_viausb}\\
        &\xrightarrow[\text{via lsb}]{\text{}}
        \begin{cases}
            a_i\left(\Delta \omega_\mrm{p};\frac{m_\mrm{p}}{2},\rho-2\phi^\mrm{sb}_{\mrm{tar},i}-\pi/2\right)
            \\
            \theta_i\left(\Delta \omega_\mrm{p};\frac{m_\mrm{p}}{2},\rho-2\phi^\mrm{sb}_{\mrm{tar},i}-\pi\right)
        \end{cases}
        % \nonumber\\
        % &\hspace{10mm} = p_i\left(\omega_{\mrm{m},i}+\Delta\omega_\mrm{p}; \frac{m_0 m_\mrm{p}}{2}, \rho-\phi^\mrm{sb}_{\mrm{tar},i}-\frac{\pi}{2}\right).
        \label{eq:pi_2fm_vialsb}
    \end{align}
\end{subequations}
This shows that the effective phase shift differs between the coupling paths via the upper modulation sideband $\mcal{S}_{+1}$ in \cref{eq:pi_2fm_viausb} or the lower modulation sideband $\mcal{S}_{-1}$ in \cref{eq:pi_2fm_vialsb}.
% Comparing \cref{eq:pi_2fm_viasb} with \cref{eq:pi2ai_2thetai}, one finds that the coupling of $p_i\left(2\omega_{\mrm{m},i}+\Delta \omega_\mrm{p}\right)$ can be written by that of $p_i(\omega_{\mrm{m},i}+\Delta\omega_\mrm{p})$.
% We express this congruence as:
This noise coupling is directly converted from $2\omega_{\mrm{m},i}+\Delta \omega_\mrm{p}$ to $\Delta \omega_\mrm{p}$, which is visualized in \cref{fig:ghz_diagram} in red.

Let us also consider the second-order modulation sidebands ($k=2$) because the modulation depth $m_0$ would not necessarily be small, unlike the noise amplitude.
In addition, second-order sidebands could cause noise coupling $\propto J_2(m_0) \sim O(m^2_0)$, unlike heterodyne-band noise coupling; see \cref{app:higher} for more discussion.
% Let us also consider the second-order modulation sidebands ($k=2$) because the modulation depth $m_0$ would not necessarily be small, unlike the noise amplitude, and could cause noise coupling $\propto J_2(m_0) \sim O(m^2_0)$.
% On the other hand, we did not consider the second-order modulation sideband in analysis of the heterodyne-band noise coupling in \cref{sec:het_band} because it would appear as the beatnote between the second-order sidebands of interfering laser beams.
% Such a beatnote could cause noise coupling, however, the coupling factor would be $\propto J^2_2(m_0) \sim O(m^4_0)$, which would be small.

Regarding the second-order coupling via the beam carrier, $n=k=2$ gives
\begin{align}
    \mcal{T}^{-2}_2/\mcal{S}_0 &+ \mcal{T}^2_{-2}/\mcal{S}_0
    \nonumber\\
    &= -j\frac{J_2(m_0)}{J_0(m_0)}m_\mrm{p}\cos(\Delta\omega_\mrm{p}\tau-2\phi^\mrm{sb}_{\mrm{tar},i}+\rho),
    \label{eq:T2-2_sum}
\end{align}
which is the pure phase noise with the amplitude of $\frac{J_2(m_0)}{J_0(m_0)}m_\mrm{p}$ at $\Delta\omega_\mrm{p}$.
Hence, e.g., if $\Delta\omega_\mrm{p}$ is the frequency in the observation band, this noise appears in the final carrier phase extraction as is through the carrier self noise coupling.

On the other hand, regarding the second-order coupling through the modulation sidebands, $k=2$ and $n=1$ for \cref{eq:Tk-n_S+1,eq:Tn-k_S-1} give
\begin{subequations}\label{eq:T12_S1}
    \begin{align}
        \mcal{T}^{-1}_2/\mcal{S}_{+1} &= \frac{J_2(m_0)J_1(m_\mrm{p})}{J_1(m_0)} \exp\left[-j(\Delta\omega_\mrm{p}\tau-\phi^\mrm{sb}_{\mrm{tar},i}+\rho+\pi)\right],
        \label{eq:T-1+2_S+1}\\
        \mcal{T}^1_{-2}/\mcal{S}_{-1} &= \frac{J_2(m_0)J_1(m_\mrm{p})}{J_1(m_0)} \exp\left[j(\Delta\omega_\mrm{p}\tau-\phi^\mrm{sb}_{\mrm{tar},i}+\rho+\pi)\right].
        \label{eq:T1-2_S-1}
    \end{align}
\end{subequations}
Similarly to \cref{eq:pi_2fm_viasb}, each produces a noise lower single sideband, which results in
\begin{subequations}\label{eq:pi_fm_viasb}
    \begin{align}
        &p_i\left(\omega_{\mrm{m},i}+\Delta \omega_\mrm{p}; m_\mrm{p}, \rho\right)
        \nonumber\\
        &\xrightarrow[\text{via usb}]{\text{}}
        \begin{cases}
            a_i\left(\Delta \omega_\mrm{p};\frac{J_2(m_0)m_\mrm{p}}{2J_1(m_0)},\rho-\phi^\mrm{sb}_{\mrm{tar},i}+\pi\right)
            \\
            \theta_i\left(\Delta \omega_\mrm{p};\frac{J_2(m_0)m_\mrm{p}}{2J_1(m_0)},\rho-\phi^\mrm{sb}_{\mrm{tar},i}+\frac{3\pi}{2}\right)
        \end{cases}
        \label{eq:pi_fm_viausb}\\
        &\xrightarrow[\text{via lsb}]{\text{}}
        \begin{cases}
            a_i\left(\Delta \omega_\mrm{p};\frac{J_2(m_0)m_\mrm{p}}{2J_1(m_0)},\rho-\phi^\mrm{sb}_{\mrm{tar},i}+\pi\right)
            \\
            \theta_i\left(\Delta \omega_\mrm{p};\frac{J_2(m_0)m_\mrm{p}}{2J_1(m_0)},\rho-\phi^\mrm{sb}_{\mrm{tar},i}+\frac{3\pi}{2}\right)
        \end{cases}
        \label{eq:pi_fm_vialsb}
    \end{align}
\end{subequations}

\begin{figure}[h]
    \centering
\includegraphics[width=8.6cm]{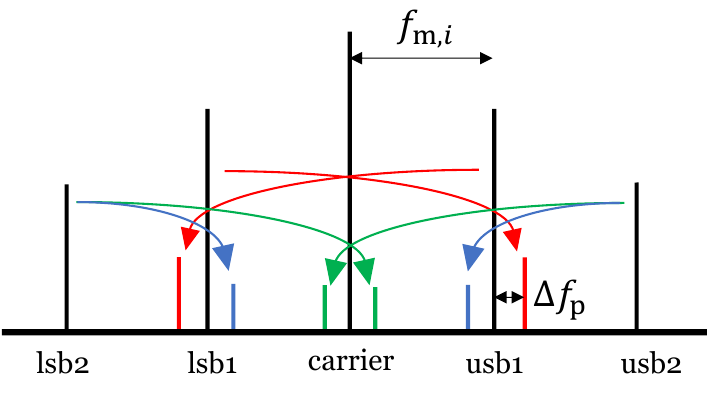}
    \caption{
    Single beam optical spectrum with the laser phase noise $p_i$ in the noise sideband picture.
    The modulation sidebands are depicted up to the second-order.
    $f_{\mrm{m},i}$ is a modulation frequency, and $\Delta f_\mrm{p}$ is an offset frequency from an integer multiple of $f_{\mrm{m},i}$, which can be either of heterodyne-band or observation-band frequencies.
    The colored noise sidebands correspond to the discussed coupling paths as follows: Red: \cref{eq:T21_S1,eq:pi_2fm_viasb}; Green: \cref{eq:T2-2_sum}; Blue: \cref{eq:T12_S1,eq:pi_fm_viasb}.
    Abbreviations: usb = upper sideband; lsb = lower sideband.
    }
    \label{fig:sideband_noise}
\end{figure}

\cref{fig:sideband_noise} visualizes the optical spectrum of a single laser beam with the carrier and modulation sidebands in black and the noise sidebands discussed in this section in color.
Their coherence properties are discussed in the next section.

\subsection{Coherence}\label{sub:mod_band_coherence}
% Contrary to the original heterodyne-band modulation noise discussed in \cref{sec:het_band}, the noises arising from the down conversions in \cref{sub:mod_band_trig,sub:mod_band_sideband} are coherent because they originate from the identical noise source.
As shown in \cref{fig:ghz_diagram}, a single modulation-band noise has various coupling paths.
Therefore, it is important to properly track individual coupling paths and consider the coherence between them to evaluate the impact on the final measured phase.
In this section, we discuss the noise coherence taking specific noises as examples: noises at $n\omega_{\mrm{m},i}+\epsilon$ where $n\in \{1,2\}$ and $\epsilon$ is a tiny angular frequency offset in the observation band as used in \cref{eq:s_add_cite,eq:IQdemod_cite}.

First, the laser phase noise $p_i$ around the modulation frequency $\omega_{\mrm{m},i}$ is, according to \cref{eq:pi2ai_2thetai}, down-converted to the modulation amplitude and phase noises in the observation band as,
\begin{widetext}
\begin{align}
    p_i(\omega_{\mrm{m},i}+\epsilon; m_\mrm{p}, \rho) &\rightarrow
    \begin{cases}
        a_i\left(\epsilon;\frac{m_\mrm{p}}{m_0},\rho-\phi^\mrm{sb}_{\mrm{tar},i}\right)
        \\
        \theta_i\left(\epsilon;\frac{m_\mrm{p}}{m_0},\rho-\phi^\mrm{sb}_{\mrm{tar},i}-\frac{\pi}{2}\right)
    \end{cases}
    \nonumber\\
    & \xrightarrow[\text{extraction}]{\text{phase}}
    \begin{cases}
    0+0 = 0 & \text{for $\phi^\mrm{car}_\mrm{read}$}
    \\
    0 + \frac{m_\mrm{p}}{m_0}\cdot\cos(\rho-\phi^\mrm{sb}_{\mrm{tar},i}-\frac{\pi}{2})
    = \frac{m_\mrm{p}}{m_0}\cdot\cos(\rho-\phi^\mrm{sb}_{\mrm{tar},i}-\frac{\pi}{2}) & \text{for $\phi^\mrm{usb}_\mrm{read}$}
    \\
    0 - \frac{m_\mrm{p}}{m_0}\cdot\cos(\rho-\phi^\mrm{sb}_{\mrm{tar},i}-\frac{\pi}{2})
    = -\frac{m_\mrm{p}}{m_0}\cdot\cos(\rho-\phi^\mrm{sb}_{\mrm{tar},i}-\frac{\pi}{2}) & \text{for $\phi^\mrm{lsb}_\mrm{read}$},
    \end{cases}
    \label{eq:pi_fmfinband}
\end{align}
where the first arrow represents the virtual down-conversion, and the second arrow represents phase extraction by the phasemeter with the right-hand side as the coupled noises.
There is no influence on the measured carrier phase $\phi^\mrm{car}_\mrm{read}$, while the sideband phases are affected off-phase; therefore, the noise cannot be suppressed by the combination in \cref{eq:sbcombi}.

Second, according to \cref{eq:pi_2fm_viausb,eq:pi_2fm_vialsb}, the laser phase noise around the double modulation frequency $2\omega_{\mrm{m},i}$ reads
\begin{align}
    p_i(2\omega_{\mrm{m},i}&+\epsilon; m_\mrm{p}, \rho)
    \nonumber\\
    &\xrightarrow[\text{extraction}]{\text{phase}}
    \begin{cases}
    0  & \text{for $\phi^\mrm{car}_\mrm{read}$}
    \\
    % p_i\left(\omega_{\mrm{m},i}+\epsilon; \frac{m_0 m_\mrm{p}}{2}, \rho-\phi^\mrm{sb}_{\mrm{tar},i}+\frac{\pi}{2}\right)
    \hat{a}_i\left(\epsilon; \frac{m_p}{2}, \rho-2\phi^\mrm{sb}_{\mrm{tar},i}+\frac{\pi}{2}\right) + \hat{\theta}_i\left(\epsilon; \frac{m_p}{2}, \rho-2\phi^\mrm{sb}_{\mrm{tar},i}\right)
    =\frac{m_\mrm{p}}{2}\cdot\cos(\rho-2\phi^\mrm{sb}_{\mrm{tar},i}) & \text{for $\phi^\mrm{usb}_\mrm{read}$}
    \\
    \hat{a}_i\left(\epsilon; \frac{m_p}{2}, \rho-2\phi^\mrm{sb}_{\mrm{tar},i}-\frac{\pi}{2}\right) + \hat{\theta}_i\left(\epsilon; \frac{m_p}{2}, \rho-2\phi^\mrm{sb}_{\mrm{tar},i} - \pi\right)
    =\frac{m_\mrm{p}}{2}\cdot\cos(\rho-2\phi^\mrm{sb}_{\mrm{tar},i})  & \text{for $\phi^\mrm{lsb}_\mrm{read}$},
    \end{cases}
    \label{eq:pi_2fmfinband}
\end{align}
where we express the coupled term of the noise $x_i$ in the measured phases by  $\hat{x}_i$.
According to \cref{tab:noise_couplings}, $\theta_i(\epsilon)$ is the self noise coupling, and the coupling to $\phi^\mrm{usb}_\mrm{read}$ ($\phi^\mrm{lsb}_\mrm{read}$) occurs through usb (lsb).
Hence, \cref{eq:pi_2fm_viausb,eq:pi_2fm_vialsb} were applied to $\phi^\mrm{usb}_\mrm{read}$ and $\phi^\mrm{lsb}_\mrm{read}$, respectively.
As the sideband phases are affected in-phase, the noise can be completely removed by the combination in \cref{eq:sbcombi}.
Regarding the carrier, $p_i(2\omega_{\mrm{m},i}+\epsilon)$ has no influence on the measured carrier phase $\phi^\mrm{car}_\mrm{read}$ either; however, if we consider the second-order modulation sidebands, there is a noise coupling as shown in \cref{eq:T2-2_sum}.

Third, considering \cref{eq:ai2pi,eq:pi_2fmfinband}, the modulation amplitude noise $a_i$ around $\omega_{\mrm{m},i}$ couples as,
\begin{align}
    a_i\left(\omega_{\mrm{m},i}+\epsilon; m_\mrm{a}, \rho\right) &\rightarrow p_i\left(\epsilon; \frac{m_0m_\mrm{a}}{2}, \rho-\phi^\mrm{sb}_{\mrm{tar},i}\right) + p_i\left(2\omega_{\mrm{m},i} + \epsilon; \frac{m_0m_\mrm{a}}{2}, \rho+\phi^\mrm{sb}_{\mrm{tar},i}\right)
    \nonumber\\
    &\xrightarrow[\text{extraction}]{\text{phase}}
    \begin{cases}
    \frac{m_0m_\mrm{a}}{2} \cos(\rho-\phi^\mrm{sb}_{\mrm{tar},i}) + 0 = \frac{m_0m_\mrm{a}}{2} \cos(\rho-\phi^\mrm{sb}_{\mrm{tar},i}) & \text{for $\phi^\mrm{car}_\mrm{read}$}
    \\
    \frac{m_0m_\mrm{a}}{2} \cos(\rho-\phi^\mrm{sb}_{\mrm{tar},i}) + \frac{m_\mrm{a}m_0}{4}\cdot\cos(\rho-\phi^\mrm{sb}_{\mrm{tar},i}) = \frac{3}{4}m_0m_\mrm{a}\cos(\rho-\phi^\mrm{sb}_{\mrm{tar},i}) & \text{for $\phi^\mrm{usb}_\mrm{read}$}
    \\
    \frac{m_0m_\mrm{a}}{2} \cos(\rho-\phi^\mrm{sb}_{\mrm{tar},i}) + \frac{m_\mrm{a}m_0}{4}\cdot\cos(\rho-\phi^\mrm{sb}_{\mrm{tar},i}) = \frac{3}{4}m_0m_\mrm{a}\cos(\rho-\phi^\mrm{sb}_{\mrm{tar},i}) & \text{for $\phi^\mrm{lsb}_\mrm{read}$},
    \end{cases}
    \label{eq:ai_fmfinband}
\end{align}
while $a_i$ around $2\omega_{\mrm{m},i}$ can be derived by \cref{eq:ai2pi2ai} as,
\begin{align}
    a_i\left(2\omega_{\mrm{m},i}+\epsilon; m_\mrm{a}, \rho\right) &\rightarrow
    \begin{cases}
        a_i\left(\epsilon; \frac{m_\mrm{a}}{2}, \rho-2\phi^\mrm{sb}_{\mrm{tar},i}\right)
        \\
        \theta_i\left(\epsilon; \frac{m_\mrm{a}}{2}, \rho-2\phi^\mrm{sb}_{\mrm{tar},i}-\frac{\pi}{2}\right)
    \end{cases}
    \nonumber\\
    &\xrightarrow[\text{extraction}]{\text{phase}}
    \begin{cases}
    0 + 0 = 0 & \text{for $\phi^\mrm{car}_\mrm{read}$}
    \\
    0 + \frac{m_\mrm{a}}{2}\cdot\cos(\rho-2\phi^\mrm{sb}_{\mrm{tar},i}-\frac{\pi}{2}) = \frac{m_\mrm{a}}{2}\cdot\cos(\rho-2\phi^\mrm{sb}_{\mrm{tar},i}-\frac{\pi}{2}) & \text{for $\phi^\mrm{usb}_\mrm{read}$}
    \\
    0 - \frac{m_\mrm{a}}{2}\cdot\cos(\rho-2\phi^\mrm{sb}_{\mrm{tar},i}-\frac{\pi}{2}) = -\frac{m_\mrm{a}}{2}\cdot\cos(\rho-2\phi^\mrm{sb}_{\mrm{tar},i}-\frac{\pi}{2}) & \text{for $\phi^\mrm{lsb}_\mrm{read}$}.
    \end{cases}
    \label{eq:ai_2fmfinband}
\end{align}
\end{widetext}
Therefore, $a_i$ at $\omega_{\mrm{m},i}+\epsilon$ in \cref{eq:ai_fmfinband} affects not only the sideband phases but also the carrier phase.
The impact on the sideband phases is in-phase; hence, it can be removed by the sideband combination in \cref{eq:sbcombi}.
On the other hand, $a_i$ at $2\omega_{\mrm{m},i}+\epsilon$ in \cref{eq:ai_2fmfinband} has noise coupling to only the sideband phases, which are off-phase and cannot be suppressed by the combination in \cref{eq:sbcombi}.

Finally, the modulation phase noise $\theta_i$ can be easily derived via comparison with $a_i$, i.e., \cref{eq:ai_fmfinband,eq:ai_2fmfinband}.
Regarding $\theta_i$ at $\omega_{\mrm{m},i}+\epsilon$, \cref{eq:ai2pi,eq:thetai2pi} suggest that we can add the $-\pi/2$ phase shift to the contribution of $p_i(\epsilon)$ and $+\pi/2$ to the contribution of $p_i(2\omega_{\mrm{m},i}+\epsilon)$ in \cref{eq:ai_fmfinband}.
This results in $\frac{1}{2}m_0m_\theta\sin(\rho-\phi^\mrm{sb}_{\mrm{tar},i})$ for $\phi^\mrm{car}_\mrm{read}$ and $\frac{1}{4}m_0m_\mrm{\theta}\sin(\rho-\phi^\mrm{sb}_{\mrm{tar},i})$ for $\phi^\mrm{usb}_\mrm{read}$ and $\phi^\mrm{lsb}_\mrm{read}$; the noise coupling to the sideband can be removed by the combination in \cref{eq:sbcombi}.
For $\theta_i$ at $2\omega_{\mrm{m},i}+\epsilon$, comparing \cref{eq:ai2pi2ai,eq:thetai2pi2thetai},we find that the result is given by shifting the phase of \cref{eq:ai_2fmfinband} by $-\pi/2$; therefore, the relative phase between $\phi^\mrm{usb}_\mrm{read}$ and $\phi^\mrm{lsb}_\mrm{read}$ does not change from \cref{eq:ai_2fmfinband}, and the combination in \cref{eq:sbcombi} cannot suppress the noise.

% As mentioned in the second last paragraph in \cref{sub:framework_beatnote}, we have calculated the coupling of the noise of Laser $i$ under the implicit assumption of $\omega_{\mrm{m},i}>\omega_{\mrm{m},j}$ such that $\Delta\omega_\mrm{m}>0$.
% However, under $\omega_{\mrm{m},i}<\omega_{\mrm{m},j}$, some noise couplings of Laser $j$ would become slightly different, especially the coupling paths that include $a_i$ and $\theta_i$ as intermediary or final quantities: e.g., $p_j$ around $\omega_{\mrm{m},j}$ or $a_j$ and $\theta_j$ around $2\omega_{\mrm{m},j}$.
% This is because $\Delta\theta=\theta_i-\theta_j$ in \cref{eq:spr} flips the sign of $\theta_j$, while $a_j$ does not change its sign; hence, the generated pair of $\theta_j$ and $a_j$ is expected to have a different relative phase from the discussion above.
% As mentioned in the second last paragraph in \cref{sub:framework_beatnote}, we have calculated the coupling of the noise of Laser $i$ under the implicit assumption of $\omega_{\mrm{m},i}>\omega_{\mrm{m},j}$ such that $\Delta\omega_\mrm{m}>0$.
% However, if $\omega_{\mrm{m},i}<\omega_{\mrm{m},j}$, the usb-usb and lsb-lsb beatnotes are swapped around the carrier-carrier beatnote.
% This means that $\theta$ has its relative sign to $\Delta\omega_\mrm{m}$, while $a_i$ has no influence from the polarity change; hence, the coherence between intermediary $a_i$ and $\theta_i$ changes.

% : ========== Verification ===============================
\section{Verification}\label{sec:verification}
\begin{figure*}
    \centering
\includegraphics[width=17.2cm]{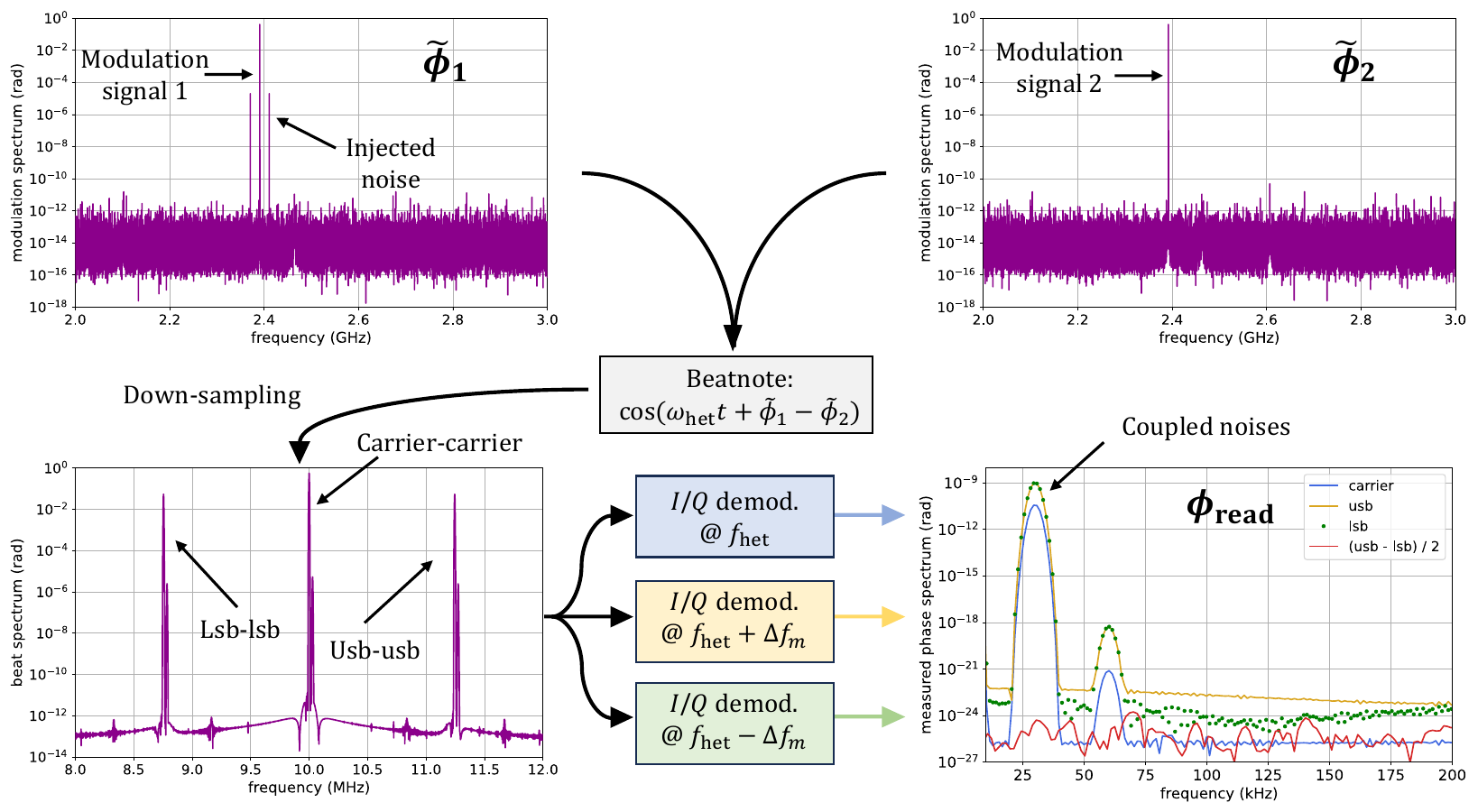}
    \caption{
    Diagram highlighting the processing steps of the numerical experiment.
    All plots are amplitude spectra, instead of ASDs, to properly assess the height of monotonic tones.
    The noise coupling factor for an individual beatnote is computed by the ratio of the coupled noise in its measured phase (bottom right) to the injected noise (top left).
    Regarding the plots of the power spectrum, the results of the modulation amplitude noise $a_i$ at $2\omega_\mrm{het}$ are taken as an example.
    }
    \label{fig:sim_flow}
\end{figure*}

We so far derived various noise couplings analytically based on some approximations like the linear approximation and the sideband expression of the phase modulation.
In order to confirm that our analytical framework really covers the main noise couplings, we report on a numerical experiment in this section.
The numerical simulation was implemented, as written below, so that it does not impose any linear approximation.
Therefore, the agreement between the analysis and the numerical simulation guaranties that the framework for the systematic search in \cref{sec:het_band,sec:mod_band} maintains sufficient precision for the primary noise couplings.
The setup of the numerical experiment is described below and is also shown in \cref{fig:sim_flow}.

Contrary to the PR signal with the first-order modulation sidebands in \cref{eq:spr}, we generate a time series according to
\begin{equation}
    \tilde s_\mrm{pr}(\tau) = \cos(\omega_\mrm{het} t + \tilde\phi_1(t) - \tilde\phi_2(t)),
    \label{eq:tildespr}
\end{equation}
where
\begin{equation}
    \tilde\phi_i(t) = p_i(t) + m_0 (1 + a_i(t)) \cos(\omega_{\mrm{m},i}t + \theta_i(t)).
    \label{eq:tildephii}
\end{equation}
We use LISA-like parameters:
The modulation frequencies $\omega_{\mrm{m},i}$ are assumed at $\omega_{\mrm{m},1}/(2\pi) = 3\times 5\times 19\times 2^{23} = \SI{2.39075328}{\giga\Hz}$, $\omega_{\mrm{m},2}/(2\pi) = 19\times 17\times 113\times 2^{16} = \SI{2.391998464}{\giga\Hz}$.
The nominal modulation depth $m_0$ was set at \SI{0.584}{\radian}, which invests \SI{16}{\percent} of the optical power for the sidebands.
The heterodyne frequency $\omega_\mrm{het}/(2\pi)$ was set at \SI{10}{\mega\Hz}.

Since we do not perform any further expansion, the highest frequency components of \cref{eq:tildespr} will be in the GHz range.
To avoid these being aliased into our measurement band, we initially use a relatively high sampling rate of \SI{1.28}{\giga\hertz} for the heterodyne-band noises and \SI{12.8}{\giga\hertz} for the modulation-band noises to produce a first instance of the signal which is then immediately filtered and downsampled to a final rate of \SI{80}{\mega\hertz}.
We then perform a standard I/Q demodulation on these downsampled signals at three frequencies: $\omega_\mrm{het}$ for the carrier-carrier beatnote, $\omega_\mrm{het} - \omega_{\mrm{m},1} + \omega_{\mrm{m},2}$ for the usb-usb beatnote, and $\omega_\mrm{het} + \omega_{\mrm{m},1} - \omega_{\mrm{m},2}$ for the lsb-lsb beatnote.
The low-pass filter in the I/Q demodulation is designed from a Kaiser window, with a transition band from \SI{500}{\kilo\hertz} to \SI{1}{\mega\hertz} and a suppression of \SI{240}{\decibel}.
Hence, \SI{500}{\kilo\hertz} defines the bandwidth of the phasemeter output.

Due to the high initial sampling rate, it is infeasible in this setup to produce sufficiently long data streams to reach the actual LISA measurement band of \SI{0.1}{\milli\hertz} to \SI{1}{\hertz}.
For each noise source, we generate a synthetic data stream with the sample size of $10^7$.
Considering the sampling rates mentioned above, the lowest frequency bin we can reach is \SI{128}{\Hz} for the heterodyne-band noise and \SI{1.28}{\kilo\Hz} for the modulation-band noise.
However, this limitation of the explored frequency band should not be a problem to verify the coupling factor.
Essential is having an offset frequency of an injected noise from a frequency under test sufficiently lower than the phasemeter bandwidth, namely \SI{500}{\kilo\Hz}; noise is generated in compliance with this rule as written below.

To verify individual noise couplings, we inject a monotonic noise at the target noise frequency with an offset frequency of \SI{30}{\kilo\Hz} and a noise amplitude of $10^{-4}$ with the units of \si{\radian} for $p_i$ and $\theta_i$ and no unit for $a_i$.
In this way, the coupled noise appears at \SI{30}{\kilo\Hz} in a phase measured by the I/Q demodulation, as shown on the bottom right in \cref{fig:sim_flow}.
Finally, comparing the height of the coupled noise peak at \SI{30}{\kilo\Hz} in the amplitude spectrum with the injected amplitude of $10^{-4}$, we can derive the magnitude of a noise coupling factor.
To compare the simulation result with analysis in the time domain, the coupling factors derived in simulation is multiplied by a factor of $\sqrt{2}$ because the height of the coupled noise peak in the amplitude spectrum corresponds to the root-mean-square value, which needs to be converted to the amplitude.

\begin{figure}[H]
    \centering
\includegraphics[width=8.6cm]{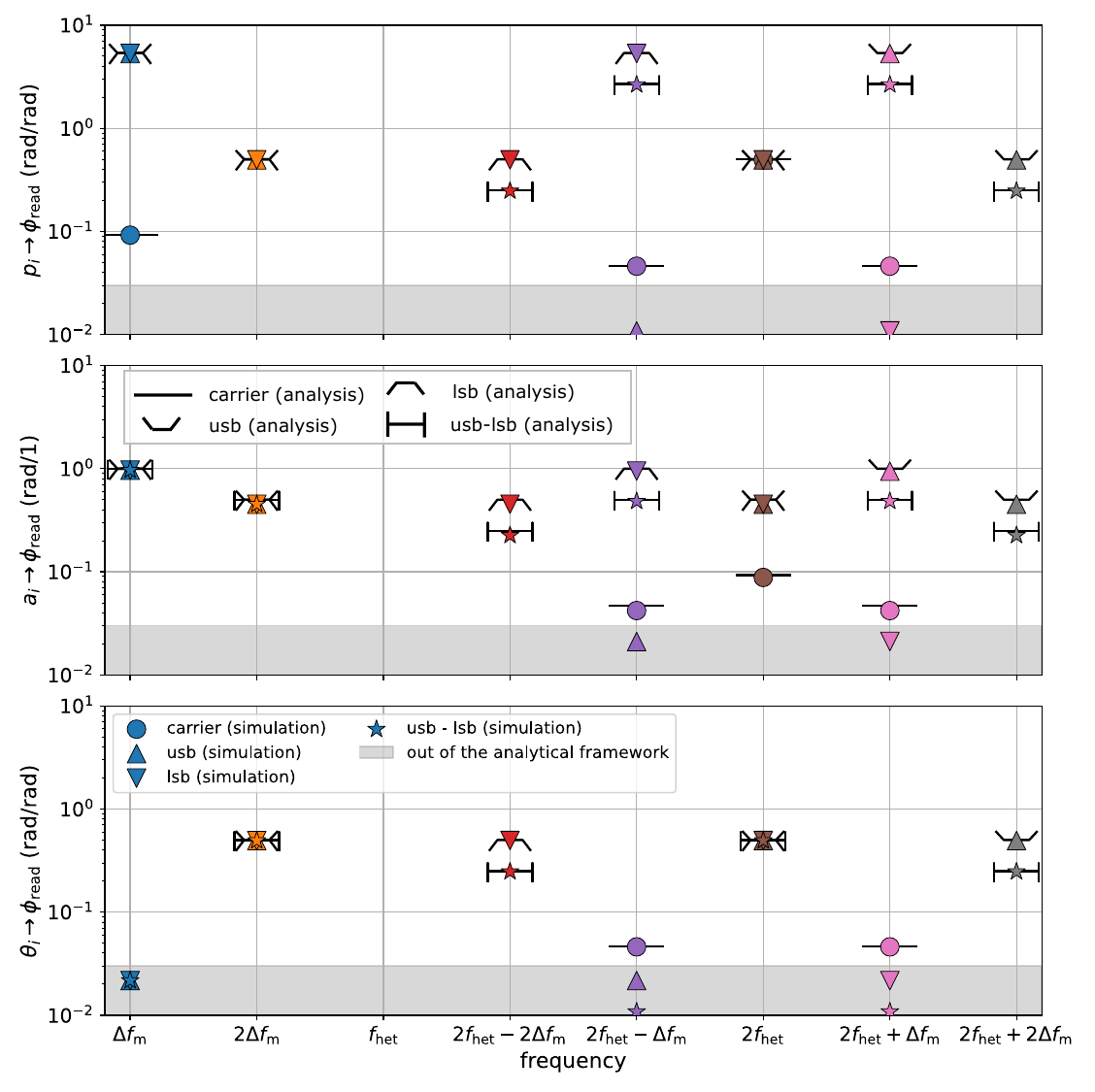}
    \caption{
    Comparison of analysis (black horizontal lines with different cap styles) from \cref{tab:noise_couplings}  and numerical simulation (circles, up triangles, down triangles, or stars) for heterodyne-band noises in terms of the noise coupling factor in the time domain.
    Top: coupling factors of the laser phase noise $p_i$;
    Middle: coupling factors of the modulation amplitude noise $a_i$;
    Bottom: coupling factors of the modulation phase noise $\theta_i$.
    % The x axis is the frequency of the noise sources with $\omega_\mrm{het}/(2\pi)=\SI{10}{\mega\Hz}$ and $\Delta\omega_\mrm{m}/(2\pi)\approx\SI{1.25}{\mega\Hz}$.
    The x axis is a different noise frequency resulting from a combination of the heterodyne frequency $f_\mrm{het}$ and the modulation offset frequency $\Delta f_\mrm{m}$.
    Our analytical framework in \cref{sec:het_band} predicts no coupling below \SI{3e-2}{}, which is grayed.
    }
    \label{fig:het_noise_sim}
\end{figure}

\cref{fig:het_noise_sim} shows the result for the heterodyne-band noise.
The coupling factors (black horizontal lines), derived from the analysis in \cref{sec:het_band}, all agree with the numerical simulation shown in the colored markers.
The simulation and the analysis are consistent also in terms of the sideband combination in \cref{eq:sbcombi}: the colored stars (simulation) always coincide with the vertically-capped black lines (analysis), while the colored stars do not appear when the black lines are not shown.
This confirms that our analysis properly calculates not only the magnitude of the coupled noises but also their phases, in other words, coherence.
There are some colored markers in the gray area below \SI{3e-2}{}.
Our analytical framework does not predict those couplings found in simulation; hence, they are expected to arise from the outside of our framework, e.g., couplings from the second- or higher-order modulation sidebands.
From these results, we conclude that we were able to capture all major couplings of the three types of noise in the heterodyne frequency band using the analytical framework in \cref{sec:het_band}.

\begin{figure}[H]
    \centering
\includegraphics[width=8.6cm]{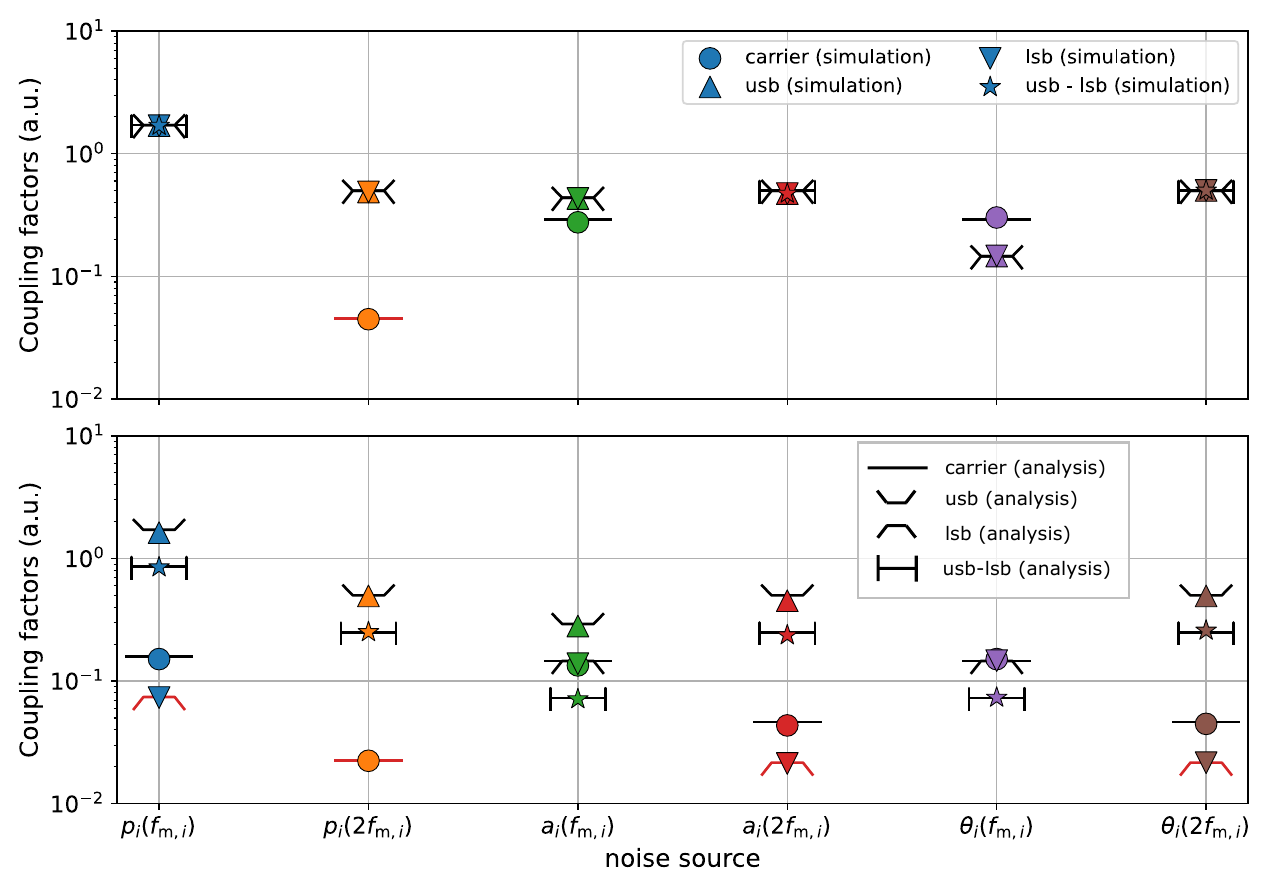}
    \caption{
    Comparison of analysis (black or red horizontal lines) and numerical simulation (circles, up triangles, down triangles,or stars) for modulation-band noises in terms of the noise coupling factor in the time domain.
    Top: noises with the offset of $\epsilon$, \crefrange{eq:pi_fmfinband}{eq:ai_2fmfinband};
    Bottom: noises with the offset of $2\omega_\mrm{het}$, \crefrange{eq:pi_fm2fhet}{eq:ai_2fm2fhet}.
    The x axis gives different noise sources in different frequency regimes.
    The units of y axis depends on the noise sources: \si{\radian/\radian} for the laser phase noise $p_i$ and the modulation phase noise $\theta_i$ and \si{\radian/1} for the modulation amplitude noise $a_i$.
    Red lines represent the analytical results of couplings from the second-order modulation sidebands, namely the cases of $k=2$ that scale with $J_2(m_0)$ in \cref{sub:mod_band_sideband}.
    }
    \label{fig:mod_noise_sim}
\end{figure}

Regarding the modulation-band noises analyzed in \cref{sec:mod_band}, we have numerous couplings around $\omega_{\mrm{m},i}$ and $2\omega_{\mrm{m},i}$ with in-band or heterodyne-band frequency offsets.
Hence, instead of showing all possible noise couplings, we take the offset frequencies of $\epsilon$ and $2\omega_\mrm{het}$ as examples.
The reason of the choices is that the former is the most fundamental example and the latter is one of the frequencies with the most variety of couplings.
The analytical solutions for the coupling of the noises with the offset of $\epsilon$ are given in \cref{sub:mod_band_coherence}.
The same type of analysis can be found for noises with the offset of $2\omega_\mrm{het}$ in \cref{app:2fhet}.
The comparison between the analysis and the numerical simulation is summarized in \cref{fig:mod_noise_sim}.
They are consistent, and the analysis could catch all the major couplings from the noises in the modulation band.
This guaranties that the diagram in \cref{fig:ghz_diagram} efficiently describes the coupling mechanisms, and we properly track the coherence between different coupling paths using the analytical framework in \cref{sec:mod_band}.
The red lines represent the analytical results of coupling from the second-order modulation sidebands, namely the cases of $k=2$ that scale with $J_2(m_0)$ in \cref{sub:mod_band_sideband}.
They are apparently secondary noises, but are too large to easily ignore.
For example, the coupling factors of $p_i(2\omega_{\mrm{m},i}+\epsilon)$ in the plot (= orange on top) are $3.2\times10^{-2}$ and $3.5\times10^{-1}$ for the carrier-carrier beatnote (circle) and the sideband-sideband beatnotes (triangles), respectively.
Particularly in the LISA case, the precision requirement on phase extraction is two-order more stringent for the carrier-carrier than for the sideband-sideband.
Hence, the influence of the second-order coupling on the carrier-carrier beatnote can become comparable to or larger than that of the first-order couplings on the sideband-sideband beatnotes.

%%%%%%%%%%%%%%%%%%%%%%%%%%%%%%%%%%%%%%%%%%%%%%%%%%%%%%%%%
\section{Application}\label{sec:application}
To demonstrate the application of this study, we consider the case where one defines noise requirements from a target sensitivity of phase extraction in ASD.
We particularly focus on heterodyne-band noise here to concisely show a use case.

As presented in \cref{app:conversion}, the time-domain coupling factors in \cref{tab:noise_couplings} can be converted to those in ASD by a factor of $\sqrt{2}$.
In addition, as mentioned at the end of \cref{sub:framework_strategy}, considering incoherent noise contributions with the same noise amplitude from the two interfering beams, a coupling factor gains another factor of $\sqrt{2}$.
Hence, a total phase noise caused by a noise source $x$ ($\in\{p,v,a,\theta\}$) in phase extraction of a beatnote $y$ ($\in\{\mrm{car},\mrm{usb},\mrm{lsb},\mrm{usb}-\mrm{lsb}\}$) can be computed by an incoherent sum:
\begin{align}
    \tilde{X}^y = \sqrt{\sum_\omega \left(2|C^y_{\tau,\omega}|\tilde{x}_\omega\right)^2} = 2\sqrt{\sum_\omega (|C^y_{\tau,\omega}|\tilde{x}_\omega)^2},
    \label{eq:Xy}
\end{align}
where $\tilde{x}_\omega$ is the ASD of the noise $x$ at a relevant frequency $\omega$ in one of the two beams, and $C^y_{\tau,\omega}$ is its time-domain coupling factor to the beatnote $y$.

In the case like LISA where the whole system comprises multiple heterodyne interferometers with different time-dependent heterodyne frequencies, a noise requirement should be defined over a wide frequency band, instead of at specific frequencies.
For this purpose, $\tilde{x}_\omega$ in \cref{eq:Xy} is replaced by a common broadband noise representation $\tilde{x}_\mrm{band}$.
If a requirement for $\tilde{X}^y$ is written by $\tilde{X}^y_\mrm{req}$, one way to formulate the requirement of the individual noise source would be the following:
\begin{align}
    \tilde{x}_\mrm{band,req} &= \min_{y\in\{\mrm{car},\mrm{usb},\mrm{lsb},\mrm{usb}-\mrm{lsb}\}} \tilde{x}^y_\mrm{band,req}
    \nonumber\\
    &= \min_{y\in\{\mrm{car},\mrm{usb},\mrm{lsb},\mrm{usb}-\mrm{lsb}\}}\frac{\tilde{X}^y_\mrm{req}}{2\sqrt{\sum_\omega |C^y_{\tau,\omega}|^2}}.
    \label{eq:x_band}
\end{align}

\begin{table}
    \centering
    \caption{\label{tab:req_param}
     Parameters used to derive requirements.
     The phase extraction requirement $\tilde{X}^\mrm{sb}_\mrm{req}$ is applied to either of usb, lsb, or their combination (``usb-lsb'') in \cref{eq:sbcombi}.
     The modulation depth of \SI{0.584}{\radian} invests \SI{16}{\percent} of the optical power for the modulation sidebands, which is the LISA nominal.
    \vspace{1.5mm}
    }
    \begin{tabular}{cc}
    \toprule
    Parameter & Value
    \\
    \hline\\
    Nominal carrier heterodyne frequency $f_\mrm{het}$ &  \SI{10}{\mega\Hz}  
    \vspace{1.5mm}\\
    Carrier heterodyne bandwidth &  \SI{30}{\mega\Hz}   
    \vspace{1.5mm}\\
    Modulation frequency offset $\Delta f_\mrm{m}$ &  \SI{1.245}{\mega\Hz}   
    \vspace{1.5mm}\\
    Modulation depth $m_0$ & \SI{0.584}{\radian}
    \vspace{1.5mm}\\
    EOM $V_\pi$ & \SI{3.5}{V}
    \vspace{1.5mm}\\
    Phase extraction requirement (carrier) $\tilde{X}^\mrm{car}_\mrm{req}$ & \SI{0.6}{\micro\radian\prtHz}
    \vspace{1.5mm}\\
    Phase extraction requirement (sidebands) $\tilde{X}^\mrm{sb}_\mrm{req}$ & \SI{60}{\micro\radian\prtHz}
    \vspace{1.5mm}\\
    \bottomrule
    \end{tabular}
\end{table}

\begin{figure}[H]
    \centering
\includegraphics[width=8.6cm]{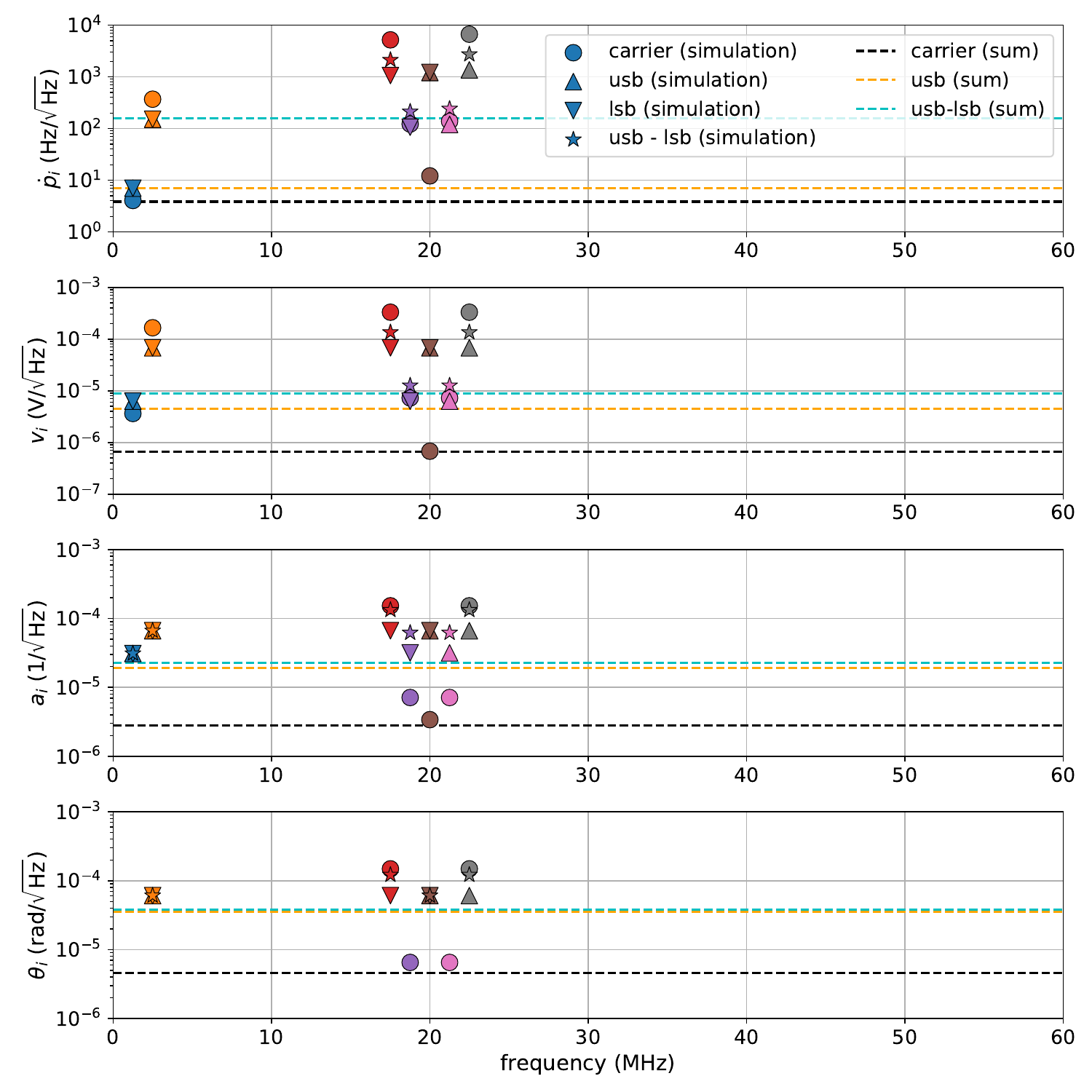}
    \caption{
    Requirements on noise sources ($p_i$, $v_i$, $a_i$, and $\theta_i$) with the parameters in \cref{tab:req_param}.
    Regarding the laser noise in the top panel, we derived the requirement in frequency instead of phase following convention.
    The modulation additive voltage noise $v_i$ in the second panel is analyzed from the laser phase noise via $V_\pi$ as shown in \cref{eq:n_i}.
    Markers show individual contributions to the requirement levels $\tilde{x}^y_{\mrm{band,req}}$ in \cref{eq:x_band}: $\tilde{X}^y_\mrm{req}/2\sqrt{|C^y_{\tau,\omega}|^2}$.
    Due to the reciprocal formulation, the individual contributions are higher than the resulting sum for each $y$.
    However, for \cref{eq:x_band} the minimum (and therefore most stringent) of each y (sum) is chosen as a requirement.
    Although the markers are computed with the nominal carrier heterodyne frequency of \SI{10}{\mega\Hz}, it varies from \SIrange{5}{30}{\mega\Hz} in the LISA case; hence, the requirement is defined up to the double heterodyne frequency, namely \SI{60}{\mega\Hz}.
    }
    \label{fig:het_noise_req}
\end{figure}

Using LISA-like parameters in \cref{tab:req_param}, we apply \cref{eq:x_band} to individual noise sources and compute heterodyne-band requirements, as shown in \cref{fig:het_noise_req}.
All markers are based on the nominal heterodyne frequency of \SI{10}{\mega\Hz}.
However, it varies up to the heterodyne bandwidth \SI{30}{\mega\Hz} over time; therefore, the requirements are defined to the double heterodyne frequency \SI{60}{\mega\Hz}.
$\tilde{x}^y_\mrm{band,req}$ in \cref{eq:x_band} for $y\in\{\mrm{car},\mrm{usb},\mrm{usb}-\mrm{lsb}\}$ is shown in black, orange, and cyan horizontal lines, and the minimum one out of the three becomes a requirement.
With the particular parameters in \cref{tab:req_param}, the noise coupling to the carrier phase extraction is dominant and determines the requirements on all noise sources: $\tilde{x}_\mrm{band,req}=\tilde{x}^\mrm{car}_\mrm{band,req}$.
Apart from the carrier, it would be notable that the combination of the sideband phases in \cref{eq:sbcombi} is particularly helpful to mitigate the coupling of the laser noise, the top panel of \cref{fig:het_noise_req}, because it could relax the requirement derived from the sideband phase extraction by more than one order of magnitude: $\tilde{p}^\mrm{usb-lsb}_\mrm{band,req} \gg \tilde{p}^\mrm{usb}_\mrm{band,req}$. 
The results are also summarized in \cref{tab:req_results}.

\begin{table}
    \centering
    \caption{\label{tab:req_results}
     Summary of the requirements shown in \cref{fig:het_noise_req} with the LISA-like parameters in \cref{tab:req_param}.
     Dominant coupling shows the coupling path in brackets, which tells if it is self or mutual noise coupling (also see \cref{tab:noise_couplings}).
    \vspace{1.5mm}
    }
    \begin{tabular}{ccc}
    \toprule
    Noise & Requirement & Dominant coupling
    \\
    \hline\\
    $\dot{p}_i$ &  \SI{3.84}{\Hz\prtHz} & $p_i(\Delta \omega_\mrm{m})$ [usb,lsb$\rightarrow$car]
    \vspace{1.5mm}\\
    $v_i$ &  \SI{659}{\nano\volt\prtHz} & $p_i(2\omega_\mrm{het})$ [car$\rightarrow$car]
    \vspace{1.5mm}\\
    $a_i$ &  \SI{2.82e-6}{1\prtHz} & $a_i(2\omega_\mrm{het})$ [car$\rightarrow$car]
    \vspace{1.5mm}\\
    $\theta_i$ &  \SI{35.0}{\micro\radian\prtHz} & $\theta_i(2\omega_\mrm{het}\pm\Delta\omega_\mrm{m})$ [usb,lsb$\rightarrow$car]
    \vspace{1.5mm}\\
    \bottomrule
    \end{tabular}
\end{table}

In this section, we demonstrate only the use case of heterodyne-band noise analysis.
Necessary requirements highly depend on experiments and their verification approach.
It would also be possible to set up a verification test in such a way that the modulation-band noise is naturally tested by looking at just a noise floor in the heterodyne band as it couples via the heterodyne band, as shown in \cref{fig:ghz_diagram}.

%%%%%%%%%%%%%%%%%%%%%%%%%%%%%%%%%%%%%%%%%%%%%%%%%%%%%%%%%
\section{Conclusion}\label{sec:conclusion}
Heterodyne interferometry is a versatile tool used in many fields of precision science.
The technology is often combined with a laser phase modulation to extract additional information of an experimental setup, e.g., intersatellite clock noise transfer for GW detection in space.

In this paper, in the format of spaceborne GW detectors, we study additional noise couplings caused by the phase modulation especially in the high frequency regimes like heterodyne or modulation bands, which were not well explored before.
In addition to the modulation noise, the laser noise is also discussed in the same framework.
In \cref{sec:het_band}, we systematically derive all possible noise couplings from the heterodyne band based on interactions between heterodyne beatnotes in a PR signal in \cref{eq:spr}.
All couplings found are listed in \cref{tab:noise_couplings}.
In \cref{sec:mod_band}, we shift the frequency band of interest to the even higher frequency regime, i.e., the modulation band.
We show that in this case, we can first rewrite the noise in a different form via trigonometric analysis or sideband analysis, focusing on a single electromagnetic field.
This enables us to transform the noise in the modulation band to the noise in the heterodyne band with amplitude scaling factors, which then can be combined with the discussion in \cref{sec:het_band} to compute the total coupling factor.
In \cref{sec:verification}, we verify the couplings derived analytically by comparison with the numerical simulation.
Contrary to the analysis, the simulation does not rely on any approximation, e.g., the first-order approximation.
Therefore, the agreement between the analysis and the simulation confirms that our analytical framework successfully captures primary noise couplings.
Finally, in \cref{sec:application}, we demonstrate a use case of this study taking the LISA-like setup as an example.
This show how we could derive requirements on each noise source ($p_i,v_i,a_i,$ and $\theta_i$) based on the target precision of phase extraction by a phasemeter and the noise coupling factors derived in this paper.

It should be noted that
although a pseudorandom-noise (PRN) signal for the intersatellite absolute ranging~\cite{Sutton2010,Esteban2011,Xie2023,Yamamoto2024,Euringer2024,Yamamoto2025,Sun2025} is not explicitly discussed, it can be considered as a constituent of the additive voltage noise $v_i$ in the phase modulation signal in \cref{eq:Vmi}.
As discussed in \cref{sec:framework}, $v_i$ turns into $n_i$ in the optical domain, which is indistinguishable from the laser phase noise $p_i$; therefore, the discussion of the laser phase noise $p_i$ is also informative for the design of binary PRN codes and their spectra.
In addition, unlike the aliased laser frequency noise~\cite{Yamamoto2022,Staab2024,Vidal2025}, the coupling of the high-frequency noise discussed in this paper cannot be suppressed by any filtering at a phasemeter.
This means that the only possible treatment is to mitigate the noise sources themselves or fine-tune a few parameters; e.g., setting $\Delta \omega_\mrm{m}$ at a PRN chip rate helps minimize the coupling from the PRN spectrum to the sideband-sideband readout.
% In addition, although a pseudorandom-noise (PRN) signal for the intersatellite absolute ranging~\cite{Sutton2010,Esteban2011,Xie2023,Yamamoto2024,Euringer2024,Yamamoto2025} is not explicitly discussed, it can be considered as a constituent of the additive voltage noise $v_i$ in the phase modulation signal in \cref{eq:Vmi}.
% As discussed in \cref{sec:framework}, $v_i$ turns into $n_i$ in the optical domain, which is indistinguishable from the laser phase noise $p_i$; therefore, the discussion of the laser phase noise $p_i$ is also informative for the design of binary PRN codes and their spectra.
In addition, we considered the EOM as an ideal linear converter from an input voltage signal to a beam phase.
However, the contribution of the EOM to the modulation noise ($a_i$, $\theta_i$, and $v_i$) is naturally covered by this study, as long as they are defined in the optical domain.
The effect of the other EOM imperfection, like residual amplitude modulation
(RAM) or nonlinearity, was beyond the scope.
RAM would cause asymmetrical noise property between the upper and lower sidebands.
However, our coupling factors were derived per sideband, hence, we conjecture that the results could be reused for such asymmetrical noise: Noise $x_i$ would be replaced by $x_i^\mrm{upper}$ and $x_i^\mrm{lower}$, and we separately apply individual coupling factors to them.
Further investigations would be needed for those effects.
%The effect of the other EOM imperfection, like residual amplitude modulation or nonlinearity, is beyond the scope and would need further investigations.
Also, we remark that if we utilize optical frequency comb for clock treatment~\cite{Tinto2015,Vinckier2020,Tan2022,Yamamoto2025B}, we could remove the EOM, which as a result eliminates any noise couplings in this paper, except for the self noise coupling for the carrier in \cref{tab:noise_couplings}: $p_i(\epsilon)$ and $p_i(2\omega_\mrm{het})$.

We expect that this study establishes a good demonstration of the systematic search for noise coupling in heterodyne interferometry or precision optical metrology systems more in general.
Besides, in the context of spaceborne GW detectors, the result is expected to provide necessary information to define requirements on essential devices, like laser transmitters and phase modulators.
A future work is to explore the high-frequency noise in the format of spaceborne GW detectors more deeply: the behaviour of these high-frequency noises in time-delay interferometry~\cite{Tinto1999,Vallisneri2005}.

%%%%%%%% Acknowledgements %%%%%%%%%%%%%%%%%%%%%%%%%%%%%%%%%%%
\begin{acknowledgments}
K.Y., H.L., K.N., and R.D. acknowledge the support from the NASA Physics of the Cosmos (PhysCOS) program.
K.Y.’s work is supported by NASA under Award No. 80GSFC24M0006.
H.L.'s work is supported by the NASA Space Technology Mission Directorate  Early Career Initiative Award.

O.H. and L.W. gratefully acknowledge the support by the Deutsches Zentrum für Luft- und Raumfahrt (DLR) with funding of the Federal Ministry for Economic Affairs and Energy based on a resolution of the Deutsche Bundestag (FKZ 50OQ1801 and FKZ 50OQ2301).
\end{acknowledgments}

%%%%%%%% Appendix %%%%%%%%%%%%%%%%%%%%%%%%%%%%%%%%%%%
% \clearpage
\appendix
\section{Additive readout noise}\label{sub:framework_readout}
Let us derive a general formula for the coupling of additive readout noise in phase extraction with a phasemeter based on its working principle~\cite{Shaddock2006}.
The phase-locked loop (PLL) on a phasemeter extracts the phase of the input sinusoidal signal with a band-pass effect around the input signal frequency.
As an input to the phasemeter, we consider the following signal:
\begin{align}
    s = k\cos(\omega \tau + \phi_k) + l\cos((\omega + \epsilon)\tau + \phi_l),
    \label{eq:s_add}
\end{align}
where the first term is a primary signal and $\phi_k$ is a target phase to extract, and the second term is the monotonic representation of the additive noise at $\omega + \epsilon$ where $\epsilon\ll\omega$.
We assume that the noise amplitude $l$ is much smaller than the nominal signal amplitude $k$: $l \ll k$.
The scope is how the second term disturbs the phase extraction of the first term.

We model the phase extraction by a simple I/Q demodulation.
The I and Q values are formulated by
\begin{subequations}\label{eq:IQ}
    \begin{align}
        I &= s\cdot \cos(\omega \tau)
        \nonumber\\
        &= k\cos(\omega \tau + \phi_k)\cdot \cos(\omega \tau) + l\cos((\omega + \epsilon)\tau + \phi_l)\cdot \cos(\omega \tau)
        \nonumber\\
        &\xrightarrow[]{\text{LPF}} (k\cos \phi_k + l\cos(\epsilon\tau + \phi_{l}))/2,
        \label{eq:I}\\
        Q &= s\cdot \sin(\omega \tau)
        \nonumber\\
        &= k\cos(\omega \tau + \phi_k)\cdot \sin(\omega \tau) + l\cos((\omega + \epsilon)\tau + \phi_l)\cdot \sin(\omega \tau)
        \nonumber\\
        &\xrightarrow[]{\text{LPF}} (k\sin \phi_k + l\sin(\epsilon\tau + \phi_{l}))/2,
        \label{eq:Q}
    \end{align}
\end{subequations}
where ``LPF" represents the low-pass filtering to filter out the second harmonic of the input signal.
The phase extraction is achieved by $\arctan$ of their ratio as
\begin{align}
    \phi_\mathrm{read} &= \arctan\left(Q/I\right)
    \nonumber\\
    &= \arctan\left(\frac{k\sin \phi_k + l\sin(\epsilon\tau + \phi_{l})}{k\cos \phi_k + l\cos(\epsilon\tau + \phi_{l})}\right)
    \nonumber\\
    &\approx \phi_k + \phi_\mrm{err},
    \label{eq:IQdemod}\\
    \phi_\mrm{err} &= l/k\cdot\sin(\epsilon\tau + \phi_l-\phi_k),
    \label{eq:IQerr}
\end{align}
where $l\ll k$ is used in the last line in \cref{eq:IQdemod}.
This suggests that the phase noise amplitude is determined by the ratio of the noise amplitude $l$ to the signal amplitude $k$ in the input signal in \cref{eq:s_add}.
Also, the resulting angular frequency of the noise is $\epsilon$; therefore, if the angular frequency offset $\epsilon$ is in the science observation band (namely \SI{0.1}{\milli\Hz} to \SI{1}{\Hz} for LISA), the noise appears in the final phase readout in the band.
Although we model phase extraction with I/Q demodulation in this section, the noise coupling does not change in the PLL-based phasemeter as long as the tracking bandwidth is sufficiently higher than the frequencies of the signal of interest.

\section{Coupling factor conversion}\label{app:conversion}
In this paper, we use a monotonic noise representation and analyze its coupling to a measured phase in the time domain.
However, in practice, the noise to deal with is stochastic, and we are interested in its coupling to the measured phase in terms of the amplitude spectral density (ASD) in the frequency domain.
In this section, the conversion of the coupling factor between the time and frequency domains is discussed.

Following \cref{eq:x_i}, let us define the monotonic noise $n$ in the time domain,
\begin{align}
    n(\tau) = m_\mrm{n}\cos(\omega_\mrm{n}\tau+\rho),
    \label{eq:noise_representation}
\end{align}
where $\omega_\mrm{n}$ is an angular frequency.
As discussed in \cref{sub:framework_readout}, of our interest is the sum of the signal frequency $\omega_s$ (or a frequency that results in $\omega_s$ after some down- or up-conversions) and a small offset $\epsilon$ in the observation band: $\omega_\mrm{n}=\omega_s+\epsilon$.
$\rho$ is a given stochastic phase.
$m_\mrm{n}$ is also random; however, considering it as a long-term rms value, it can be replaced by
\begin{align}
    m_\mrm{n} &= \sqrt{2}\cdot\sqrt{2n_0b},
    \label{eq:m_n_n0}
\end{align}
as discussed around Eq. (26) in \cite{Wissel2022}.
Assuming a white noise, $n_0$ is the noise floor in power spectral density (PSD), and $b=\SI{1}{\Hz}$, which is the white noise bandwidth we consider.
The factor of $\sqrt{2}$ in \cref{eq:m_n_n0} reflects the fact that the mixing process in PLL of a phasemeter wraps the uncorrelated noise powers below and above the primary signal frequency $\omega_s$ (namely, at $\omega_\mrm{n} = \omega_s\pm\epsilon$) around $f=\SI{0}{\Hz}$, which result in the phase noise at $\epsilon$ with the PSD noise floor increased by $2$ in the phasemeter output.
Therefore, to represent the total coupling of the noise at $\omega_\mrm{n}$ by a single monotonic tone in \cref{eq:noise_representation}, we applied a factor of $\sqrt{2}$ to $m_\mrm{n}$ in \cref{eq:m_n_n0}.

An important question is how to relate a noise coupling factor of $n(\tau)$ to the one of the PSD (or ASD) noise floor $n_0$ (or $\sqrt{n_0}$).
In other words, we need to drive the relation between the coupling factors of $C_{\tau}$ in the time domain and $C_\mrm{asd}$ in units of ASD in the following expressions of the phasemeter output phases,
\begin{align}
    % \phi_\mrm{read} &= \phi + m_\mrm{n},
    \phi_\mrm{read}(\tau) &= \phi(\tau) + C_{\tau}\cdot m_\mrm{n}\sin(\epsilon_n\tau + \rho-\phi),
    \label{eq:phi_read_time}\\
    S_{\phi_\mrm{read}}(\omega) &= S_{\phi}(\omega) + C^2_\mrm{asd}\cdot n_0.
    \label{eq:phi_read_PSD}
    \end{align}
\cref{eq:phi_read_time} is written in the time domain, while \cref{eq:phi_read_PSD} is in units of PSD.
$\sin(\epsilon_n\tau + \rho-\phi)$ comes from the down-conversion of the noise in \cref{eq:noise_representation} by a given target signal represented by $\cos(\omega_s+\phi)$.
This happens in the mixing process on a phasemeter.
$\phi$ is the target phase to extract, and $S_{\phi}$ is its PSD.

We can now derive the noise coupling in PSD from the time-domain expression in \cref{eq:phi_read_time} by calculating its variance and normalizing it by a \SI{1}{\Hz} bandwidth ($=b$),
\begin{align}
    S_n \approx E\left[\left(C_{\tau}\cdot m_\mrm{n}\sin(\epsilon_n\tau + \rho-\phi)\right)^2\right]/b = 2C^2_\tau n_0.
    \label{eq:S_n}
\end{align}
By equating it to the noise term in \cref{eq:phi_read_PSD}, we end up with
\begin{align}
    2C^2_\tau n_0 &= C^2_\mrm{asd}\cdot n_0
    \nonumber\\
    &\Leftrightarrow C_\mrm{asd} = \sqrt{2}|C_\tau|.
    \label{eq:Casd_Ctau}
\end{align}
Therefore, we can easily switch between the noise coupling factor in the time and frequency domain via a factor of $\sqrt{2}$.
Considering this fact, the discussion in \cref{sec:het_band,sec:mod_band}, including the summary in \cref{tab:noise_couplings}, sticks to the time domain.

\section{Single sideband}\label{app:ssb}
We recall the formulation of a single sideband here.
Let us consider an electromagnetic field at $\omega$ with a single sideband at $\Delta\omega$
\begin{align}
    E &= A(1 + a \e^{j\Delta\omega t})\e^{j\omega t},
    \label{eq:E_ussb}
\end{align}
where we assume $\Delta\omega>0$; hence, this is an upper single sideband.
\cref{eq:E_ussb} suggests that the single sideband modulates both the phase and the amplitude of the field.
If we decompose this into amplitude and phase, we get
\begin{align}
    E &= A\left(1 + \frac{a}{2} \e^{j\Delta\omega t} + \frac{a}{2} \e^{j\Delta\omega t} + \frac{a}{2} \e^{-j\Delta\omega t} - \frac{a}{2} \e^{-j\Delta\omega t}\right)\e^{j\omega t}
    \nonumber\\
    &= A\left(1 + \frac{a}{2} (\e^{j\Delta\omega t} + \e^{-j\Delta\omega t}) + \frac{a}{2} (\e^{j\Delta\omega t} - \e^{-j\Delta\omega t})\right)\e^{j\omega t}
    \nonumber\\
    &\approx A\left(1 + \frac{a}{2} (\e^{j\Delta\omega t} + \e^{-j\Delta\omega t})\right)\cdot\left(1 + \frac{a}{2} (\e^{j\Delta\omega t} - \e^{-j\Delta\omega t})\right)\e^{j\omega t}
    \nonumber\\
    &\approx A\left(1 + a\cos(\Delta\omega t) \right)\e^{j(\omega t + a\cos(\Delta\omega t-\pi/2))}.
    \label{eq:E_ussb2}
\end{align}
Therefore, the upper single sideband can be interpreted as the combination of an amplitude modulation and a coherent phase modulation with the phase delay of $\pi/2$.
The lower single sideband can be easily derived with the replacement as $\Delta\omega\rightarrow-\Delta\omega$,
\begin{align}
    E &= A(1 + a \e^{-j\Delta\omega t})\e^{j\omega t}
    \nonumber\\
    &\approx A\left(1 + a\cos(-\Delta\omega t) \right)\e^{j(\omega t + a\sin(-\Delta\omega t))}
    \nonumber\\
    &= A\left(1 + a\cos(\Delta\omega t) \right)\e^{j(\omega t + a\cos(\Delta\omega t+\pi/2))}.
    \label{eq:E_lssb}
\end{align}
Hence, the lower single sideband can be interpreted as the combination of the amplitude modulation and the coherent phase modulation with the phase advancement of $\pi/2$.
Note that the amplitude of the single sideband $a$ directly becomes the amplitude of the amplitude and phase modulations.

\section{modulation-band noise with the $2\omega_\mrm{het}$ offset}\label{app:2fhet}
Similarly as \cref{sub:mod_band_coherence}, we provide the coherence analysis of various coupling paths from the modulation-band noise with the $2\omega_\mrm{het}$ offset from the integer multiple of the modulation frequency $\omega_{m,i}$ in this section.
According to \cref{eq:pi2ai_2thetai}, the laser phase noise $p_i$ at $\omega_{m,i}+2\omega_\mrm{het}$ becomes
\begin{widetext}
\begin{align}
    p_i(\omega_{\mrm{m},i}+2\omega_\mrm{het}; m_\mrm{p}, \rho)
    &\rightarrow
    \begin{cases}
        a_i\left(2\omega_\mrm{het};\frac{m_\mrm{p}}{m_0},\rho-\phi^\mrm{sb}_{\mrm{tar},i}\right)
        \\
        \theta_i\left(2\omega_\mrm{het};\frac{m_\mrm{p}}{m_0},\rho-\phi^\mrm{sb}_{\mrm{tar},i}-\frac{\pi}{2}\right)
    \end{cases}
    \nonumber\\
     &\xrightarrow[\text{extraction}]{\text{phase}}
     \begin{cases}
    -\frac{m_\mrm{p}}{m_0}\cdot\frac{m^2_0}{4J_0(m_0)}\sin(\rho-\phi^\mrm{sb}_{\mrm{tar},i}-2\phi^\mrm{car}_\mrm{tar}) + 0
    =-\frac{m_\mrm{p} m_0}{4J_0(m_0)}\sin(\rho-\phi^\mrm{sb}_{\mrm{tar},i}-2\phi^\mrm{car}_\mrm{tar}) & \text{for $\phi^\mrm{car}_\mrm{read}$}
    \\
    \frac{m_\mrm{p}}{m_0}\cdot\frac{1}{2}\sin(\rho-\phi^\mrm{sb}_{\mrm{tar},i}-2\phi^\mrm{car}_\mrm{tar}) + \frac{m_\mrm{p}}{m_0}\cdot\frac{1}{2}\cos(\rho-\phi^\mrm{sb}_{\mrm{tar},i}-2\phi^\mrm{car}_\mrm{tar}-\frac{\pi}{2})
    = \frac{m_\mrm{p}}{m_0}\sin(\rho-\phi^\mrm{sb}_{\mrm{tar},i}-2\phi^\mrm{car}_\mrm{tar}) & \text{for $\phi^\mrm{usb}_\mrm{read}$}
    \\
    \frac{m_\mrm{p}}{m_0}\cdot\frac{1}{2}\sin(\rho-\phi^\mrm{sb}_{\mrm{tar},i}-2\phi^\mrm{car}_\mrm{tar}) - \frac{m_\mrm{p}}{m_0}\cdot\frac{1}{2}\cos(\rho-\phi^\mrm{sb}_{\mrm{tar},i}-2\phi^\mrm{car}_\mrm{tar}-\frac{\pi}{2})
    = 0 & \text{for $\phi^\mrm{lsb}_\mrm{read}$}.
    \end{cases}
    \label{eq:pi_fm2fhet}
\end{align}
Following \cref{eq:pi_2fm_viausb,eq:pi_2fm_vialsb}, the laser phase noise $p_i$ at $2\omega_{\mrm{m},i}+2\omega_\mrm{het}$ becomes
\begin{align}
    p_i(2\omega_{\mrm{m},i}&+2\omega_\mrm{het}; m_\mrm{p}, \rho)
    \nonumber\\
    &\xrightarrow[\text{extraction}]{\text{phase}}
    \begin{cases}
    0 & \text{for $\phi^\mrm{car}_\mrm{read}$}
    \\
    \hat{a}_i\left(2\omega_\mrm{het}; \frac{m_p}{2}, \rho-2\phi^\mrm{sb}_{\mrm{tar},i}-\frac{\pi}{2}\right) + \hat{\theta}_i\left(2\omega_\mrm{het}; \frac{m_p}{2}, \rho-2\phi^\mrm{sb}_{\mrm{tar},i}-\pi\right) = -\frac{m_\mrm{p}}{2}\cos(\rho-2\phi^\mrm{sb}_{\mrm{tar},i}-2\phi^\mrm{car}_\mrm{tar}) & \text{for $\phi^\mrm{usb}_\mrm{read}$}
    \\
    \hat{a}_i\left(2\omega_\mrm{het}; \frac{m_p}{2}, \rho-2\phi^\mrm{sb}_{\mrm{tar},i}+\frac{\pi}{2}\right) + \hat{\theta}_i\left(2\omega_\mrm{het}; \frac{m_p}{2}, \rho-2\phi^\mrm{sb}_{\mrm{tar},i}\right)
    = 0  & \text{for $\phi^\mrm{lsb}_\mrm{read}$}.
    \end{cases}
    \label{eq:pi_2fm2fhet}
\end{align}
Similarly to \cref{eq:pi_2fmfinband}, we express the coupled term of the noise $x_i$ in the measured phases by  $\hat{x}_i$.
According to \cref{tab:noise_couplings}, $a_i(2\omega_\mrm{het})$ and $\theta_i(2\omega_\mrm{het})$ are the mutual noise coupling between the usb-usb and lsb-lsb beatnotes: the lsb-lsb beatnote affects $\phi^\mrm{usb}_\mrm{read}$, or vice versa.
Hence, \cref{eq:pi_2fm_viausb,eq:pi_2fm_vialsb} are applied to $\phi^\mrm{lsb}_\mrm{read}$ and $\phi^\mrm{usb}_\mrm{read}$, respectively.

On the other hand, from \cref{eq:ai2pi,eq:pi_2fm2fhet}, the modulation amplitude noise at $\omega_{\mrm{m},i}+2\omega_\mrm{het}$ and $2\omega_{\mrm{m},i}+2\omega_\mrm{het}$ couples as,
\begin{align}
    a_i\left(\omega_{\mrm{m},i}+2\omega_\mrm{het}; m_\mrm{a}, \rho\right) &\rightarrow p_i\left(2\omega_\mrm{het}; \frac{m_0m_\mrm{a}}{2}, \rho-\phi^\mrm{sb}_{\mrm{tar},i}\right) + p_i\left(2\omega_{\mrm{m},i} + 2\omega_\mrm{het}; \frac{m_0m_\mrm{a}}{2}, \rho+\phi^\mrm{sb}_{\mrm{tar},i}\right)
    \nonumber\\
    & \xrightarrow[\text{extraction}]{\text{phase}}
    \begin{cases}
    -\frac{m_0m_\mrm{a}}{4} \cos(\rho-\phi^\mrm{sb}_{\mrm{tar},i}-2\phi^\mrm{car}_\mrm{tar}) + 0 = -\frac{m_0m_\mrm{a}}{4} \cos(\rho-\phi^\mrm{sb}_{\mrm{tar},i}-2\phi^\mrm{car}_\mrm{tar}) & \text{for $\phi^\mrm{car}_\mrm{read}$}
    \\
    -\frac{m_0m_\mrm{a}}{4} \cos(\rho-\phi^\mrm{sb}_{\mrm{tar},i}-2\phi^\mrm{car}_\mrm{tar}) - \frac{m_0m_\mrm{a}}{4}\cos(\rho-\phi^\mrm{sb}_{\mrm{tar},i}-2\phi^\mrm{car}_\mrm{tar})
    = -\frac{m_0m_\mrm{a}}{2}\cos(\rho-\phi^\mrm{sb}_{\mrm{tar},i}-2\phi^\mrm{car}_\mrm{tar}) & \text{for $\phi^\mrm{usb}_\mrm{read}$}
    \\
    -\frac{m_0m_\mrm{a}}{4} \cos(\rho-\phi^\mrm{sb}_{\mrm{tar},i}-2\phi^\mrm{car}_\mrm{tar}) + 0
    = -\frac{m_0m_\mrm{a}}{4} \cos(\rho-\phi^\mrm{sb}_{\mrm{tar},i}-2\phi^\mrm{car}_\mrm{tar}) & \text{for $\phi^\mrm{lsb}_\mrm{read}$},
    \end{cases}
    \label{eq:ai_fm2fhet}\\
    a_i\left(2\omega_{\mrm{m},i}+2\omega_\mrm{het}; m_\mrm{a}, \rho\right) &\rightarrow
    \begin{cases}
        a_i\left(2\omega_\mrm{het}; \frac{m_\mrm{a}}{2}, \rho-2\phi^\mrm{sb}_{\mrm{tar},i}\right)
        \\
        \theta_i\left(2\omega_\mrm{het}; \frac{m_\mrm{a}}{2}, \rho-2\phi^\mrm{sb}_{\mrm{tar},i}-\frac{\pi}{2}\right)
    \end{cases}
    \nonumber\\
    & \xrightarrow[\text{extraction}]{\text{phase}}
    \begin{cases}
    -\frac{m^2_0m_\mrm{a}}{8J_0(m_0)}\sin\left(\rho-2\phi^\mrm{car}_\mrm{tar}-2\phi^\mrm{sb}_{\mrm{tar},i}\right) + 0 =-\frac{m^2_0m_\mrm{a}}{8J_0(m_0)}\sin\left(\rho-2\phi^\mrm{car}_\mrm{tar}-2\phi^\mrm{sb}_{\mrm{tar},i}\right) & \text{for $\phi^\mrm{car}_\mrm{read}$}
    \\
    \frac{m_\mrm{a}}{4}\sin(\rho-2\phi^\mrm{sb}_{\mrm{tar},i}-2\phi^\mrm{car}_\mrm{tar}) + \frac{m_\mrm{a}}{4}\cos(\rho-2\phi^\mrm{sb}_{\mrm{tar},i}-2\phi^\mrm{car}_\mrm{tar}-\frac{\pi}{2})
    = \frac{m_\mrm{a}}{2}\sin(\rho-2\phi^\mrm{sb}_{\mrm{tar},i}-2\phi^\mrm{car}_\mrm{tar}) & \text{for $\phi^\mrm{usb}_\mrm{read}$}
    \\
    \frac{m_\mrm{a}}{4}\sin(\rho-2\phi^\mrm{sb}_{\mrm{tar},i}-2\phi^\mrm{car}_\mrm{tar}) - \frac{m_\mrm{a}}{4}\cos(\rho-2\phi^\mrm{sb}_{\mrm{tar},i}-2\phi^\mrm{car}_\mrm{tar}-\frac{\pi}{2})
    = 0 & \text{for $\phi^\mrm{lsb}_\mrm{read}$}.
    \end{cases}
    \label{eq:ai_2fm2fhet}
\end{align}
\end{widetext}
Regarding the modulation phase noise, $\theta_i$($\omega_{\mrm{m},i}+2\omega_\mrm{het}$) effectively has the additional minus sign in the coupling from $p_i$($2\omega_{\mrm{m},i}+2\omega_\mrm{het}$) to that of \cref{eq:ai_fm2fhet} because \cref{eq:thetai2pi} suggests the $\pi$ phase shift between $p_i$($2\omega_\mrm{het})$ and $p_i$($2\omega_{\mrm{m},i}+2\omega_\mrm{het})$.
Concerning $\theta_i$($2\omega_{\mrm{m},i}+2\omega_\mrm{het}$), the result (to be precise, the coupled noise amplitude) is effectively the same as $a_i$($2\omega_{\mrm{m},i}+2\omega_\mrm{het}$) above because both \cref{eq:ai2pi2ai,eq:thetai2pi2thetai} show the same relative phase shift between the resulting $a_i$ and $\theta_i$: the $\pi/2$ delay of $\theta_i$ against $a_i$.

\section{Relation to laser RIN}\label{app:rin}
In this paper, we focus on laser phase noise $p_i$ and phase modulation noises $v_i$, $a_i$, and $\theta_i$.
Nevertheless, as the modulation relative amplitude noise $a_i$ changes the amplitude of individual optical fields as shown in \cref{eq:Ei_phim_sb}, many couplings in \cref{tab:noise_couplings} are closely related to the laser-RIN-coupling mechanism.
The description of laser RIN below references to \cite{Wissel2022}.

First, any noise discussed in this work does not change the power of the non-interfering optical components at all: $(P_i+P_j)/2$ in \cref{eq:spr}.
Therefore, none of noise couplings found here has the same mechanism as that of ``1f RIN" that comes from the jitter in the non-interfering terms.

Second, the source of ``2f RIN" is the jitter of the amplitude of the heterodyne beatnotes, which is caused also by the modulation relative amplitude noise $a_i$.
Laser RIN can be modeled by power fluctuation: $P'_i=P_i(1+r_i)$ where $P_i$ is the averaged power of beam $i$ and $r_i$ is its RIN.
This is related to the laser-relative-amplitude noise $\alpha_i$ by $\alpha_i\approx r_i/2$: the field amplitude is $A'_i=A_i(1 + r_i/2)$ where $A_i$ is the averaged amplitude.
Focusing on the noise of one of the two interfering beams, we write the beatnote amplitude as follows: $A_iA_jJ_0^2(m_0)\left(1-\frac{m_0^2a_i}{2J_0(m_0)}\right)$ for the car-car beatnote in \cref{eq:scar_prifo}, and $A_iA_jJ_1^2(m_0)(1+a_i)$ for the sideband-sideband beatnotes in \cref{eq:ssb_prifo}.
Therefore, we conjecture that any $a_i$ coupling in \cref{tab:noise_couplings} can be translated to laser's 2f-RIN-like coupling at each noise frequency by the following treatments: the additional factor of $-\frac{J_0(m_0)}{m_0^2}$ for the couplings from the car-car beatnote, and the additional factor of $\frac{1}{2}$ for the couplings from the usb-usb and lsb-lsb beatnotes.

Regarding the self noise coupling in \cref{tab:noise_couplings} (i.e., $a_i(2\omega_\mathrm{het})$, $a_i(2(\omega_\mathrm{het}+\Delta\omega_m))$, and $a_i(2(\omega_\mathrm{het}-\Delta\omega_m))$), we can validate this translation as follows: the above translations result in the coupling factor of $1/4$ for all.
\cref{tab:noise_couplings} shows the time-domain coupling factor from a single-beam noise source.
To calculate the coupling factor in ASD considering noise from both beams, we need to apply an additional factor of $\sqrt{2}\cdot\sqrt{2}=2$.
This results in $1/4\cdot\sqrt{2}\cdot\sqrt{2}=1/2$, which is consistent with \cite{Wissel2022}.

Regarding the mutual noise coupling in \cref{tab:noise_couplings}, laser RIN has never been discussed for the interaction between the optical fields; hence, this would be considered a new result.

\section{Higher-order sidebands}\label{app:higher}
Although most of our calculations focus on the interaction between the carrier and the first-order sidebands, there of course exists the noise coupling from the higher-order sidebands.
We provide a rough estimate of the modulation depth threshold for considering the second-order sidebands below.
We can focus on the mutual noise coupling because we never measure the phase of the higher-order sidebands.

As shown in \cref{tab:noise_couplings}, heterodyne-band noise from the order $n$ to the order $k$ ($k=0$ represents the carrier) is proportional to $J_n(m_0)^2/J_k(m_0)^2 \sim m_0^{2(n-k)}$.
On the other hand, as discussed in \cref{sub:mod_band_sideband}, modulation-band noise can be proportional to $J_n(m_0)/J_k(m_0) \sim m_0^{n-k}$.
Hence, the ratio of the second-order coupling magnitude to the total, neglecting even higher-order sidebands, would be $J_2^2(m_0)/\sqrt{J_n^4(m_0)+J_2^4(m_0)}$ for heterodyne-band noise and $J_2(m_0)/\sqrt{J_n^2(m_0)+J_2^2(m_0)}$ for modulation-band noise.

\begin{figure}
    \centering
    \includegraphics[width=8.6cm]{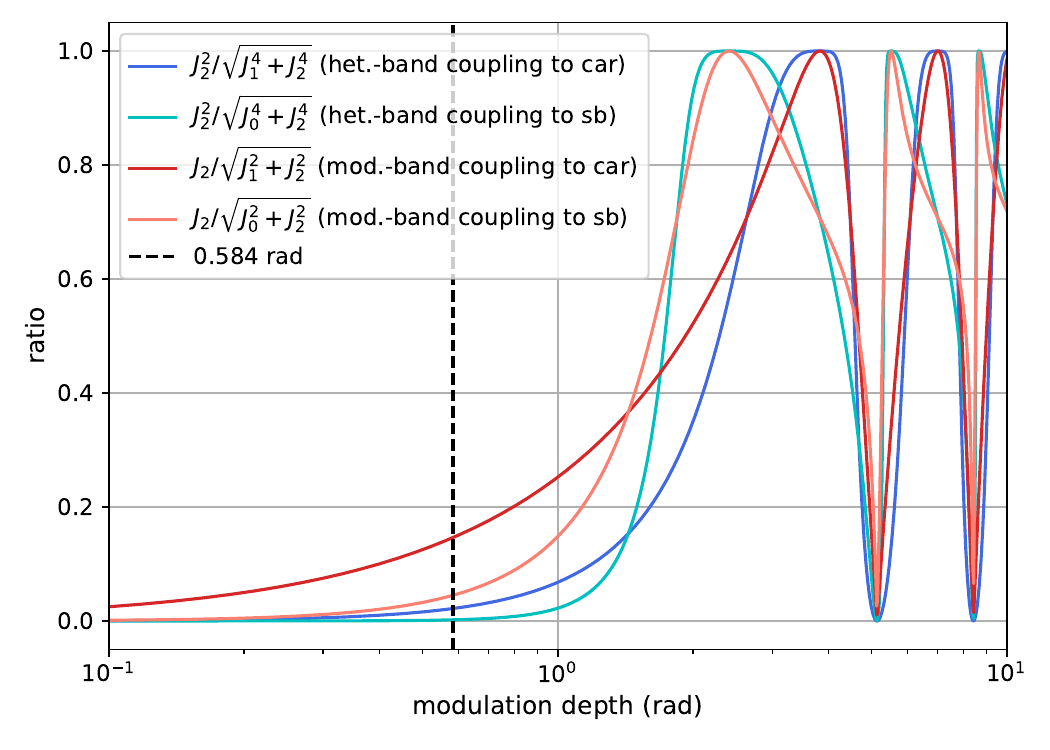}
    \caption{Rough estimate on how the noise contribution of the second-order sidebands grows with the increase of the modulation depth.
    Blue and cyan curves are the estimations in the heterodyne-band coupling to phase extraction of the carrier-carrier beatnote and the sideband-sideband beatnotes, while red and pink curves are the modulation-band-noise counterparts.
    }
    \label{fig:j2_ratio}
\end{figure}

This is plotted in \cref{fig:j2_ratio}.
Under $m_0$ of \SI{0.584}{\radian} used in \cref{sec:verification}, the second-order modulation-band coupling to the carrier (red) accounts for about \SI{15}{\percent} the total mutual coupling, while that of heterodyne-band coupling (blue) is only \SI{2}{\percent}.
Hence, we neglect the second-order sidebands for heterodyne-band noise coupling but consider it for modulation-band noise coupling in the main texts.
If we consider \SI{10}{\percent} as a minimum non-negligible contribution, the modulation depth threshold for considering second-order sidebands would be \SI{0.5}{\radian} for modulation-band noise coupling and \SI{1.2}{\radian} for heterodyne-band noise coupling.

\bibliography{main}

\end{document}